\documentclass[preprint,12pt]{article}
\usepackage{graphicx}
\usepackage{amssymb}
\usepackage{amsmath}
\usepackage{siunitx}
\usepackage[T1]{fontenc}
\usepackage{physics}
\usepackage{hyperref}
\usepackage[authoryear,round]{natbib}
\usepackage{authblk}
\usepackage[lofdepth,lotdepth]{subfig}
\usepackage{lineno}
\usepackage{soul}
\hypersetup{pdftitle={ },
	pdfsubject={  },
	pdfauthor={J.M. Scherer},
	pdfkeywords={ }
	bookmarks=false,
	pdfpagemode=UseNone,
	linkcolor=blue,
	citecolor=blue,
	colorlinks=true,
	bookmarksopen=false,
	bookmarksopenlevel=1,
	urlcolor=blue}

\title{Grain-size dependence of plastic-brittle transgranular fracture}

\author[1]{Jean-Michel Scherer\thanks{Corresponding author at: Mines Paris, Universit\'e PSL, Centre des Mat\'eriaux (MAT), UMR7633 CNRS, 91003 Evry, France. \texttt{jean-michel.scherer@minesparis.psl.eu}}\footnote{This work was started when JMS held a position at the California Institute of Technology}}
\author[2]{Mythreyi Ramesh}
\author[3]{Blaise Bourdin}
\author[4]{Kaushik Bhattacharya}
\affil[1]{Mines Paris, Universit\'e PSL, Centre des Mat\'eriaux (MAT), UMR7633 CNRS, Evry, 91003, France}
\affil[2]{Northwestern University, Evanston, IL 60208 USA}
\affil[3]{McMaster University, Hamilton, Ontario L8S 4K1, Canada}
\affil[4]{California Institute of Technology, Pasadena, CA 91125, USA}

\date{}

\begin{document}

\maketitle

\begin{abstract}
\noindent The role of grain size in determining fracture toughness in metals is incompletely understood with apparently contradictory experimental observations.  We study this grain-size dependence computationally by building a model that combines the phase-field formulation of fracture mechanics with dislocation density-based crystal plasticity.  We apply the model to cleavage fracture of body-centered cubic materials in plane strain conditions, and find non-monotonic grain-size dependence of plastic-brittle transgranular fracture.  We find two mechanisms at play.  The first is the nucleation of failure due to cross-slip in critically located grains within transgranular band of localized deformation, and this follows the classical Hall-Petch law that predicts a higher failure stress for smaller grains.  The second is the resistance to the propagation of a mode I crack, where grain boundaries can potentially pin a crack, and this follows an inverse Hall-Petch law with higher toughness for larger grains.  The result of the competition between the two mechanisms gives rise to non-monotonic behavior and reconciles the apparently contradictory experimental observations.
\end{abstract}

\section*{Keywords}
phase-field fracture, crystal plasticity, transgranular fracture, Hall-Petch size effect, strength, fracture toughness

\section*{Highlights}
\begin{itemize}
    \item Explanation of the non-monotonic grain-size dependence of plastic-brittle transgranular fracture. 
    \item A phase-field model of fracture is coupled to dislocation density-based crystal plasticity.
    \item The model predicts a Hall-Petch size effect on the yield, failure stress and crack nucleation.
    \item The model predicts an inverse Hall-Petch size effect on fracture toughness.
    \item Large grains significantly increase the fracture toughness in bimodal microstructures.
\end{itemize}


\section{Introduction}
\label{sec:introduction}

Many metals and alloys with body-centered cubic, hexagonal closed-packed and non-face-centered cubic crystal structures fracture in a so-called plastic-brittle manner, where the yield behavior is followed by limited macroscopic plastic  deformation before fracture (e.g.,~\cite{pineau2016failure1,ashby1983mechanisms}).  Further, in many of these metals and alloys, the fracture surface is not dimpled as in the classical ``ductile'' fracture of face-centered cubic metals and alloys, but characterized by transgranular cleavage or relatively smooth fracture surfaces along certain crystallographic planes.  An important question in these materials is to understand their fracture toughness and how this fracture toughness depends on the grain size.

The literature is unclear on this dependence.  Based on an approximate stress analysis,~\cite{curry1978grain} used the model developed by~\cite{ritchie1973relationship} to predict the grain-size dependence of fracture toughness. This analysis suggests that the fracture toughness would be a non-monotonic function of grain size, with a peak toughness at an intermediate grain size of $\sim$\SI{5}{\micro\meter} for mild steel.    Other arguments suggests different trends.  Since fracture follows yield, and the yield strength increases with decreasing grain size due to the Hall-Petch effect, we expect that the toughness would also increase with decreasing grain size.  This is supported by the so-called Zener-Stroh and Cotrell mechanisms~\citep{zener1948micro,stroh1954formation,stroh1955formation,cottrell1958theory}.  Alternately, we anticipate propagating cracks to be pinned at the grain boundaries due to elastic and plastic heterogeneity~\citep{ming1989crack}, and therefore expect the toughness to increase with increasing grain size.

\begin{figure}
	\centering
	\subfloat[]{
		\includegraphics[width=.48\textwidth]{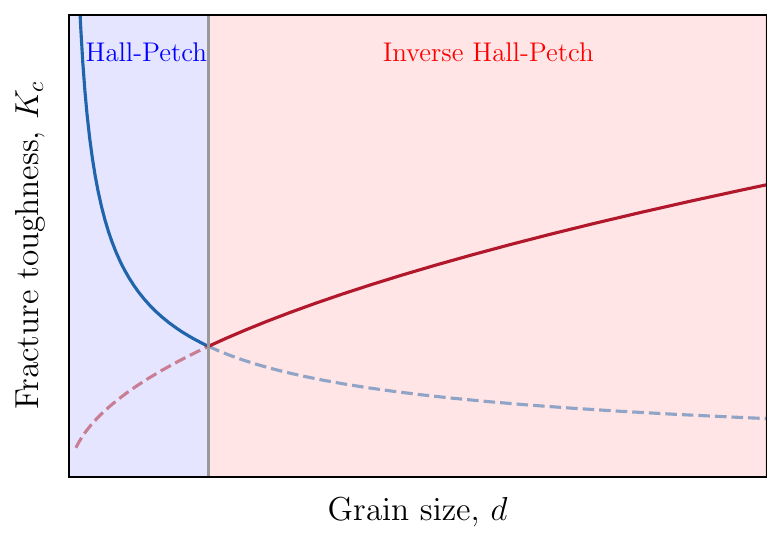}
		\label{subfig:reiser_hartmaier}
	}
	\subfloat[]{
		\includegraphics[width=0.49\textwidth]{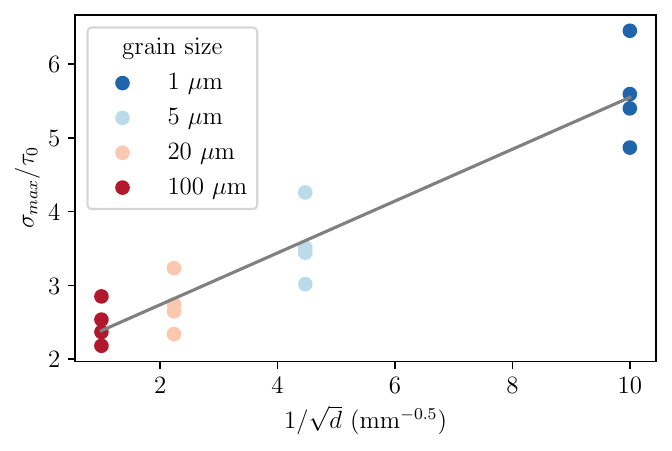}
		\label{subfig:hall_petch}
	}\\
	\subfloat[]{
		\includegraphics[width=0.85\textwidth]{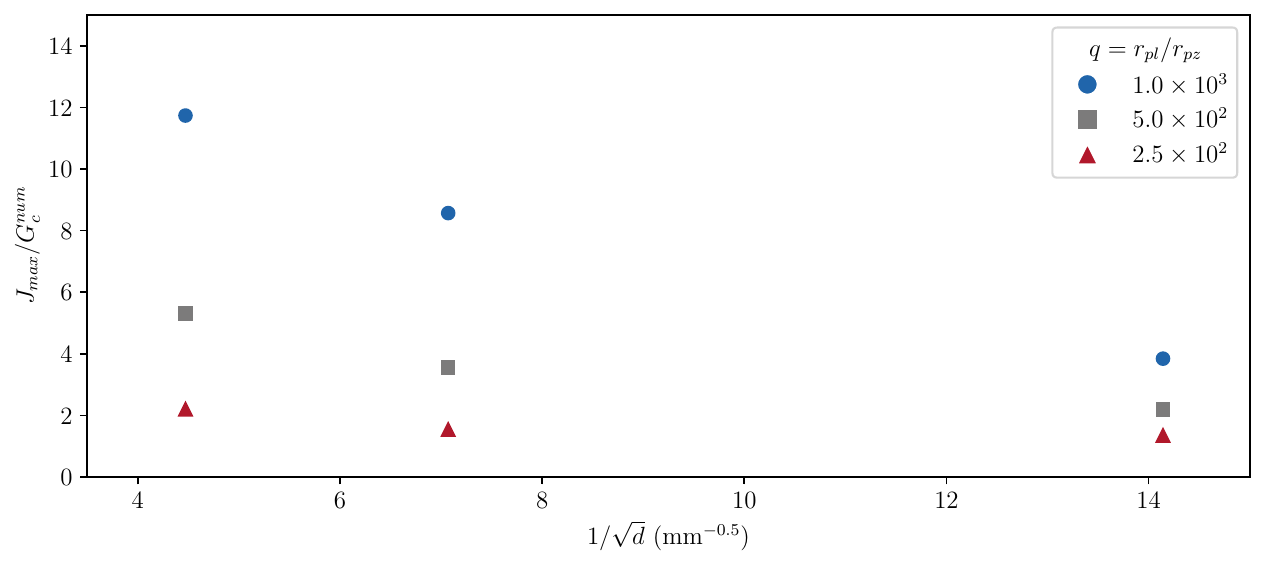}
		\label{subfig:inverse_hall_petch}
	}	
	\caption{(a) Fracture toughness as a function of grain size displaying a Hall-Petch regime for small grain sizes and an inverse Hall-Petch regime for large grain sizes. The fracture toughness reaches a minimum value at an intermediate grain size~(adapted from~\cite{reiser2020elucidating}). (b) Normalized peak stress $\sigma_{max}$ as a function of the inverse square root of the mean grain diameter. (c) Normalized $J$-integral as a function of the inverse square root of the mean grain diameter.}
	\label{fig:surfing_reiser_hartmeier}
\end{figure}
Recently,~\cite{reiser2020elucidating} reviewed the literature on the grain size effect on the fracture toughness and the ductile-brittle transition temperature (DBTT) of polycrystalline metals. The data that they collected shows that the fracture toughness can either decrease, increase or remain unaffected by the grain size depending on the material and the  range of grain sizes investigated. The DBTT shows a similar non-monotonic behavior with respect to the grain size, but with an opposite trend. For small grains, the DBTT increases when the grain size increases. Then, it reaches a maximum value at an intermediate grain size (between 20 and \SI{200}{\micro\meter}), before decreasing again for larger grains. Conversely, as shown in Figure~\ref{subfig:reiser_hartmaier}, the fracture toughness displays a minimum value at an intermediate grain size.  Notice that this is the opposite of the trend predicted by~\cite{curry1978grain}, and suggests a cross-over between the Hall-Petch effect and grain boundary pinning.  Finally,~\cite{qiao2003cleavage} measured the fracture toughness of polycrystalline Fe-2\%Si alloy with grain size from 1 to \SI{10}{\milli\meter}. No discernable grain size effect was observed on the fracture toughness for such large grain sizes. This suggests a saturation of the grain size effect for sufficiently large grains.

\cite{reiser2020elucidating} attribute this non-monotonic dependence to the competition between two competing mechanisms. The first mechanism is the confinement of the plastic zone at the crack tip by the grain boundaries, which act as obstacles for gliding dislocations. The smaller the grain size, the earlier the dislocations pile-up at grain boundaries and the sooner the nucleation of new dislocations at the crack tip gets inhibited. This mechanism thus induces a decrease in the fracture toughness when the grain size decreases (later referred to as \textit{inverse Hall-Petch size effect}, right domain in Figure~\ref{subfig:reiser_hartmaier}). The second mechanism is the nucleation of dislocations where the crack front intersects with the grain boundaries. The smaller the grain size, the more intersection points there are and the more dislocations that can be nucleated. This mechanism thus induces an increase in the fracture toughness when the grain size decreases, which is consistent with the Hall-Petch size effect (left domain in Figure~\ref{subfig:reiser_hartmaier}). According to~\cite{reiser2020elucidating} the former mechanism dominates at larger grain sizes, while the latter prevails at lower grain sizes.  However, it is unclear why this is the case.

In this work, we use a computational model involving both fracture and  plasticity in polycrystalline body centered cubic materials to examine the role of grain size and texture in plastic-brittle fracture.  We treat fracture using 
the variational phase field approach~\citep{francfort1998revisiting,bourdin2000numerical,bourdin2008variational}.  We combine this with crystal visco-plasticity through dislocation density evolution equations~\citep{hoc2001polycrystal}.  We study these in finite deformation.  Our formulation builds on that of~\cite{brach2019phase} who considered fracture of elastic-perfectly plastic solids in small strains.

Our main results are summarized in Figures~\ref{subfig:hall_petch} and~\ref{subfig:inverse_hall_petch}.  We conduct two sets of calculations.  In the first, we subject a representative volume consisting of a number of grains to uniaxial tension, and study the stress-strain behavior.  We observe an initial elastic regime, which is followed by yield and hardening till the stress reaches a peak followed by a load drop.  The peak stress is associated with the nucleation of cracks.  We observe, as shown in Figure~\ref{subfig:hall_petch} that the peak stress follows the Hall-Petch relation and decreases with the (square-root of the) grain size.  Briefly, the  heterogeneity in slip activity due to plastic anisotropy and cross-hardening leads to stress concentration and further, crack nucleation.  The heterogeneity in slip is larger in larger grains and this leads to easier crack nucleation.  Thus, the crack nucleation threshold decreases with the (square-root of the) grain size.  Once a crack has nucleated it has to propagate through multiple grains.  This is studied in the second set of calculations, where we compute the critical energy release rate necessary for crack propagation over multiple grains in a polycrystalline specimen using the surfing boundary conditions following~\cite{hossain2014effective}.  We observe, as shown in Figure~\ref{subfig:inverse_hall_petch}, that the toughness as measured by the critical energy release rate necessary for crack propagation increases with increasing grain size. In other words, crack propagation follows the inverse Hall Petch relationship.   This is true for various ratios of the relative strength to toughness ratios of the single crystal, though the effect decreases with increasing brittle behavior.   Briefly, crack propagation is intragranular (especially in larger grain specimens), and the crack gets pinned at the grain boundaries as it tries to transition from one grain to another due to the heterogeneity induced by anisotropy. 

In summary, our results show that crack nucleation follows the Hall Petch relationship whereas crack propagation follows the inverse Hall Petch relationship.  This leads to non-monotonic fracture resistance, with increased resistance at smaller and larger grain sizes and a minimum value at intermediate values.  These results are consistent with the experimental observations described above.

The paper is organized as follows. We provide an extended literature review on the fracture of metals in Section~\ref{sec:background}.  In Section~\ref{sec:model}, we present the variational phase-field model of fracture coupled to crystal plasticity. In Section~\ref{sec:nucleation}, we discuss the nucleation of cleavage cracks in polycrystals loaded in plane strain tension for grain sizes ranging from \SI{1}{\micro\meter} to \SI{100}{\micro\meter}. In Section~\ref{sec:propagation}, we study the propagation of cleavage cracks in polycrystals with grain sizes ranging from \SI{5}{\micro\meter} to \SI{50}{\micro\meter}. 
We conclude in Section~\ref{sec:conclusion}. 

\section{Background}
\label{sec:background}

\begin{figure}
	\centering
	\includegraphics[width=0.9\textwidth]{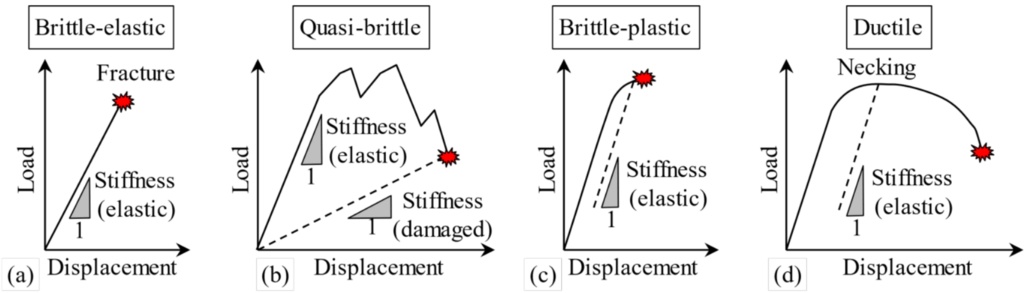}
	\caption{Macroscopic failure modes as viewed on a load vs.\ displacement curve. (Reproduced from~\cite{gourges2023mealor}.)}
	\label{fig:failure_modes}
\end{figure}
Fracture or rupture of materials is classified into various categories as shown in Figure~\ref{fig:failure_modes}, based on the macroscopic load vs.\ displacement curve (e.g.,~\cite{gourges2023mealor}).  Typically, ceramics fail in a brittle or quasi-brittle manner, while metals and alloys fail in a plastic-brittle or ductile manner.  Ductile fracture, typical in metals with face-centered cubic crystal structure, involves significant macroscopic plastic deformation and the fracture surface is dimpled.  On the other hand, plastic-brittle failure, often observed in non-FCC metals, especially at low temperatures, involves very little macroscopic plastic deformation and the fracture surface is not dimpled.

The mechanisms of the fracture of metals has been a recurrent topic in materials science over the last century. These mechanisms were reviewed by~\cite{hollomon1946problems,zener1948micro,orowan1949fracture,hall1953brittle,petch1954fracture,stroh1957theory,friedel1959propagation,cottrell1963bakerian,low1963fracture,ashby1983mechanisms,becker2002mechanisms,pineau2016failure1,pineau2016failure2,pineau2016failure3}.  Ductile fracture proceeds by the \textit{nucleation, growth and coalescence} of voids leading to the final rupture of the material. On the other hand, plastic-brittle fracture does not involve voids even though there is significant plasticity at the subgranular level, and is the result of complex interplay of local plastic deformation, crack nucleation and propagation. This can be either intergranular or transgranular.

\begin{figure}
	\centering
	\subfloat[Cleavage 1]{
		\includegraphics[width=0.23\textwidth]{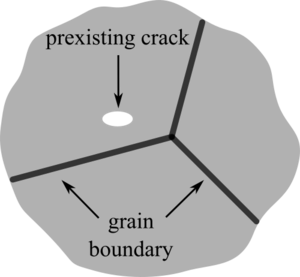}
		\label{subfig:preexisting_defect}
	}
	\subfloat[Zener-Stroh and Cottrell mechanisms (Cleavage 2)]{
		\includegraphics[width=0.23\textwidth]{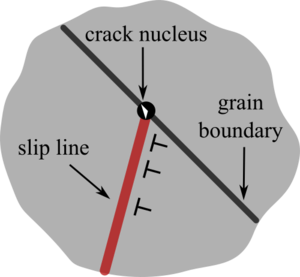}
		\includegraphics[width=0.23\textwidth]{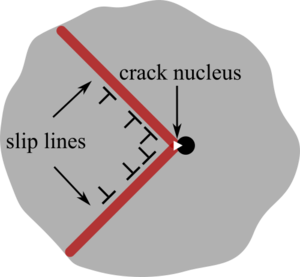}
		\label{subfig:stroh_cottrell}
	}
	\subfloat[Cleavage 3]{
		\includegraphics[width=0.23\textwidth]{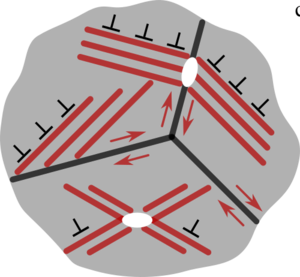}
		\label{subfig:creep}
	}
	\caption{Three regimes of cleavage fracture according to~\cite{ashby1983mechanisms}. (a) Cleavage fracture initiated from a preexisting crack. (b) Zener-Stroh crack nucleation at intersecting slip line and grain boundary~\citep{zener1948micro,stroh1954formation,stroh1955formation} and Cottrell crack nucleation at intersecting slip lines~\citep{cottrell1958theory} (Cleavage 2). (c) Generalized plasticity or creep preceeding cleavage fracture (Cleavage 3).}
	\label{fig:fracture_mechanisms}
\end{figure}

As shown in Figure~\ref{fig:fracture_mechanisms},~\cite{ashby1983mechanisms} classifies transgranular cleavage mechanisms into three \textit{fields} denoted as cleavage 1, 2 and 3, respectively. Cleavage 1, corresponds to the situation where the material contains preexisting cracks (growth defects, corrosion, abrasion, \textit{etc}) which are large enough to trigger fracture upon loading without the need of plastic deformation. The failure stress is then given by
\begin{align}
	\sigma_f \simeq \sqrt{\frac{EG_c}{\pi c}}
\end{align} 
where $E$ is the Young's modulus, $G_c$ the critical energy release rate~\citep{griffith1921vi} and $2c$ the preexisting crack length. Cleavage 2, involves the nucleation of cracks in regions of stress concentrations induced by plastic deformation. In the absence of sufficiently large preexisting cracks, the stress may reach the yield strength of the material and trigger plastic slip and/or twinning. While the stresses are relaxed above and below slip bands, or twins, particularly large stresses may be generated at their leading edges. Several mechanisms have been proposed. The Zener-Stroh mechanism~\citep{zener1948micro,stroh1954formation,stroh1955formation} (see Figure~\ref{subfig:stroh_cottrell}) corresponds to the pile-up of dislocations at a grain boundary. The Cottrell mechanism~\citep{cottrell1958theory} (see Figure~\ref{subfig:stroh_cottrell}) corresponds to the pile-up of dislocations at the intersection of two slip lines. In both cases, a crack nucleus might be formed due to the large opening stresses at the root of the pile-up, which can exceed the cohesive strength of the material. Other mechanisms, such as the termination of a tilt boundary (dislocation wall) within a single crystal~\citep{cohen1954dislocations,stroh1958cleavage} as observed experimentally in zinc under compressive load by~\cite{gilman1954mechanism} or the interaction of twins~\citep{hull1960twinning,honda1961cleavage,mcmahon1965initiation} have also been proposed as crack nucleation mechanisms. In polycrystals, the size of such cracks is proportional to the grain size $d$, because this is the characteristic size of the internal stresses fluctuations. A crack nucleation stress can then be defined as
\begin{align}
	\sigma^* \simeq \sqrt{\frac{EG_c}{\pi d}}
	\label{eq:cleavage_2}
\end{align}
A crack propagates as soon as it is nucleated if the flow or twinning stress ($\sigma_y$) exceeds $\sigma^*$. Conversely, if $\sigma^* > \sigma_y$, new cracks formed by slip or twinning do not lead to immediate failure. The stress needs to be raised further to lead to failure. Cleavage 3, as shown in Figure~\ref{subfig:creep}, corresponds to the regime generally at higher temperature, where the material undergoes generalized plasticity or creep before cleavage fracture. In this case, general plasticity and/or grain boundary sliding can lead to the growth of a preexisting crack or nucleation of a grain boundary crack. Once the crack reaches a critical size, on increasing loading by work hardening, it grows as a cleavage crack.

In this work, we focus on the cleavage 2 regime which involves two competing mechanisms. Plastic strains are necessary to produce the stress concentration required for cleavage onset, but large plastic deformations relax the stresses at the crack tip and inhibit crack growth~\citep{friedel1959propagation,rice1974ductile,ohr1985electron}. This synergy between plasticity and fracture is particularly interesting in polycrystals as it is associated with size effects. These size effects can be leveraged in order to enhance fracture properties such as strength, ductility or toughness. For an ideal single-phase material, the main characteristic length of the polycrystalline microstructure is the grain size. Since~\cite{eshelby1951xli}, it has been known that this size plays a crucial role on the yield point of a polycrystal, because it limits the length of dislocation pile-ups. These pile-ups are formed as dislocations gliding in the same direction are arrested by a grain boundary.~\cite{eshelby1951xli} showed that, by limiting the length of a pile-up, which is proportional to the number of dislocations in the pile-up, the stress concentration at the root of the pile-up is reduced. Therefore, the applied stress required to let the dislocation pile-up overcome the grain boundary obstacle increases as the grain size decreases. This analysis gives the well known relation
\begin{align}
	\sigma_y = \sigma_0 + \frac{K}{\sqrt{d}}
	\label{eq:hall_petch}
\end{align} 
which was later validated experimentally by~\cite{hall1951deformation} and~\cite{petch1953cleavage,petch1958ductile}. The yield stress of the single crystal is denoted $\sigma_0$, and $K$ is a material constant with units $\SI{}{\pascal\sqrt{\meter}}$. Eq.~\eqref{eq:hall_petch} closely resembles the definition of the failure stress for cleavage 2 in Eq.~\eqref{eq:cleavage_2}. Both the yield and the failure stress involve an inverse square root dependence on the grain size. This is due to the underlying mechanism, which in both cases involves a dislocation pile-up, the size of which is constrained by the grain size.~\cite{hull1961effect} and \cite{worthington1966slip} showed that, in addition to an increase of the yield and failure stresses, a decrease of the grain size also leads to an increase in the strain at fracture in a 3\% silicon-iron alloy for grain sizes between 9.8 and \SI{625}{\micro\meter}. The same observation was made later in NiAl by~\cite{schulson1983brittle} in a grain size range from 8 to \SI{125}{\micro\meter}. This effect apparently contradicts the strength-ductility trade-off~\citep{ritchie2011conflicts} frequently encountered in the design of high-strength materials.

The numerical modelling of the Hall-Petch size effect in polycrystals has been the subject of numerous studies. Following~\cite{kocks2003physics,de2010grain,HAOUALA201872}, the Hall-Petch effect can be captured in crystal plasticity models by introducing a term $K_s/\delta$ in the Kocks-Mecking evolution equation of dislocation densities, where $\delta$ is the distance of a material point to the closest grain boundary. This approach results in an increased strain hardening capacity in the vicinity of grain boundaries. The lower the grain size, the larger the grain boundary surface area to volume ratio and hence, the greater the grain boundary effect on the macroscopic behaviour. Alternative approaches have also been proposed to capture grain size effects in crystal plasticity models.~\cite{cordero2010size} proposed higher-order (generalised) continuum crystal plasticity models to predict scaling laws such as the Hall-Petch size effect (scaling exponent equal to -0.5). In their analysis, three concurrent models where compared based on the scaling exponent range they could predict. Their \textit{micro-curl} model can produce scaling laws with exponent ranging from -2 to 0. Their Cosserat model leads to scaling exponents between -1 and 0. Finally, their \textit{curl-H$^p$} model invariably leads to a scaling law exponent of -2. Grain size effects involving similar strain gradient plasticity models were obtained by~\cite{evers2003strain,borg2007strain,counts2008predicting,hamid2018dislocation,sun2019strain,haouala2020simulation,pai2022study}. In the context of brittle fracture,~\cite{clayton2015phase} were able to capture Hall-Petch size effect in a phase-field model of anisotropic fracture in the absence of plasticity.~\cite{giang2018dislocation} proposed a strain gradient plasticity model coupled with a cohesive zone model to study size effects in the nucleation of micro-cracks in a ferritic steel.

Once a crack is nucleated and reaches a critical size, it propagates following cleavage planes.   As the crack tip reaches a grain boundary and seeks to pass from one grain to the neighbor, the cleavage plane and elastic modulus changes due to the change in crystallographic orientation.  This can potentially pin the crack.  In the brittle setting,~\cite{ming1989crack} showed that cracks can get pinned at interfaces across which the elastic stiffness changes (from compliant to stiff). ~\cite{hossain2014effective} built on this using variational phase field calculations and showed that in a layered material with varying elastic modulus, the overall fracture toughness increases with increasing layer spacing.  Briefly, when the layer spacing is very small (compared to the K-dominant zone), the crack does not sense the elastic heterogeneity and is not pinned.  However, when the layer spacing grows, the crack begins to sense the elastic heterogeneity and is pinned at the interface.  This effect eventually saturates for large spacing.  This suggests that crack propagation will follow an inverse Hall-Petch relationships. ~\cite{qiao2003cleavage} also suggest that the transition from one cleavage plane to another across a grain boundary provides additional resistance to crack growth, though how this depends on grain size is unclear.  Similarly, all of this is in the brittle setting, and how crystal plasticity affects these considerations is also unclear.

\section{Variational phase-field approach to fracture coupled to crystal plasticity}
\label{sec:model}

\subsection{Kinematics in the finite strain setting}

Following~\cite{lee1969elastic,mandel1973thermodynamics}, the kinematics is described by the deformation gradient $\boldsymbol{F} = \partial\boldsymbol{x}/\partial\boldsymbol{X}$, where $\boldsymbol{x}$ and $\boldsymbol{X}$ represent the position vector in the current and initial configurations, respectively. The deformation gradient is multiplicatively decomposed into an elastic deformation gradient $\boldsymbol{F}^e$ and a plastic deformation gradient $\boldsymbol{F}^p$ such that
\begin{align}
	\boldsymbol{F} = \boldsymbol{F}^e \cdot \boldsymbol{F}^p.
\end{align}
The Green-Lagrange elastic strain tensor $\boldsymbol{E}$ is related to the elastic deformation gradient by $\boldsymbol{E} = (\boldsymbol{F}^e \cdot \boldsymbol{F}^{e^T} - \boldsymbol{1})/2$. The strain rate tensor $\boldsymbol{L}$ is defined as $\boldsymbol{L} = \dot{\boldsymbol{F}} \cdot \boldsymbol{F}^{-1}$ and can be expressed in terms of the elastic and plastic strain rates as follows
\begin{align}
	&\boldsymbol{L} = \boldsymbol{L}^e + \boldsymbol{F^e}\cdot\boldsymbol{L}^p\cdot\boldsymbol{F}^{e^{-1}}, \\
	&\mathrm{with} \quad \boldsymbol{L}^e = \dot{\boldsymbol{F}}^e \cdot \boldsymbol{F}^{e^{-1}} \quad \mathrm{and} \quad \boldsymbol{L}^p =  \dot{\boldsymbol{F}}^p \cdot \boldsymbol{F}^{p^{-1}} \label{eq:Le}.
\end{align}
In this context, the Mandel stress tensor $\boldsymbol{\Pi}$, lying in the isoclinic intermediate configuration~\citep{mandel1973equations} is defined with respect to the Cauchy stress tensor $\boldsymbol{\sigma}$ as follows
\begin{align}
	\boldsymbol{\Pi} = \det(\boldsymbol{F}^e)\boldsymbol{F}^{e^T} \cdot \boldsymbol{\sigma} \cdot \boldsymbol{F}^{e^{-T}}.
\end{align}

\subsection{Variational setting}

The variational approach to fracture proposed by~\citep{francfort1998revisiting,bourdin2000numerical,bourdin2008variational} is based on the minimization of a functional that combines the elastic energy and the fracture energy. It was extended to the case of elastic-plastic materials by~\cite{brach2019phase,alessi_gradient_2014}. The phase-field variable $\alpha$ is introduced to regularize the crack surface. It takes the value 1 in the crack, 0 in the intact material and intermediate values near the crack lips. In this study, we extend this approach to the case of crystal visco-plasticity at finite strains. The variational formulation of the problem is given by the following functional
\begin{align}
	\label{eq:functional}
	\mathcal{F}_\ell(\alpha,\pmb{u}) = \int_{\Omega} \left( \frac{1}{2}\boldsymbol{E} : a(\alpha)\mathbb{C} : \boldsymbol{E} + \frac{3G_c}{8} \left( \ell |\nabla \alpha|^2 + \frac{\alpha}{\ell} \right) + b(\alpha)\int_{0}^ {\bar{t}}\pi \left(\boldsymbol{L}^p, \boldsymbol{\Pi}\right)\, \mathrm{d}t \right) \, \mathrm{d}V
\end{align}  
where $a(\alpha)$ is the stiffness degradation function $a(\alpha) = (1-\alpha)^2 + \eta_e$ and $\mathbb{C}$ the fourth order elastic modulus (which has  cubic symmetry in what follows). The numerical parameter $\eta_e = 10^{-6}$ is chosen such that a small residual stiffness remains inside the crack ($\alpha=1$). The critical energy release rate is denoted by $G_c$ and the phase-field length scale  is $\ell$. The visco-plastic dissipation is noted $\pi$ and depends on the plastic strain rate $\boldsymbol{L}^p$. The expression of the visco-plastic dissipation will be discussed in the next section. The function $b(\alpha)$ is the plastic degradation function and is chosen as $b(\alpha) = (1-\alpha)^2$.

At equilibrium, the displacement and phase-field variables are the solutions of the following minimization problem
\begin{align}
	(\alpha^*, \pmb{u}^*) = \underset{\pmb{u}\in\mathcal{K}_u,\, \dot{\alpha}\geq 0}{\text{argmin}}\mathcal{F}_\ell(\alpha,\pmb{u})\label{eq:minimization}
\end{align}
where $\mathcal{K}_u$ is the set of kinematically admissible displacements fields subject to the boundary conditions. The phase-field variable $\alpha$ is subject to the constraint $\dot{\alpha}\geq 0$ to ensure the irreversibility of fracture. In the context of crystal visco-plasticity, the minimization problem~\eqref{eq:minimization} requires the computation of the Green-Lagrange elastic strain tensor $\boldsymbol{E}$, the Mandel stress tensor $\boldsymbol{\Pi}$ and the plastic strain rate $\boldsymbol{L}^p$. The constitutive evolution equations are derived in the next section.

\subsection{Crystal visco-plasticity model}
We adopt a dislocation density-based crystal plasticity model for body-centered (BCC) crystals. To showcase the coupling between phase-field fracture and crystal plasticity, while keeping the number of material parameters to a minimum, we consider a set of crystal plasticity constitutive equations similar to the one used by~\cite{hoc2001polycrystal}. We consider two families of slip systems, $\langle111\rangle\{110\}$ and $\langle111\rangle\{112\}$, with a total of 24 slip systems. Throughout the paper, a slip plane normal of a given slip system $s$ is noted $\boldsymbol{n}^s$ and the associated slip direction is noted $\boldsymbol{m}^s$. The resolved shear stress $\tau^s$ is computed from the Mandel stress tensor $\boldsymbol{\Pi}$ as follows
\begin{align}
	\tau^s = \boldsymbol{\Pi}:(\boldsymbol{m}^s\otimes\boldsymbol{n}^s).
\end{align}
The plastic strain rate $\boldsymbol{L}^p$ is given by the sum of plastic strain rates $\dot{\gamma}^s$ on each slip system $s$ as follows
\begin{align}
	\boldsymbol{L}^p = \dot{\boldsymbol{F}}^p\cdot\boldsymbol{F}^{p^{-1}} = \sum_{s=1}^{24} \dot{\gamma}^s \boldsymbol{m}^s\otimes\boldsymbol{n}^s.
\end{align}
From Eq.~\eqref{eq:Le}, the rate of change of the elastic part of the deformation gradient $\boldsymbol{F}^e$ is given by
\begin{align}
	\dot{\boldsymbol{F}}^e = \dot{\boldsymbol{F}}\cdot\boldsymbol{F}^{-1}\cdot\boldsymbol{F}^e - \boldsymbol{F}^e\cdot \boldsymbol{L}^p.
	\label{eq:Fe}
\end{align}
A visco-plastic flow rule is used to describe the evolution of the plastic strain rates $\dot{\gamma}^s$ as a function of the resolved shear stresses $\tau^s$ as follows
\begin{align}
	\dot{\gamma}^s = \text{sign}(\tau^s) \left\langle \frac{|\tau^s| - \tau_c^s}{K} \right\rangle^n
	\label{eq:viscoplastic_flow}
\end{align}
where $\langle\bullet\rangle=\max(\bullet,0)$ and $K$ and $n$ are material parameters characterizing the viscosity of the material. The critical resolved shear stress $\tau_c^s$ is given by an extended Taylor law~\citep{franciosi1980latent} as follows
\begin{align}
	\tau_c^s = \tau_0 + \mu b \sqrt{\sum_{u=1}^{24}a^{su}\rho^u}
\end{align}
where $\tau_0$ is the initial critical resolved shear stress, $\mu$ is the shear modulus, $b$ is the magnitude of the Burgers vector, $a^{su}$ is a $24 \times 24$ interaction matrix and $\rho^u$ is the dislocation density on slip system $u$. Following~\cite{hoc2001polycrystal}, the interaction matrix is decomposed in 8 independent terms as described in Table~\ref{tab:interaction}.
\renewcommand{\arraystretch}{1.5}
\begin{table}
	\centering
	\caption{Decomposition of the interaction matrix $a^{su}$.}
	\label{tab:interaction}
	\begin{tabular}{cccc}
        \hline
		Plane & $\{110\}\cap\{110\}$ & $\{110\}\cap\{112\}$ & $\{112\}\cap\{112\}$ \\
		\hline \hline
		Self & $a_0$ & -- & $k_{s_0} a_0$ \\
		Colinear & $k_1 a_0$ & $k_{p_1}a_0$ & $k_{s_0} k_1 a_0$ \\
		Non-colinear  & $k_2 k_1 a_0$ & $k_{p_2}k_{p_1}a_0$ & $k_{s_0} k_2 k_1 a_0$ \\
		\hline
	\end{tabular}
\end{table}

The dislocation densities evolve according to the following equations
\begin{align}
	\dot{\rho}^s = \frac{|\dot{\gamma}^s|}{b} \left( \max \left( \frac{K_s}{\delta}, \frac{\sqrt{\sum_{u\notin\mathrm{coplanar}(s)}a^{su}\rho^u}}{K_{obstacle}} + \frac{\sqrt{\sum_{u\in\mathrm{coplanar}(s)}a^{su}\rho^u}}{K_{coplanar}} \right) - yb\rho^s \right).
	\label{eq:dislocation_density}
\end{align}
Following~\cite{haouala2020effect}, the Hall-Petch effect on the yield stress can be modeled with the term $K_s / \delta$, where $K_s$ is a non-dimensional material parameter and $\delta (\pmb{x})$ is the distance between the material point located at $\pmb{x}$ and the closest grain boundary. Therefore $\delta$ is a field which depends on the position $\boldsymbol{x}$. For the sake of simplicity $\delta$ is not updated during the simulation. The term $K_s / \delta$ increases significantly close to a grain boundary and is thus responsible for a sharp increase of dislocation densities in the vicinity of grain boundaries. This effect is a phenomenological way of modelling dislocation pile-ups at grain boundaries. One could also consider using strain-gradient plasticity models to account for this size effect~\citep{cordero2010size}. The terms $K_{obstacle}$ and $K_{coplanar}$ are non-dimensional material parameters and $\mathrm{coplanar}(s)$ is the set of slip systems coplanar with the slip system $s$. The negative term in equation~\eqref{eq:dislocation_density} models the dynamic recovery of dislocations with $y$, a non-dimensional material parameter.

The visco-plastic dissipation $\pi$ entering the energy functional in equation~\eqref{eq:functional} is given by the expression
\begin{align}
	\pi\left(\dot{\boldsymbol{F}}^p\cdot\boldsymbol{F}^{p^{-1}},\boldsymbol{\Pi}\right) = \sum_{s=1}^{24} \dot{\gamma}^s \tau^s.
\end{align}

\subsection{Non-dimensionalization of physical quantities}
In practice, it is convenient to work with non-dimensional quantities to avoid ill-conditioned systems of equations. Furthermore, working with non-dimensional quantities allows comparison between results for different materials. We introduce the following non-dimensional quantities
\begin{align}
	\tilde{\boldsymbol{x}} = \frac{\boldsymbol{x}}{\eta L_0},\quad \tilde{\mathbb{C}} = \frac{\mathbb{C}}{E_0},\quad \tilde{\tau_0} = \frac{\tau_0}{E_0},\quad \tilde{G_c} = \frac{G_c}{G_{c_0}} = \frac{G_c}{E_0\eta L_0},\quad \tilde{\ell} = \frac{\ell}{\eta L_0},\quad \tilde{\delta} = \frac{\eta \delta}{b},\quad \tilde{\rho}^s = \rho^s b^2
	\label{eq:non-dimensional}
\end{align}
where $L_0$ is a characteristic length, $\eta$ is a non-dimensional scaling factor and $E_0$ is a characteristic Young's modulus. 

We introduce a scaling factor $\eta$ which may be regarded as an inverse grain size.  A challenge of studying specimens with various grain sizes is that as we change the grain size, we also need to vary either the number of grains or the physical domain. These make computation difficult. So, we rescale each problem with the inverse grain size so that the number of grains and computational cell size remains the same even as we vary the grain size. One consequence of this is that the length scale in the phase field fracture is also scaled. We ensure that our mesh is fine enough in order to resolve the crack in all cases including the case with the largest grain size (and smallest rescaled phase field fracture size).
To simplify notations, the tilde is omitted in the following.

Following~\cite{anderson2005fracture}, we define the plastic zone and fracture process zone radii and their ratio as follows
\begin{align}
	&r_{pl} = \alpha \frac{E_0G_{c_0}}{\tau_0^2} = \alpha \frac{E_0^2\eta L_0}{\tau_0^2}, \quad r_{pz} = \beta \frac{G_{c_0}}{\tau_0} = \beta \frac{E_0\eta L_0}{\tau_0}, \label{eq:process_zone}\\
	&q = \frac{r_{pl}}{r_{pz}} = \frac{\alpha}{\beta}\frac{E_0}{\tau_0}, \quad \left (\frac{\alpha}{\beta}\sim 1 \right ) \label{eq:quotient}
\end{align}
where $\alpha$ and $\beta$ are geometrical factors which depend on the mechanical state (\textit{e.g.}, plane stress, plane strain).  These two geometrical parameters are neglected with the assumption that $\alpha/\beta \sim 1$. The plastic zone radius $r_{pl}$ characterizes the size of the region where plastic deformation occurs while the fracture process zone radius $r_{pz}$ characterizes the size of the region where fracture mechanisms are active. The ratio $q$ is a measure of the relative size of these two regions. The quantity $q$ is intimately correlated with the ductility of the material (as will be clear in the later sections).  

\begin{figure}
	\centering
	\includegraphics[width=0.5\textwidth]{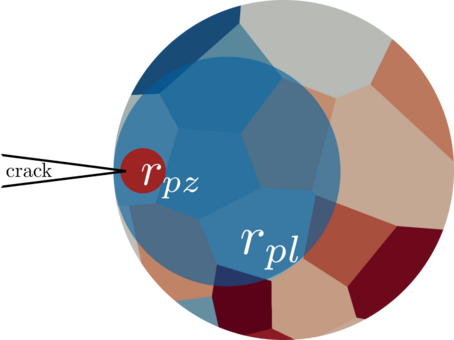}
	\caption{Schematic representation of the plastic and process zones sizes, $r_{pl}$ and $r_{pz}$, at the crack tip in a polycrystal. The plastic zone is the region where dislocations can glide due to the stress field at the crack tip. The process zone corresponds to the volume in which damage mechanisms take place. The process zone size has been exaggerated for the sake of clarity.}
	\label{fig:pl_zone}
\end{figure}

Figure~\ref{fig:pl_zone} shows a schematic representation of the plastic and process zones within a polycrystal microstructure. The process zone size is usually smaller than the grain size. However, depending on the yield strength and grain size, the plastic zone can either engulf multiple grains or be contained within a single grain. This is a reason for taking the characteristic length $L_0$ to be of the order of magnitude of the grain size. In practice, for many metallic alloys, the ratio $q$ is in the range of $10^2$ to $10^4$. 

\subsection{Numerical implementation}
\label{subsec:numerical_implementation}

The model is implemented with the finite element software \texttt{FEniCS}~\citep{logg2012automated}. A fixed point algorithm is used to solve the mechanical equilibrium and the phase-field evolution problems. At each iteration, both problems are solved independently. The phase-field $\alpha$ is fixed when solving for $\pmb{u}$. Conversly, $\pmb{u}$ is fixed when solving for $\alpha$. The nonlinear mechanical equilibrium problem is solved with a Newton solver. The phase-field evolution problem is solved with the bound constrained optimization solver \texttt{PETScTao}~\citep{balaypetsc}. The \texttt{gradam} code~\citep{jean_michel_scherer_2021_5764329,scherer2022assessment} was used to setup both problems, handle the iterations of the fixed point algorithm, apply boundary conditions and run the simulations. The crystal plasticity material integration is performed with a semi-implicit time integration scheme implemented in \texttt{MFront}~\citep{helfer2015introducing}. A Newton solver is used to find the increment of integration variables, namely $\Delta F^e$, $\Delta \gamma^s$ and $\Delta \rho^s$, with the set of evolution equations given by Eq.~\eqref{eq:Fe},~\eqref{eq:viscoplastic_flow} and~\eqref{eq:dislocation_density}. The crystal plasticity implementation was adapted from~\cite{KAFO01_2023} and is available as open-source code~\citep{IUXPA3_2025}. The interface \texttt{mgis.fenics}~\citep{helfer2020mfrontgenericinterfacesupport} is used to handle the communication between \texttt{FEniCS} and \texttt{MFront}. 

\section{Crack nucleation in polycrystals}
\label{sec:nucleation}

\subsection{Setting}
\label{subsec:nucleation_setting}

\begin{figure}
	\centering
	\subfloat{  
	    \includegraphics[width=0.7\textwidth,trim=18cm 6cm 13.8cm 6cm,clip]{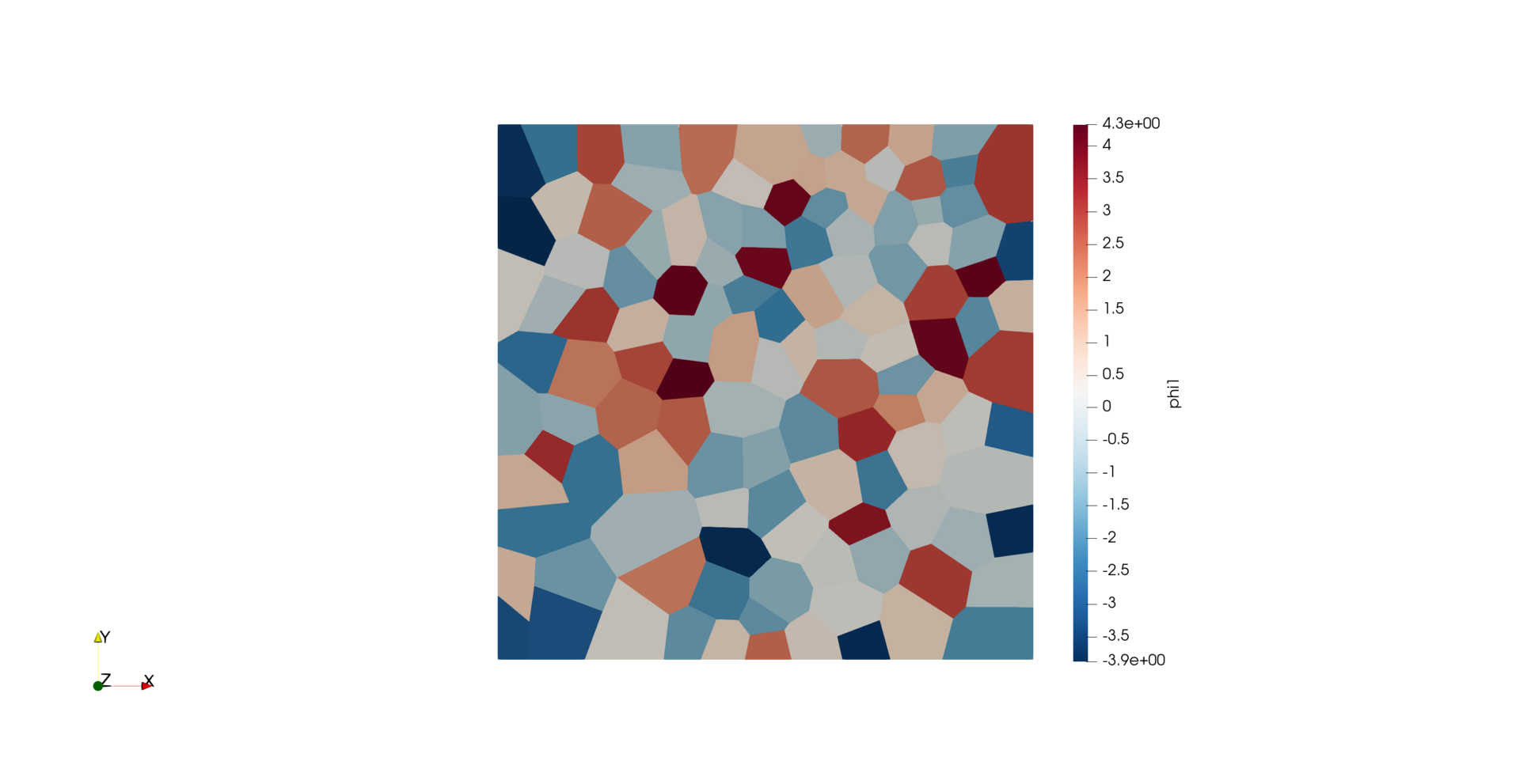}
	}
	\caption{One of the polycrystalline microstructures used to simulate crack nucleation. Tension is applied along the horizontal direction with plane strain conditions. The colorscale represents the first Euler angle $\phi_1$.}
	\label{fig:tension_microstructure}
\end{figure}

We first investigate the effect of grain size on the yield strength and the nucleation of cracks in polycrystals. We consider four realisations of 2D polycrystal microstructures with a random distribution of grains and uniform distribution of orientations. These synthetic microstructures are generated using \texttt{Neper}~\citep{quey2011large,quey2022neper}. For each microstructure, the grain morphology is fixed and the grain size is scaled over two orders of magnitude, from \SI{1}{\micro\meter} to \SI{100}{\micro\meter}. Each microstructure is composed of approximately 100 grains. The norm of the Burgers vector is fixed to $b=2.54\times$\SI{e-10}{\meter}. The first realisation is shown in Figure~\ref{fig:tension_microstructure}. Tension is applied along the horizontal direction in 2D plane strain conditions.
The non-dimensionalized material parameters are given in Table~\ref{tab:parameters}. With this set of parameters, the ratio $q$ is equal to $10^3$.
\renewcommand{\arraystretch}{1.5}
\begin{table}
    \centering
	\caption{Non-dimensionalized material parameters.}
    \label{tab:parameters}
    \begin{tabular}{ccccccccccccccc}
        \hline
        $E$ & $\nu$ & $G$ & $\tau_0$ & $G_{c}$ & $\ell$ & $\rho^s_0$ & $n$ & $K$ & $K_{obs}$ & $K_{cop}$ \\
        \hline \hline
        1 & 0.3 & 1.154 & $10^{-3}$ & $10^{-5}/\eta$ & $10^{-2}/\eta$ & $6.4516\times 10^{-8}$ & 10 & $5\times10^{-5}$ & 33 & 100 \\
        \hline \\ \hline
        $K_s$ & $y$ & $\mu$ & $a_0$ & $k_1$ & $k_2$ & $k_{p_1}$ & $k_{p_2}$ & $k_{s_0}$ \\
        \hline \hline
        5 & 3 & 1.154 &  0.4 & 1 & 1.15 & 1.05 & 1.05 & 1.3 \\
        \hline
    \end{tabular}
\end{table}

\subsection{Hall-Petch size effect}
\label{subsec:tension_hall_petch}

The polycrystalline microstructures are initially free of cracks. Given the material parameters in Table~\ref{tab:parameters}, the four grain sizes considered lead to a plastic-brittle behaviour with varying ductility on loading under unaxial tension in the horizontal direction.

\begin{figure}
	\centering
	\includegraphics[width=0.8\textwidth]{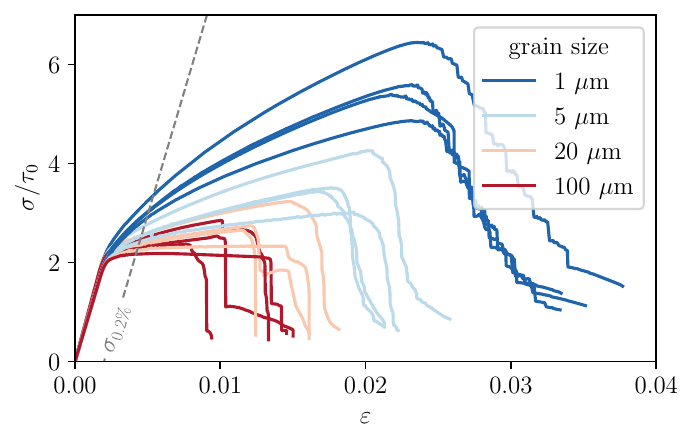}
	\caption{Stress-strain curves for four different grain sizes. Four different microstructures are considered for each grain size. The intersection between the dashed line and the solid lines is used to compute the $0.2\%$ plastic strain yield stress. The peak stress $\sigma_{max}$ is the maximum stress reached during the simulation.}
	\label{fig:tension_stress}
\end{figure}

The stress strain curves for the four grain sizes are shown in Figure~\ref{fig:tension_stress}. 
Note that we have four specimens (realisations) of polycrystals at each grain size. We see some  variation among the samples at all grain sizes, but the overall trend is clear.
The nominal yield strength, or the stress at 0.2\% plastic strain, increases with decreasing grain size as does the overall ductility. This kind of behaviour is comparable to experimental data on plastic-brittle materials~\citep{low1963fracture} such as in 3\% silicon-iron~\citep{hull1961effect,worthington1966slip}.  The variability that is observed across the four realisations for each grain size  is due to the random distribution of grains and the limited number of grains in each microstructure ($\sim$100). 

\begin{figure}
	\centering
	\subfloat[$d = \SI{20}{\micro\meter}$]{
	    \includegraphics[width=0.245\textwidth,trim=18cm 0cm 18cm 0cm,clip]{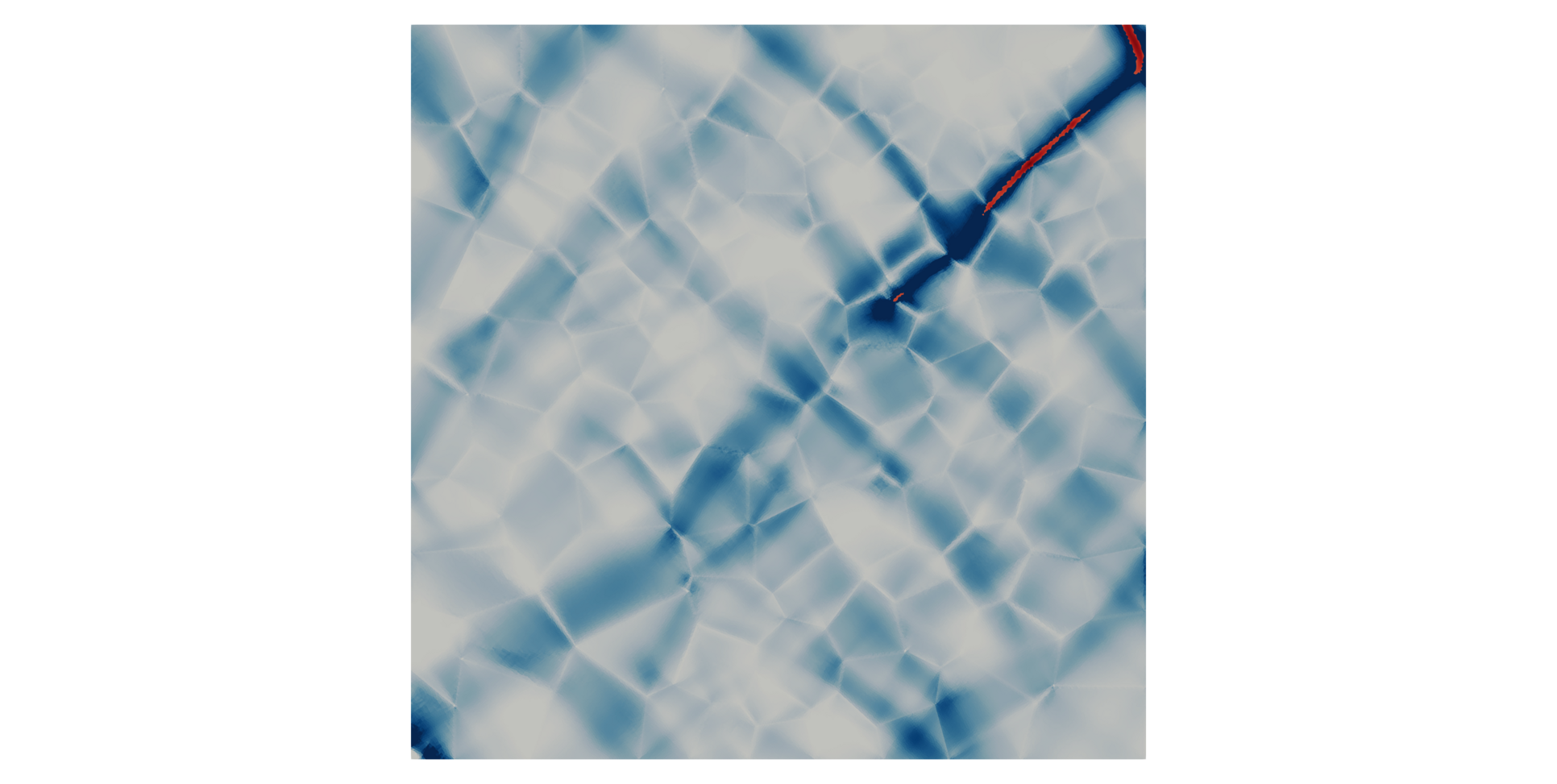}\hspace{.1cm}
		\includegraphics[width=0.245\textwidth,trim=18cm 0cm 18cm 0cm,clip]{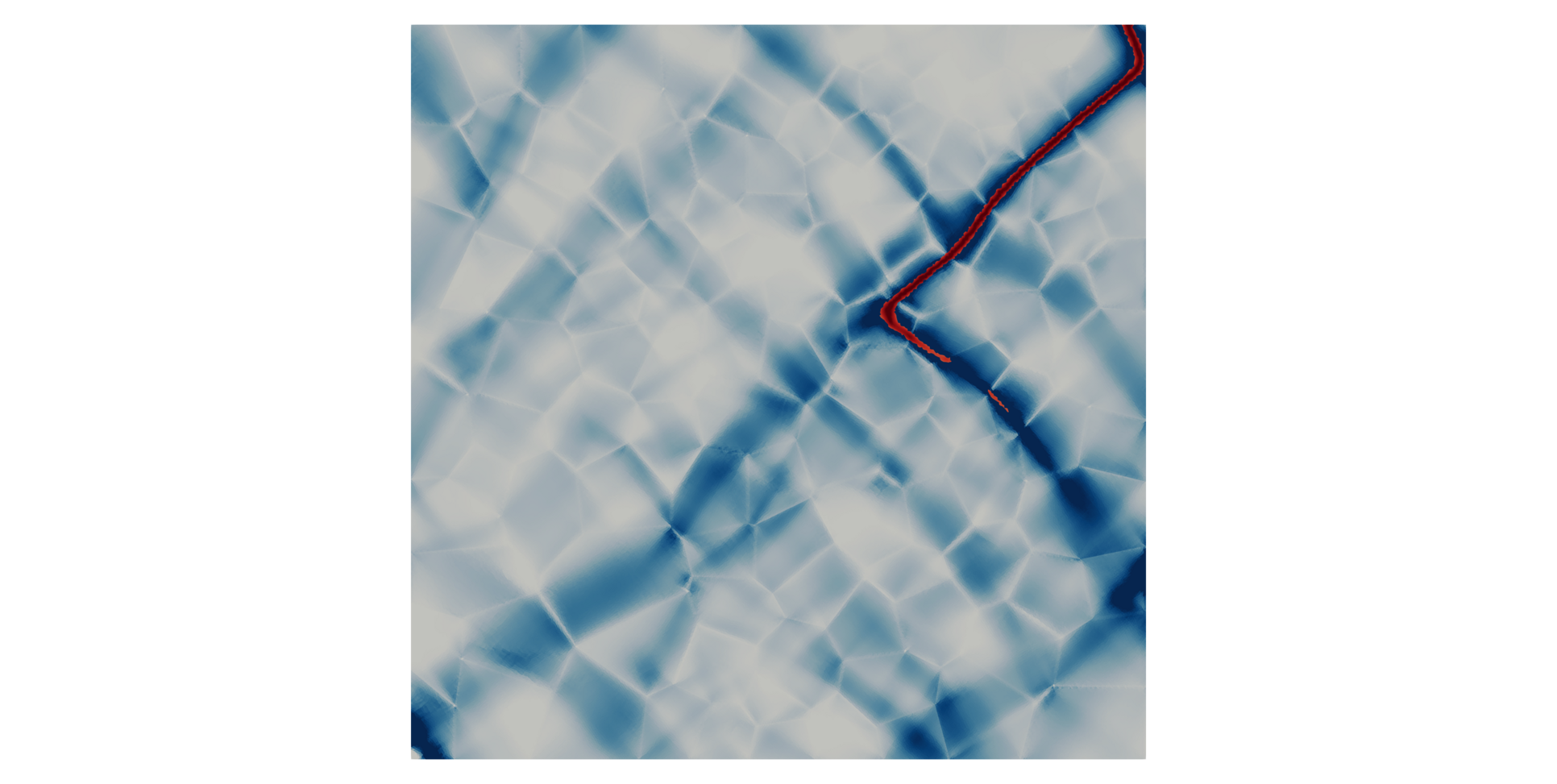}\hspace{.1cm}
		\includegraphics[width=0.245\textwidth,trim=18cm 0cm 18cm 0cm,clip]{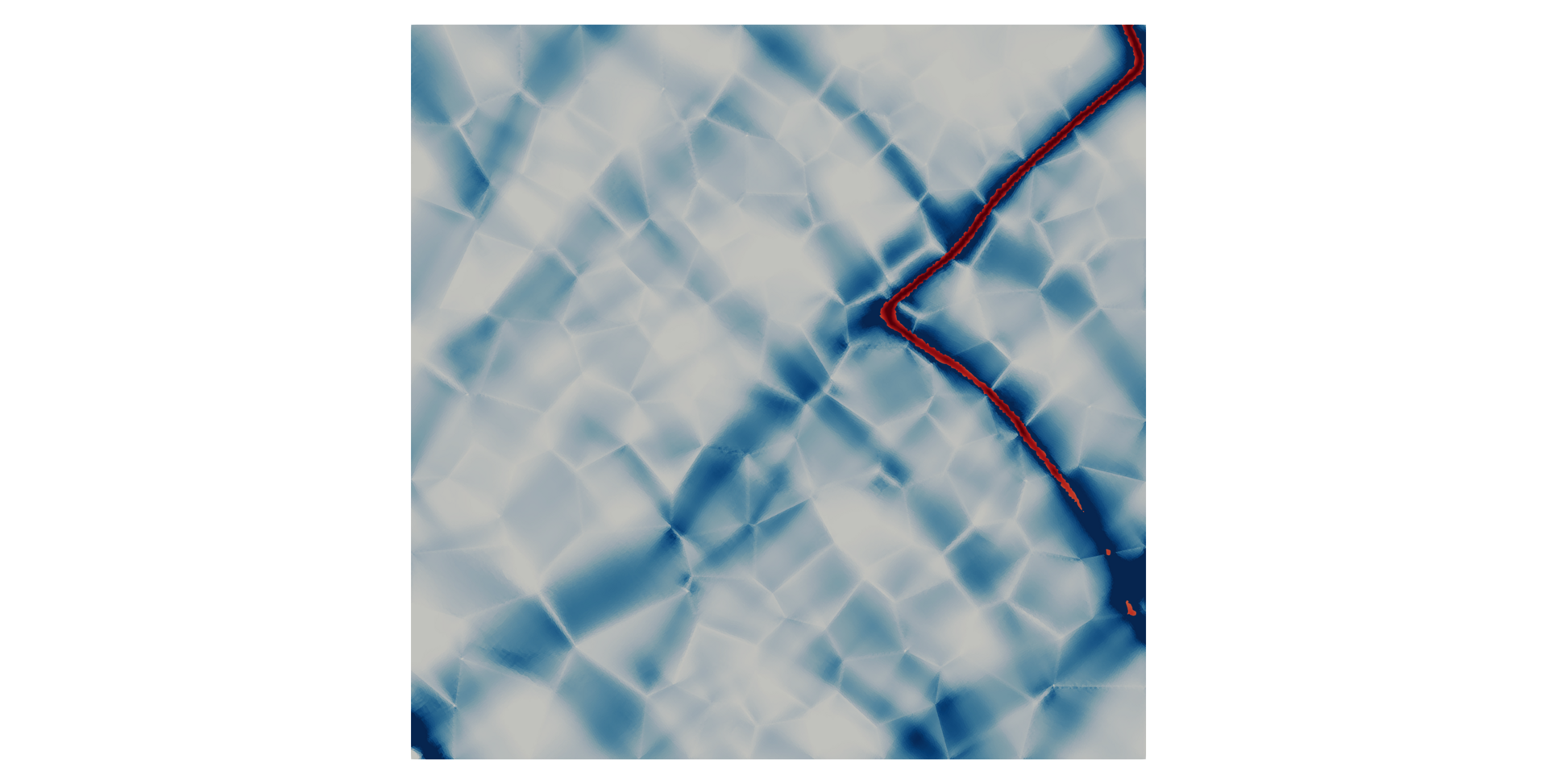}\hspace{.1cm}
		\includegraphics[width=0.245\textwidth,trim=18cm 0cm 18cm 0cm,clip]{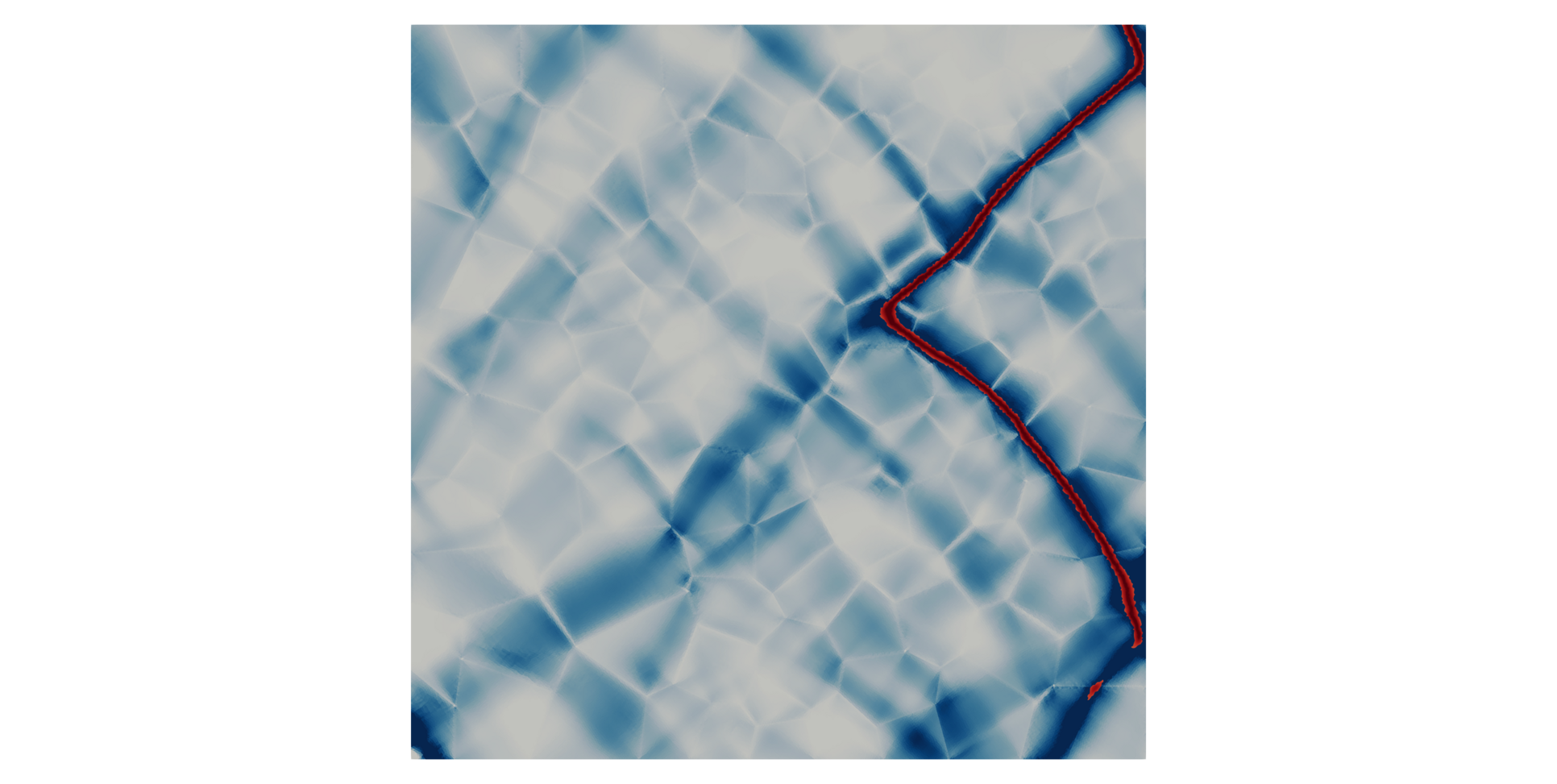}
		\label{subfig:tension_crack_nucleation_scale_2.0}
	}\\
	\subfloat[$d = \SI{1}{\micro\meter}$]{
	    \includegraphics[width=0.245\textwidth,trim=18cm 0cm 18cm 0cm,clip]{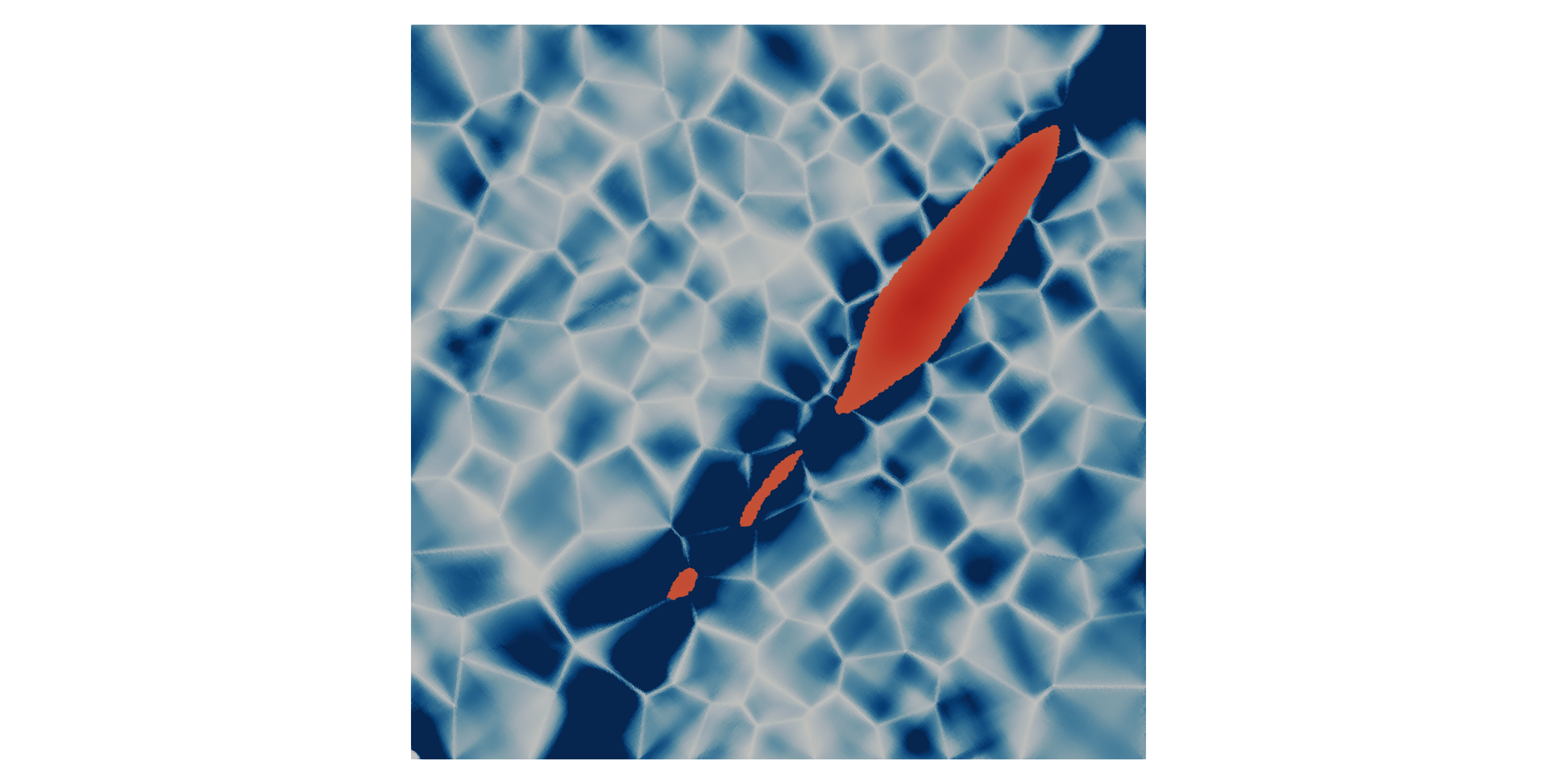}\hspace{.1cm}
		\includegraphics[width=0.245\textwidth,trim=18cm 0cm 18cm 0cm,clip]{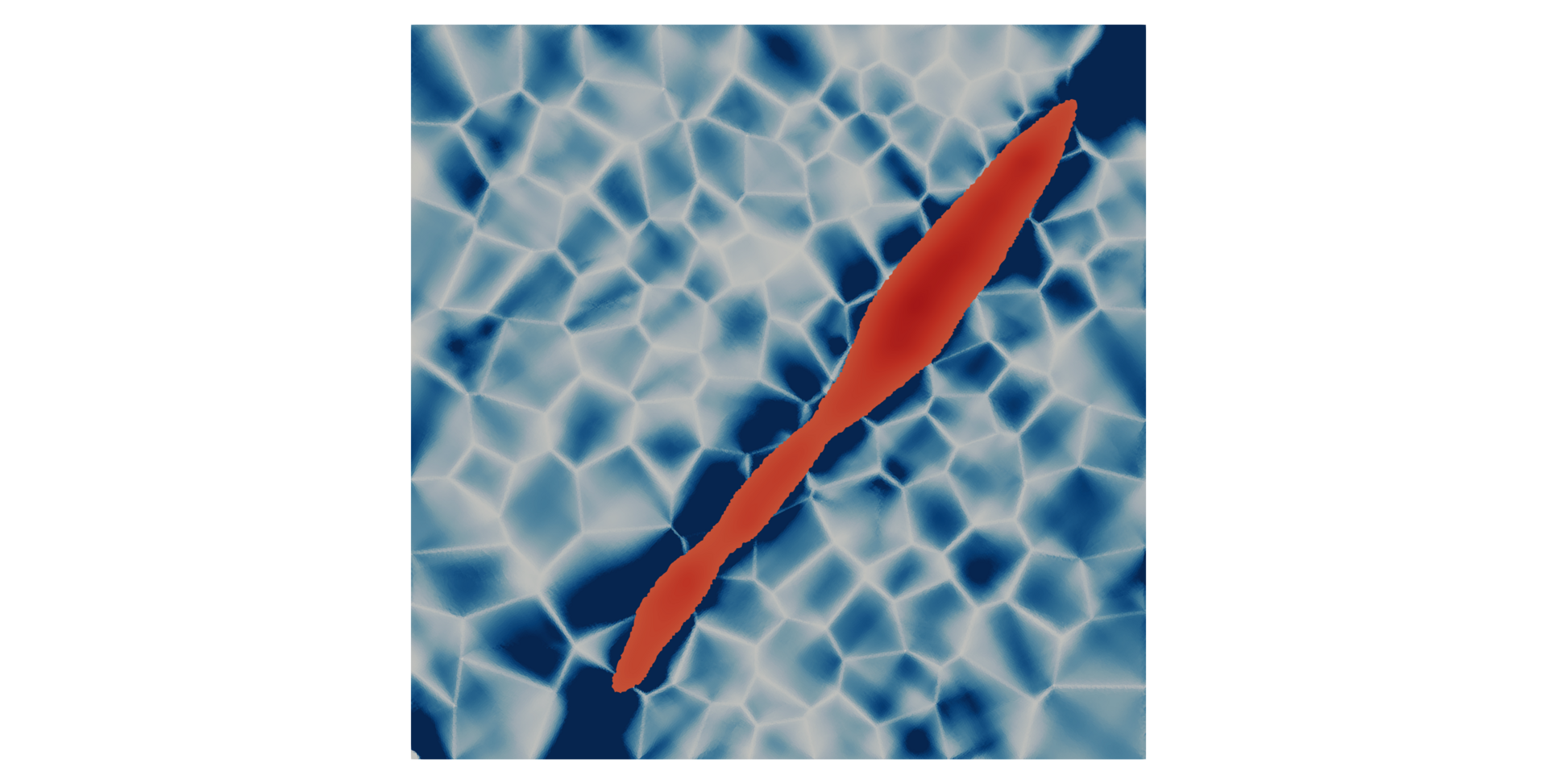}\hspace{.1cm}
		\includegraphics[width=0.245\textwidth,trim=18cm 0cm 18cm 0cm,clip]{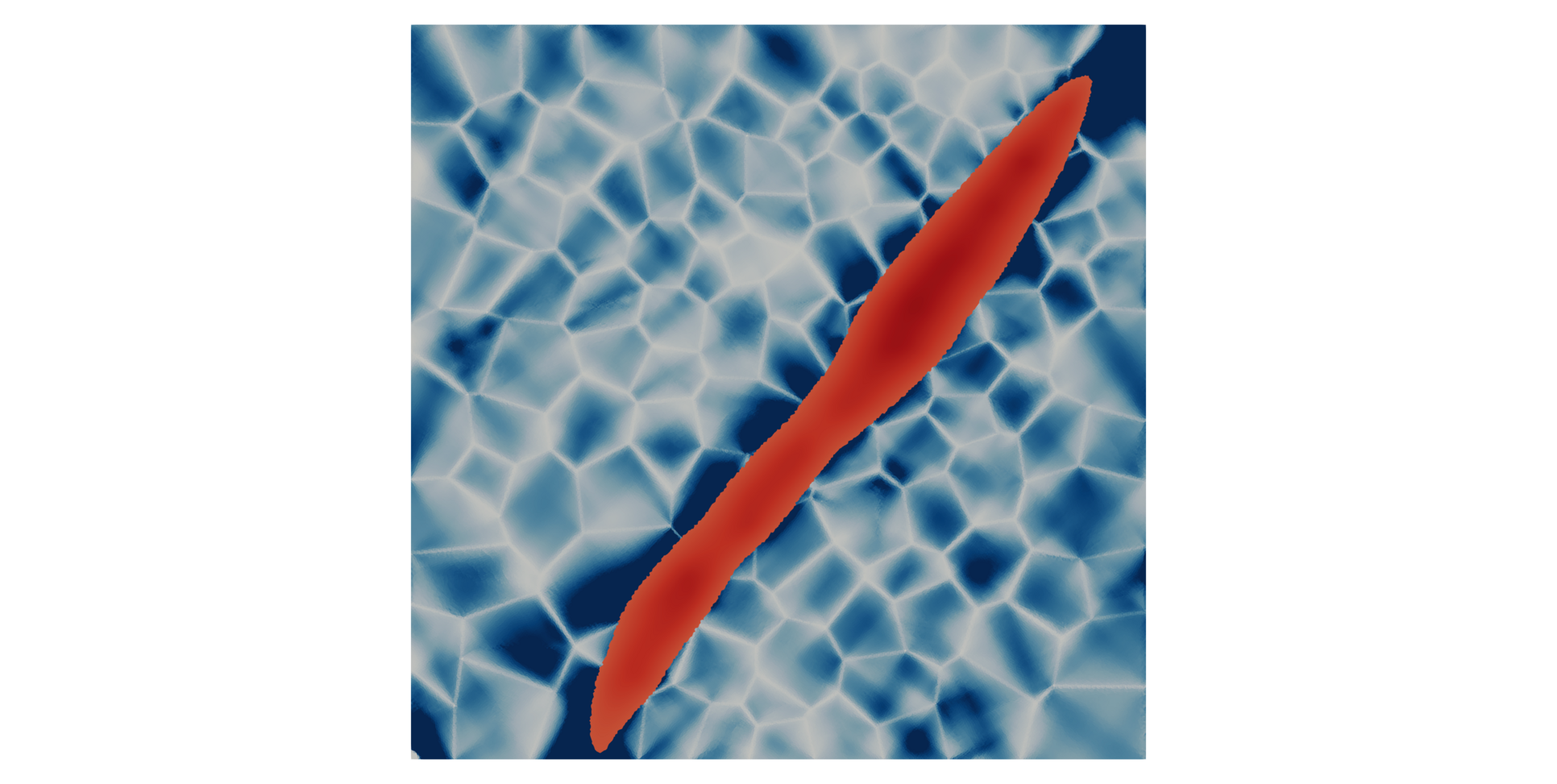}\hspace{.1cm}
		\includegraphics[width=0.245\textwidth,trim=18cm 0cm 18cm 0cm,clip]{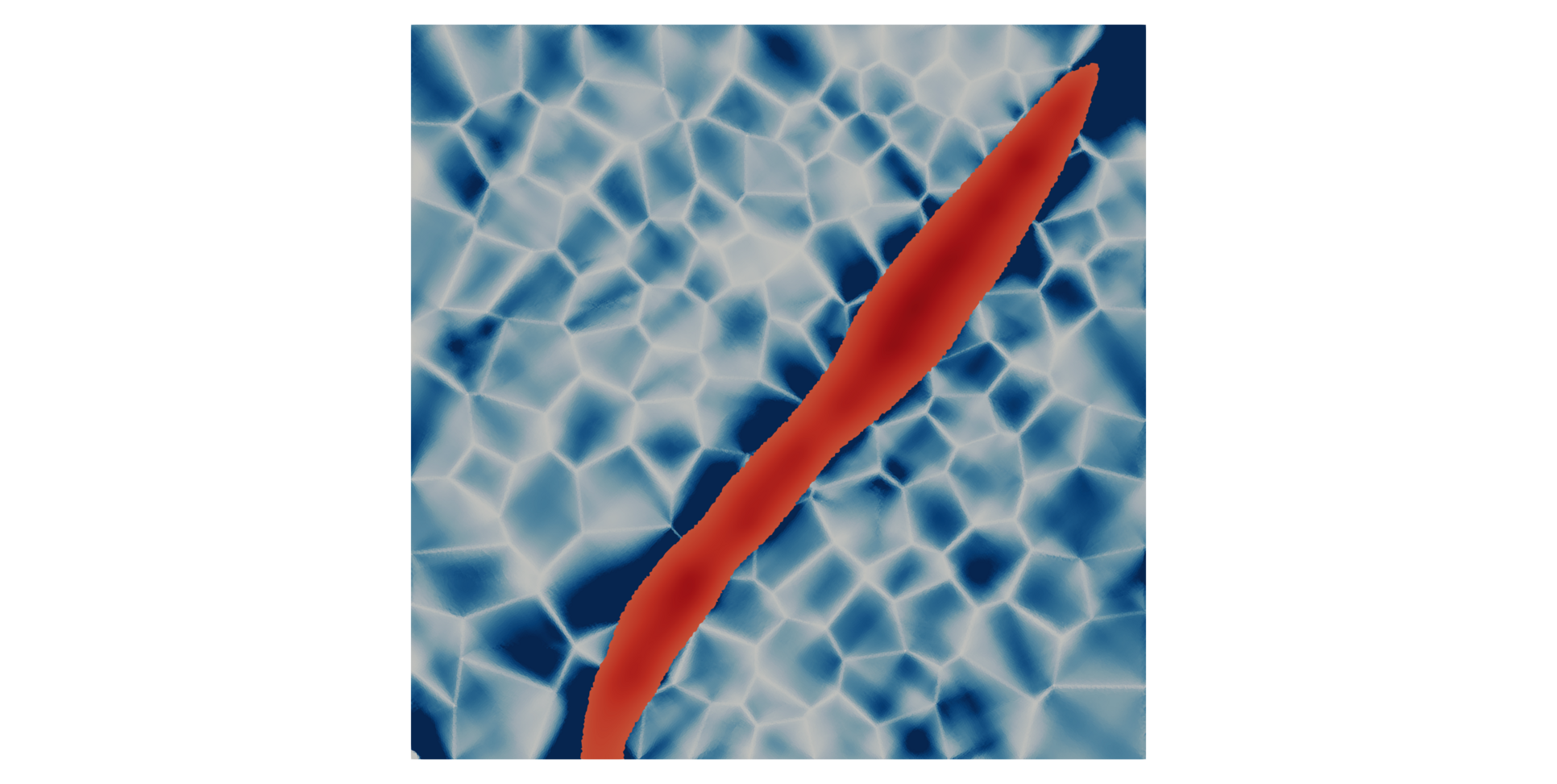}
		\label{subfig:tension_crack_nucleation_scale_0.1}
	}
	\caption{Crack nucleation history for two different grain sizes. The phase field variable $\alpha$ and accumulated plastic slip are shown for four different loading increments.}
	\label{fig:tension_crack_nucleation_histories}
\end{figure}

The microstructures with the largest grain size display the most abrupt decrease of the load past the peak stress. On the other hand, microstructures with smaller grains display a more gradual decrease of the load. These distinct stress evolutions are associated with different crack nucleation histories. In microstructures with larger grains, the cracks first nucleate  at several locations in an unstable way, \textit{i.e} the phase field variable jumps from 0 to 1 within a short loading increment. This is shown in the first snapshot of the crack history in Figure~\ref{subfig:tension_crack_nucleation_scale_2.0} for the grain size $d = \SI{20}{\micro\meter}$. In this figure, the contours where the phase field variable  is greater than 0.5 are plotted (red colorscale) on top of the accumulated plastic strain field (blue colorscale). After nucleation, the crack grows rapidly over a short period. During the course of crack propagation, new crack nuclei are formed at a distance ahead of the crack tip (see the second and third snapshots in Figure~\ref{subfig:tension_crack_nucleation_scale_2.0}). These crack nucleation events occur preferentially at grain boundaries. Ultimately, the crack propagates towards these new nucleation sites till it has crossed the whole polycrystal. As shown in Figure~\ref{subfig:tension_crack_nucleation_scale_0.1}, for $d = \SI{1}{\micro\meter}$, in microstructures with smaller grains, cracks nucleate at several locations almost simultaneously. Compared to larger grains, the increase of the phase field variable from 0 to 1 is more progressive. In fact, the different nucleation sites merge together even before $\alpha$ reaches 1 anywhere in the microstructure. This observation is consistent with the analysis presented in Section~\ref{sec:introduction}, where equation~\eqref{eq:cleavage_2} shows that increasing the grain size reduces $\sigma^*$ and thus favors crack propagation immediately after nucleation. 

\begin{figure}
	\subfloat{ 
	    \includegraphics[width=0.2\textwidth,trim=18cm 0cm 18cm 0cm,clip]{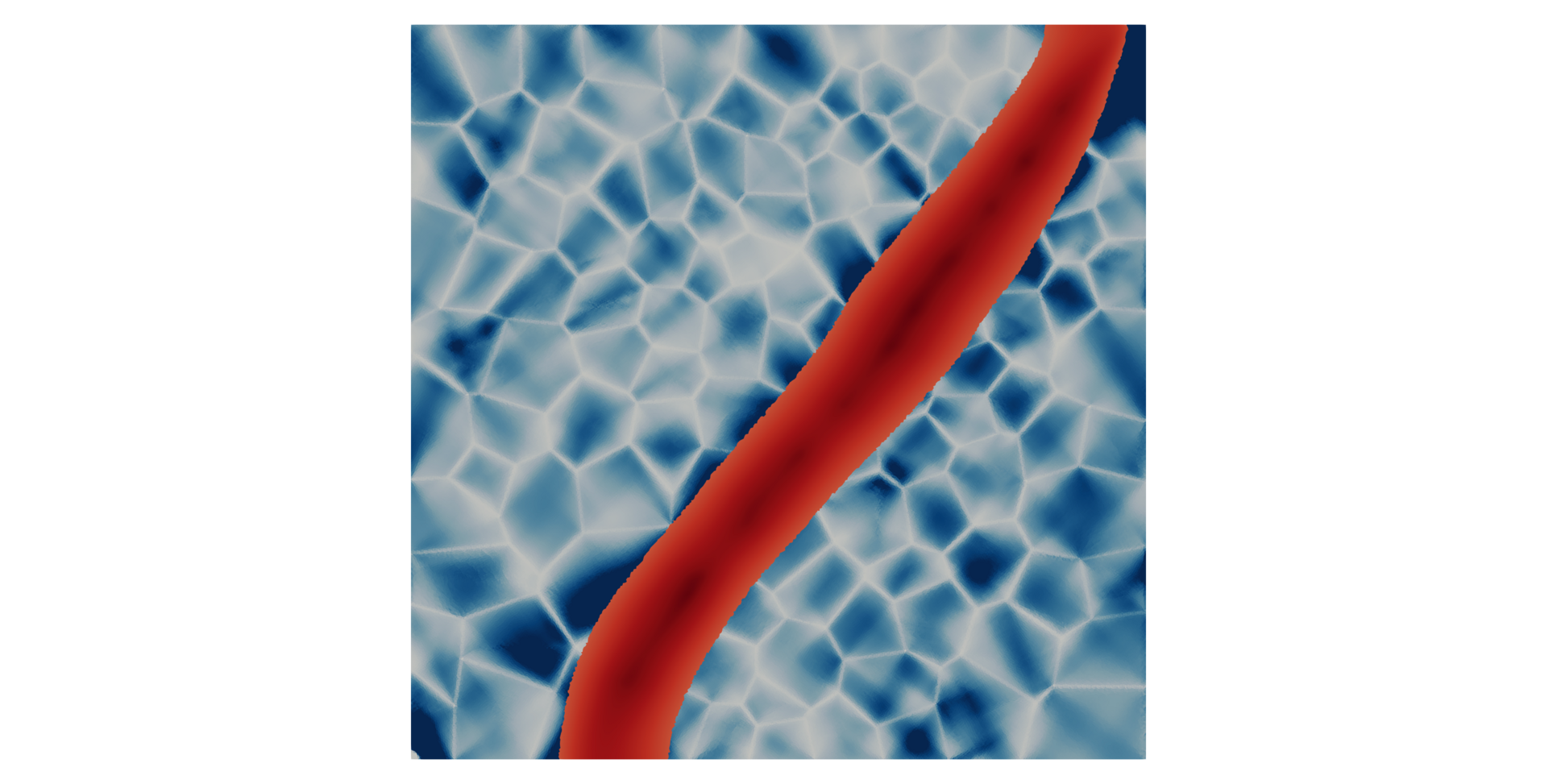}
	}
	\subfloat{ 
	    \includegraphics[width=0.2\textwidth,trim=18cm 0cm 18cm 0cm,clip]{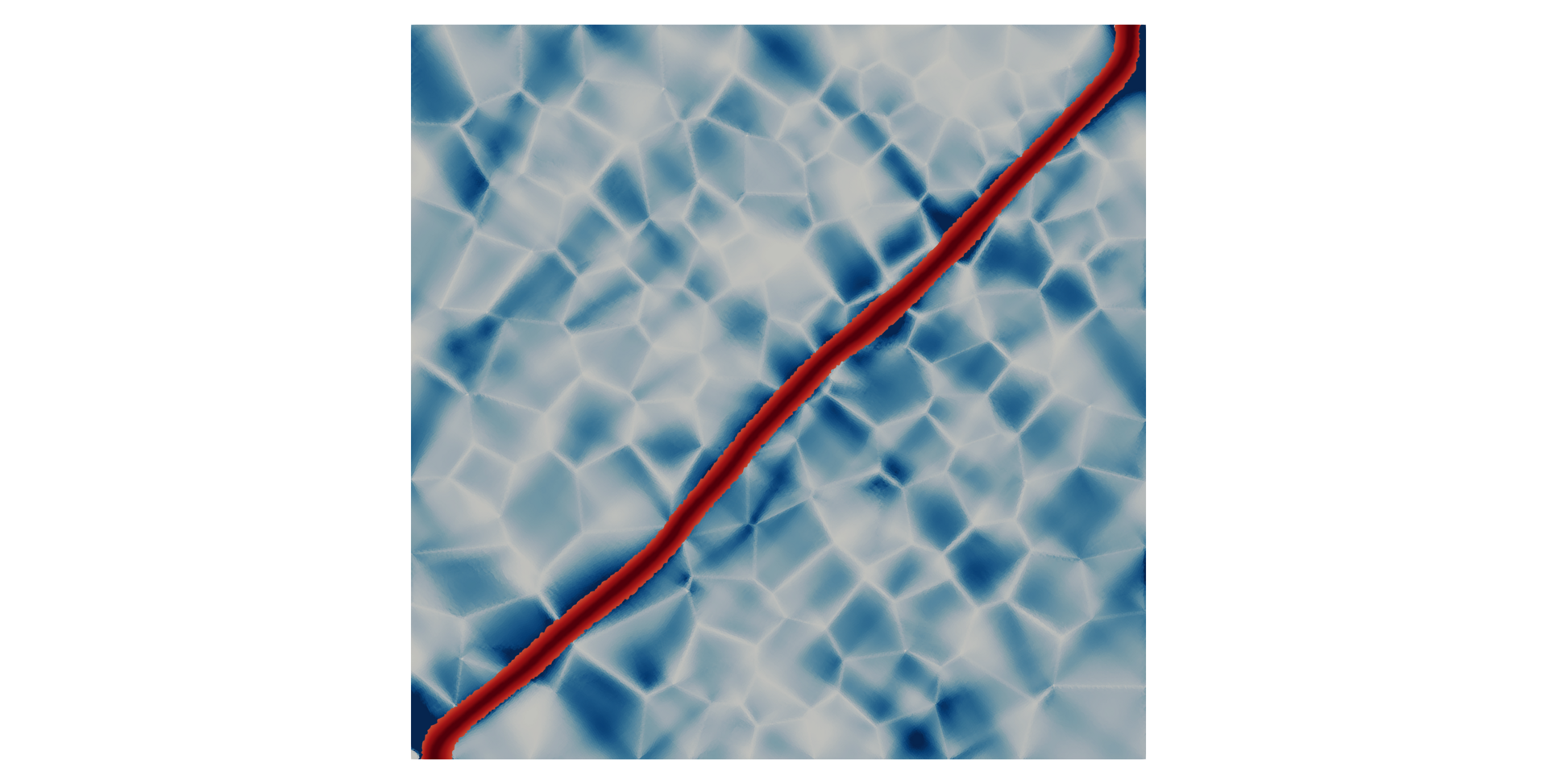}
	}
	\subfloat{ 
	    \includegraphics[width=0.2\textwidth,trim=18cm 0cm 18cm 0cm,clip]{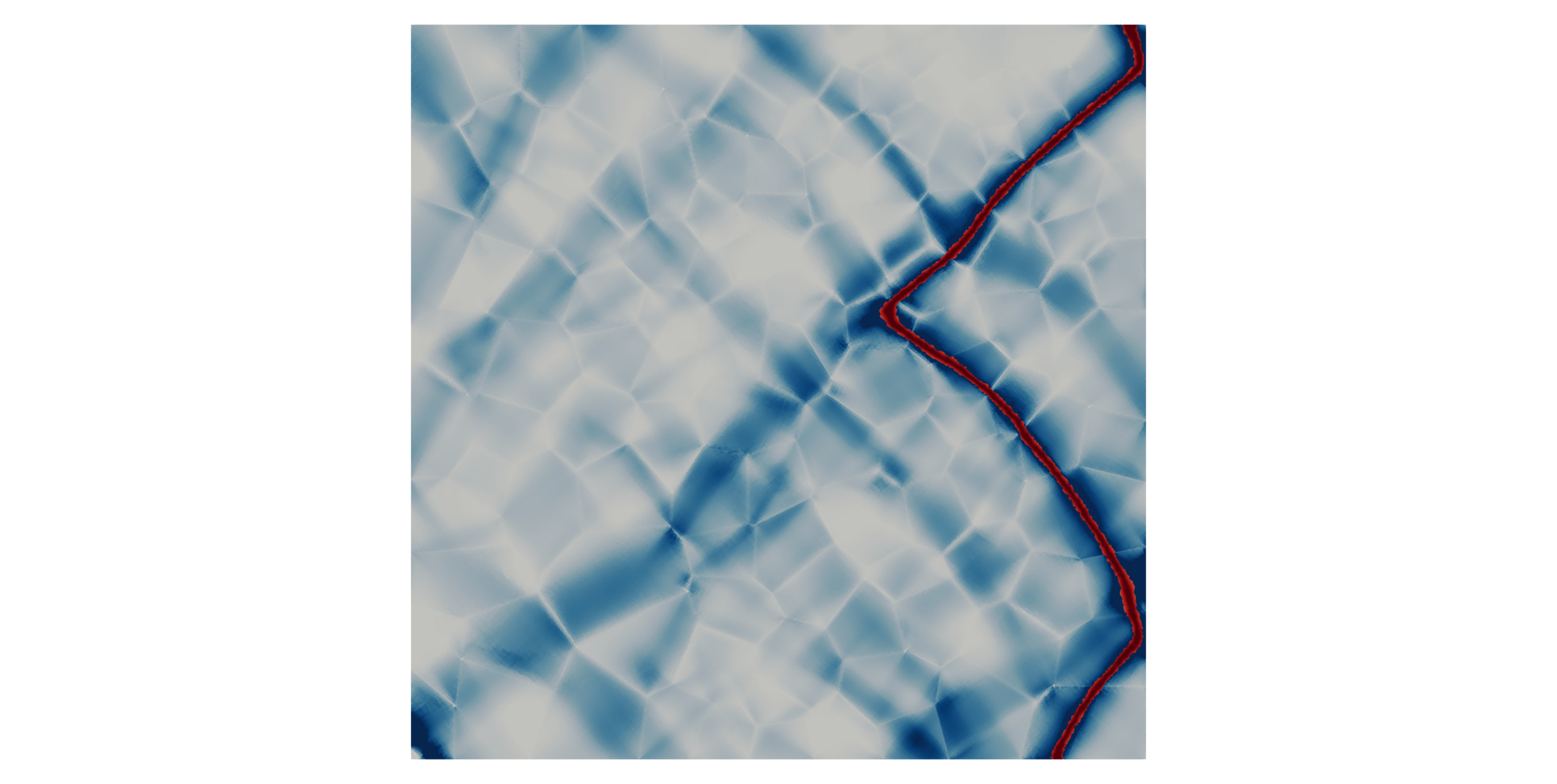}
	}
	\subfloat{ 
	    \includegraphics[width=0.2\textwidth,trim=18cm 0cm 18cm 0cm,clip]{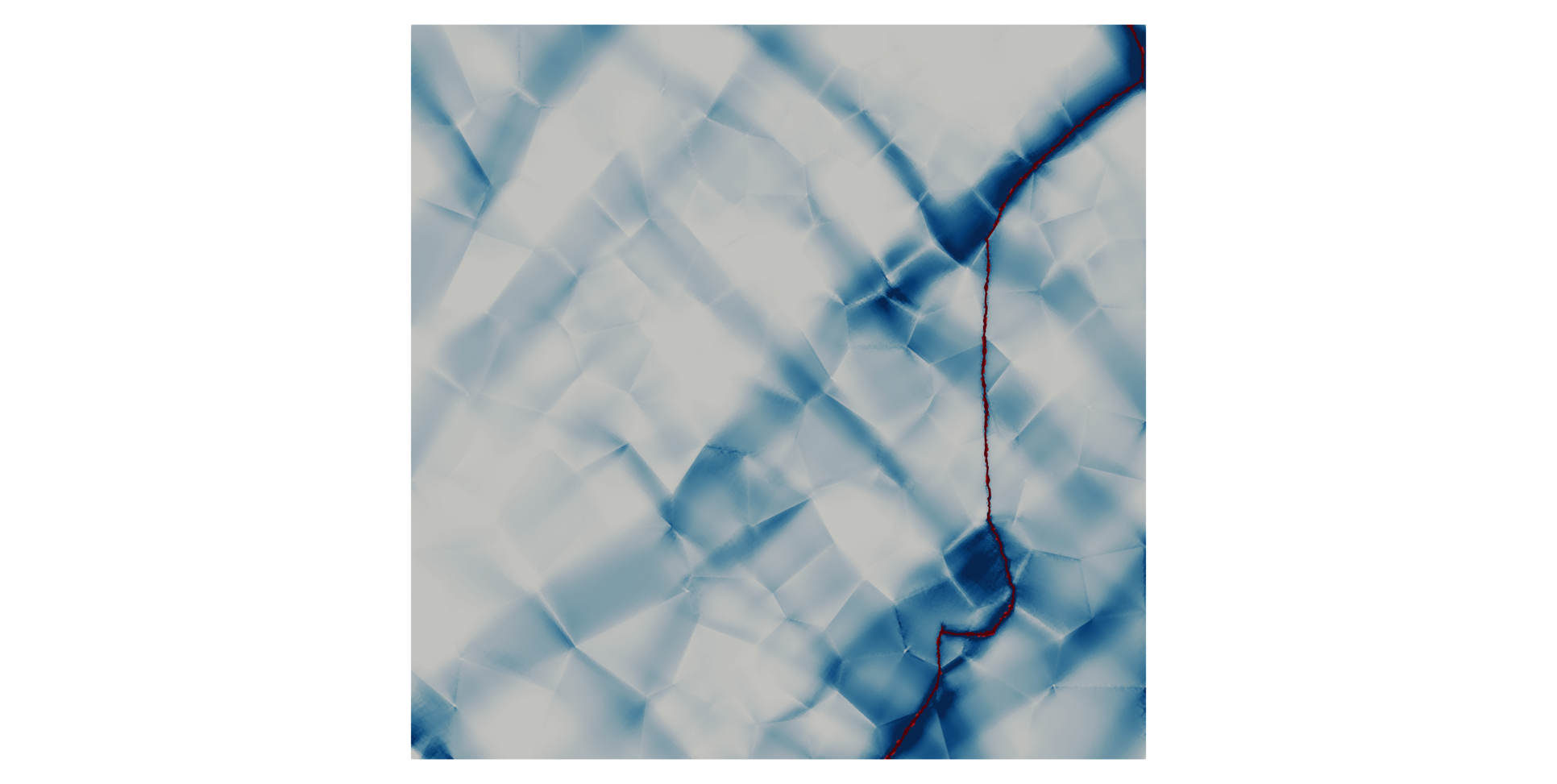}
	}
	\subfloat{  
	    \includegraphics[width=0.085\textwidth]{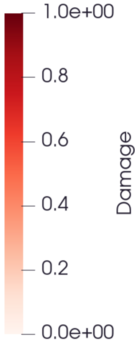}
	}
	\subfloat{  
	    \includegraphics[width=0.085\textwidth]{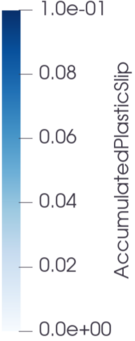}
	}\\
	\addtocounter{subfigure}{-6}
	\subfloat[$d = \SI{1}{\micro\meter}$]{  
	    \includegraphics[width=0.2\textwidth,trim=18cm 0cm 18cm 0cm,clip]{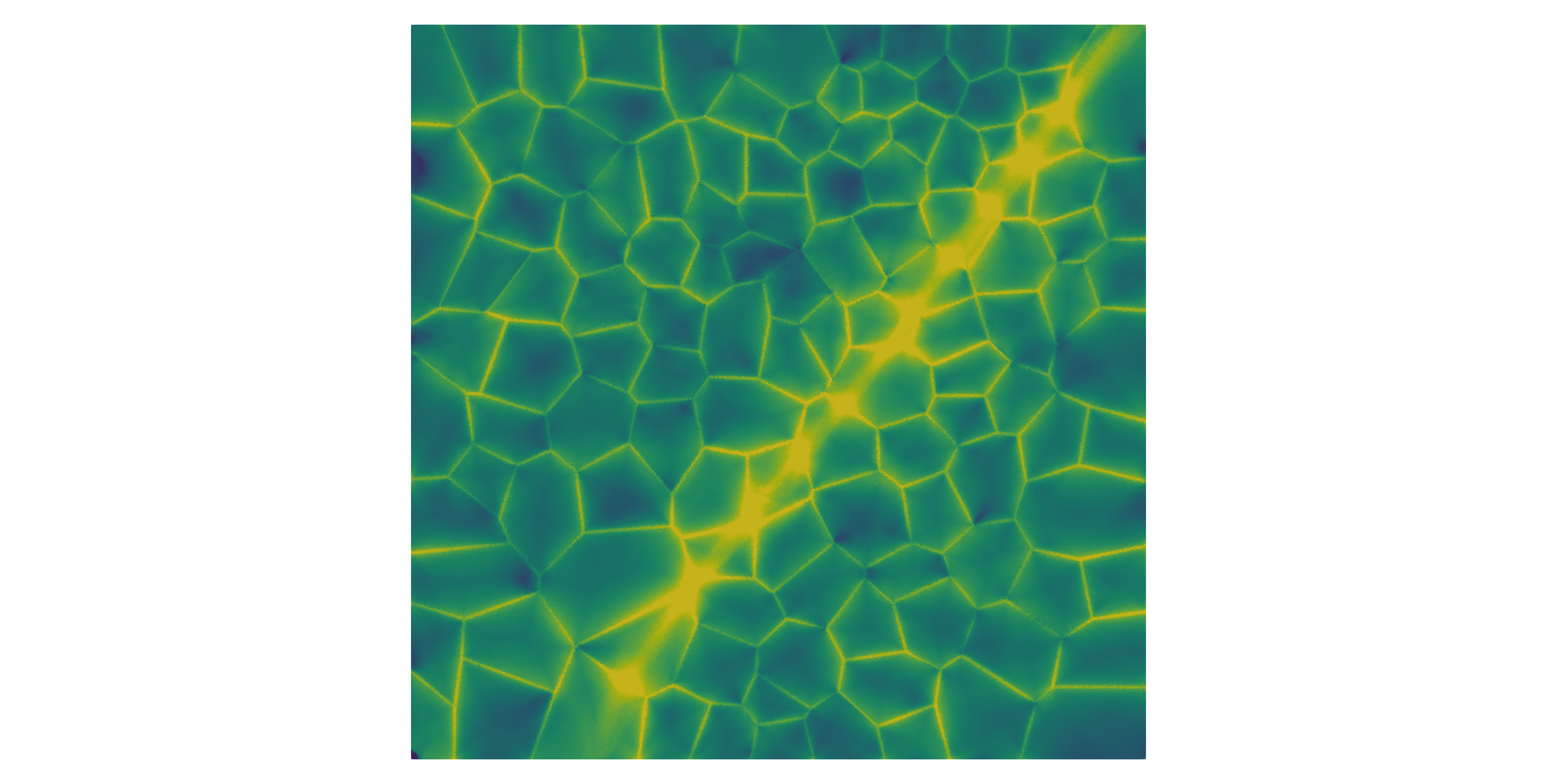}
	}
	\subfloat[$d = \SI{5}{\micro\meter}$]{  
	    \includegraphics[width=0.2\textwidth,trim=18cm 0cm 18cm 0cm,clip]{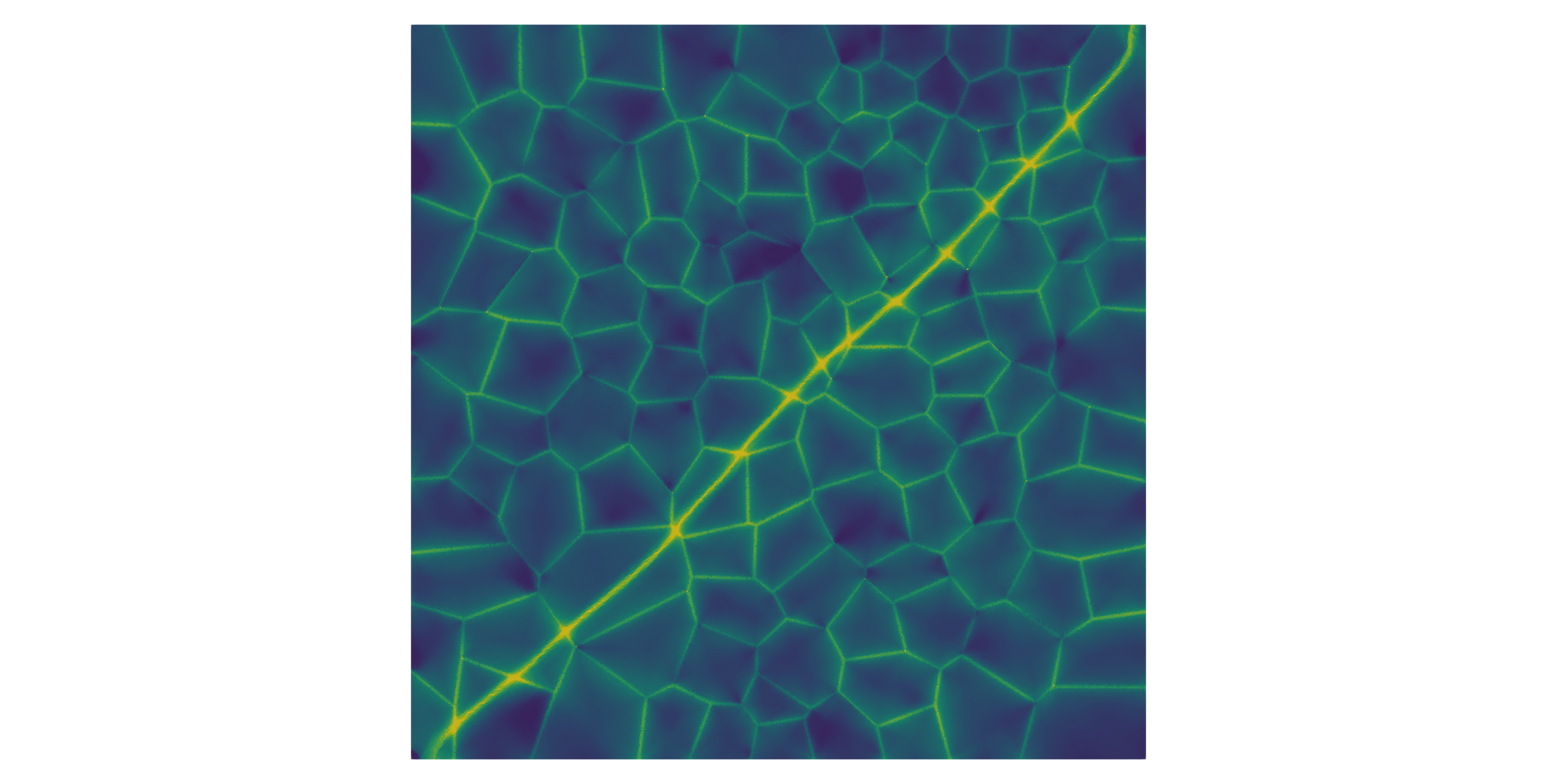}
	}
	\subfloat[$d = \SI{20}{\micro\meter}$]{  
	    \includegraphics[width=0.2\textwidth,trim=18cm 0cm 18cm 0cm,clip]{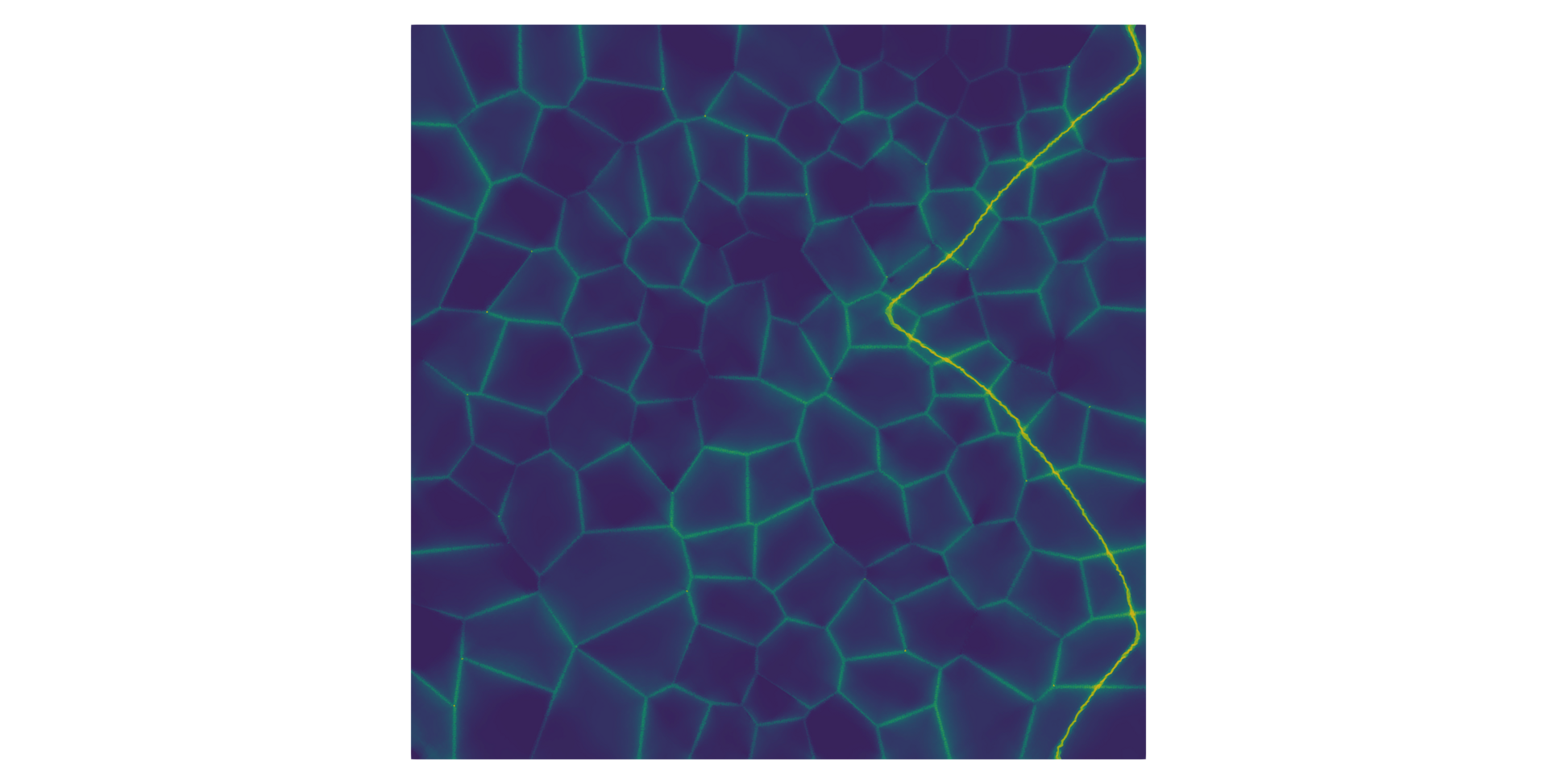}
	}
	\subfloat[$d = \SI{100}{\micro\meter}$]{  
	    \includegraphics[width=0.2\textwidth,trim=18cm 0cm 18cm 0cm,clip]{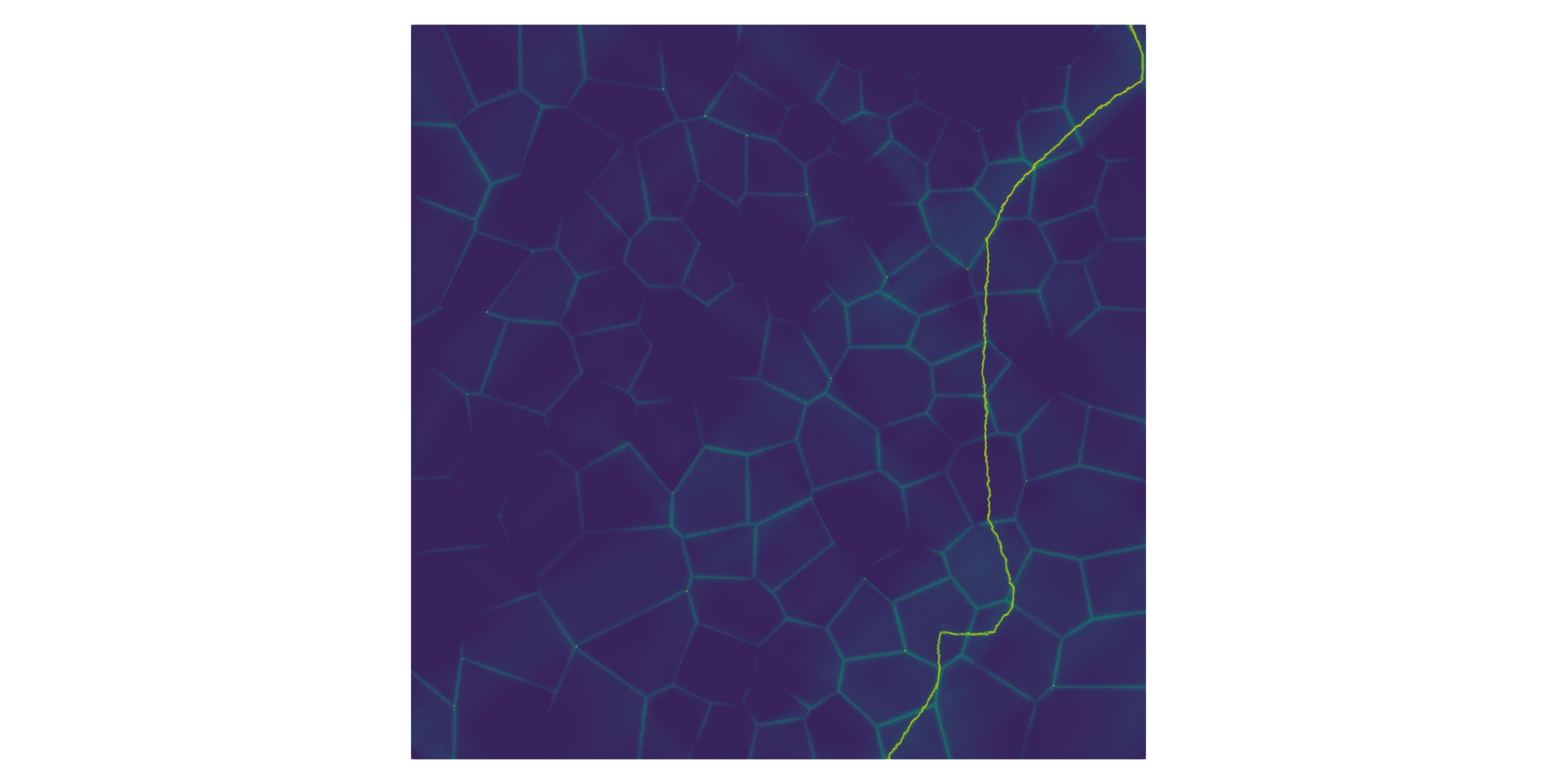}
	}
	\subfloat{  
	    \includegraphics[width=0.085\textwidth]{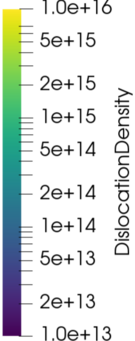}
	}
	\caption{Accumulated plastic slip field (blue colorscale), phase field variable $\alpha$ (red colorscale) and total dislocation density (logarithmic colorscale) for four different grain sizes. The contours of the phase field variable where it is lower than 0.5 are not displayed. Tension is applied along the horizontal direction.}
	\label{fig:tension_damage_fields}
\end{figure}

Figure~\ref{fig:tension_damage_fields} shows the phase field variable $\alpha$ and accumulated plastic slip after crack nucleation in the microstructure shown in Figure~\ref{fig:tension_microstructure}  for the four different grain sizes. As the grain size decreases, the apparent width of the crack increases because the non-dimensionalized phase field length $\ell_0$ is scaled with the factor $\eta L_0$. The coupling introduced in Eq.~\eqref{eq:functional}, between elastic energy density, plastic dissipation and the phase field, induces crack nucleation in regions undergoing severe plastic strains or high stress concentration. The crystal plasticity model leads to heterogeneous plastic strain fields with high plastic strain gradients in the vicinity of grain boundaries. Triple junctions also act as stress concentrators. The nucleation of cracks is hence favored in these regions. Although grain boundaries and triple junctions are preferential sites for crack nucleation, the crack can propagate through the grains. Abrupt changes in the direction of crack propagation can be observed for the two largest grain sizes. This is because the relative crack width decreases as the grain size increases. Hence, the crack  interacts with fewer grains and is therefore more likely to bifurcate towards (i) one of its neighbours which undergoes severe plastic deformations or (ii) a neighbouring triple junction concentrating the stresses.

Figure~\ref{fig:tension_damage_fields} also shows the total dislocation density field $\rho$ for the same microstructures with a logarithmic colorscale. The dislocation density is particularly high in the vicinity of grain boundaries. This is due to the Hall-Petch term $K_s / \delta$ in Eq.~\eqref{eq:dislocation_density} which increases significantly close to grain boundaries. The greater hardening induced by this term explains why the plastic slip intensity is lower in these regions. The overall dislocation density is also higher for smaller grain sizes as expected from the Hall-Petch effect. The dislocation density apparently becomes artificially large inside the crack. This is however a chimera with no bearing on the physics, because the coupling introduced in Eq.~\eqref{eq:functional} makes the plastic dissipation vanish inside the crack. From Figure~\ref{fig:tension_damage_fields} it is clear that, despite the larger dislocation density  in the vicinity of grain boundaries, the crack can propagate through the grains. It is also observed that the dislocation density increases over a wider area where the crack intersects grain boundaries. This is the sign of crack pinning at grain boundaries. The stiffness can radically change from one side of a grain boundary to the other, and so does the plastic activity. The crack can hence be arrested and deflected~\citep{ming1989crack} at interfaces.

\begin{figure}
	\centering
	\subfloat{  
	    \includegraphics[width=0.49\textwidth]{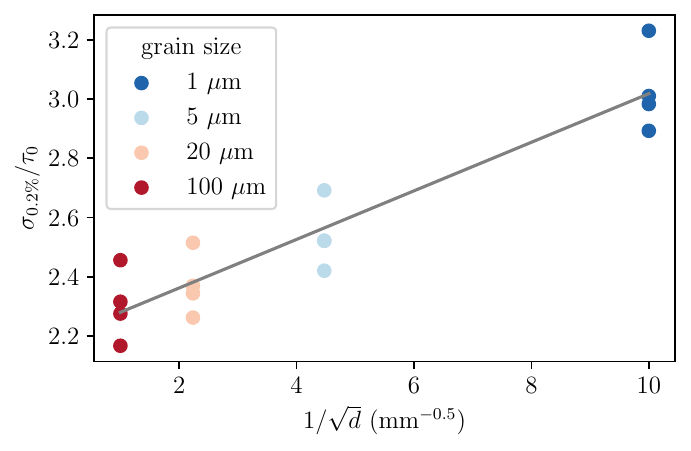}
		\label{subfig:tension_hall_petch_R02}
	}
	\subfloat{  
	    \includegraphics[width=0.49\textwidth]{fig37.pdf}
		\label{subfig:tension_hall_petch_Rmax}
	}
	\caption{Normalized yield stress at $0.2\%$ plastic strain ($\sigma_{0.2\%}$) (a) and peak stress $\sigma_{max}$ (b) as a function of the inverse square root of the mean grain diameter. The gray solid lines are linear fits indicating the Hall-Petch size effect.}
	\label{fig:tension_hall_petch}
\end{figure}

The normalized yield stress at $0.2\%$ plastic strain ($\sigma_{0.2\%}$) and peak stress are shown in Figure~\ref{subfig:tension_hall_petch_R02} and Figure~\ref{subfig:tension_hall_petch_Rmax} respectively. Both quantities follow a linear relationship with the inverse square root of the mean grain diameter. This corresponds to the well known Hall-Petch effect~\citep{hall1951deformation,petch1953cleavage,petch1958ductile}. This effect on the yield stress is captured through the grain size dependent evolution equations of dislocation densities in Eq.~\eqref{eq:dislocation_density}. We show that the simple crystal plasticity-fracture coupling adopted in Eq.~\eqref{eq:functional} is able to capture the Hall-Petch effect frequently obeserved on the peak stress as well~\citep{sasaki1975grain,schulson1983brittle}. The increase of ductility with decreasing the grain size is also consistent with experimental evidence~\citep{hull1961effect,worthington1966slip,schulson1983brittle}.

\section{Crack propagation in polycrystals}
\label{sec:propagation}

\subsection{Setting}
\label{subsec:propagation_setting}

\begin{figure}
	\centering
	\subfloat{  
	    \includegraphics[width=0.7\textwidth,trim=0cm 0cm 0cm .6cm,clip]{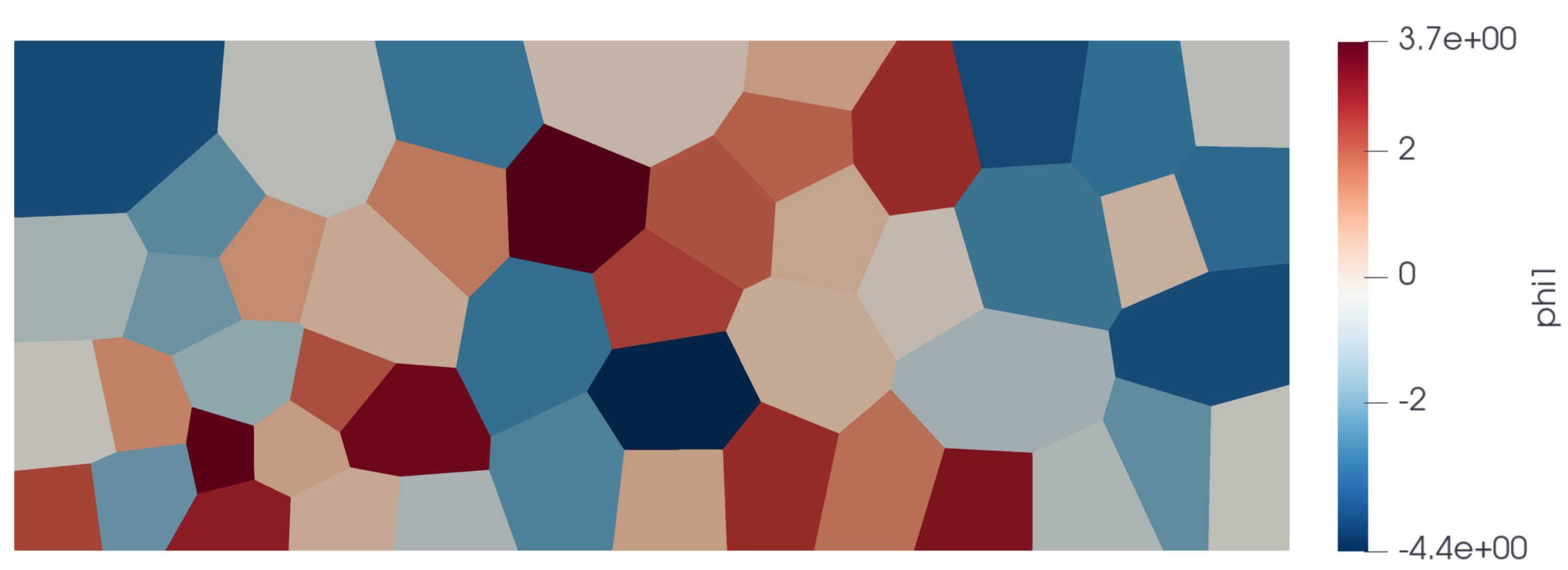}
	}
	\caption{Polycrystalline microstructure used to simulate crack propagation in plane strain conditions. The colorscale represents the first Euler angle $\phi_1$.}
	\label{fig:surfing_microstructure}
\end{figure}

We now investigate the effect of grain size on crack propagation in polycrystals. We consider a 2D polycrystalline microstructure with a random distribution of grains and uniform distribution of orientation as shown in Figure~\ref{fig:surfing_microstructure}. The grain morphology is fixed (one realisation) and the grain size is scaled over one order of magnitude, from \SI{5}{\micro\meter} to \SI{50}{\micro\meter}. To do so, we use the same scaling as before and then work on a fixed microstructure and computational domain. The microstructure is composed of 49 grains.

The non-dimensionalized material parameters are identical to the previous section and are given in Table~\ref{tab:parameters}. In the subsequent sections, the ratio $q$ between the plastic and process zone sizes as defined in Eq.~\eqref{eq:quotient} is varied by using different values of the non-dimensional yield stress $\tau_0 \in [10^{-3}\, ; 2\times 10^{-3}\, ; 4\times 10^{-3}]$. The ratio $q$ is thus varied from $10^3$ to $2.5\times 10^2$, where lower values of $q$ correspond to a smaller plastic zone and thereby, a more brittle behaviour.

A pre-crack of infinitesimal length is introduced in the middle of the left edge of the polycrystal domain. Following~\cite{hossain2014effective,brach2019anisotropy,brach2019phase}, time-dependent displacement conditions corresponding to plane strain asymptotic mode I crack propagation~\citep{anderson2005fracture} are applied on the boundary of the domain. These so-called surfing boundary conditions are defined as follows
\begin{align}
	&\displaystyle\boldsymbol{u}(\boldsymbol x, t) = \boldsymbol{U}(\boldsymbol{x}-v_0t\boldsymbol{e}_1) = \psi\sqrt{\frac{(1+\nu)G_c}{2E}}\left(3-4\nu - \cos{\theta}\right)\sqrt{\frac{r}{2\pi}}\left(\cos\left(\frac{\theta}{2}\right)\boldsymbol{e}_1+\sin\left(\frac{\theta}{2}\right)\boldsymbol{e}_2\right),\\
	&\text{where}\quad r (\boldsymbol{x}, t) = \sqrt{(x - x_0 - v_0 t)^2 + (y - y_0)^2} \quad \mathrm{and} \quad 
    \theta (\boldsymbol{x}, t) = \arctan \left( \frac{y - y_0}{x - x_0 - v_0 t} \right)  \label{eq:surfing_bc}
\end{align}
and where $\psi$ is an arbitrary non-dimensional scaling parameter taken equal to 1 in the following. The coordinates $(x_0, y_0)$ correspond to the initial position of the crack tip, \textit{i.e.} the middle of the left edge of the domain. The boundary condition defined in Eq.~\eqref{eq:surfing_bc} drives the crack from left to right at a velocity $v_0$ at the macroscopic scale.

The $J$-integral is defined here as
\begin{equation}
	J = \int_{\Gamma} \pmb{t}\cdot\left((\boldsymbol{E}:a(\alpha)\mathbb{C}:\boldsymbol{E})\boldsymbol{1} - \nabla\boldsymbol{u}^T\cdot a(\alpha)\mathbb{C}:\boldsymbol{E}\right)\cdot\boldsymbol{n}\, \mathrm{d}s,
	\label{eq:jintegral}
\end{equation}
where $\Gamma$ is a contour around the crack tip, $\boldsymbol{t}$ is the unit vector tangent to the crack at the crack tip and $\boldsymbol{n}$ is the outward unit normal to the contour. The $J$-integral describes the energy release rate if the crack-tip contained in the region within the contour $\Gamma$ were to advance. In a homogeneous elastic material, the $J$-integral is path-independent since there is no energy released away from the crack tip and thus, the far field $J$-integral also provides the driving force at the crack tip. In contrast, in a heterogeneous material or in an elastic-plastic material, the energy is released due to heterogeneity in elastic modulus and due to plastic dissipation as the crack advances. Physically, energy is released due to the heterogeneity and plastic processes as the crack propagates. Therefore, it is not independent of the contour. However, the $J$-integral becomes path-independent for contours that are large compared to the grains and the plastic process zone~\citep{hsueh2016homogenization}. This far-field $J$-integral provides the true driving force to advance the crack. In other words, it not only accounts for crack propagation but also the shielding effects of heterogeneity and plastic deformation in the vicinity of the crack tip. In this work, we compute the $J$-integral on the boundary of the computational domain (see also~\cite{hossain2014effective} for discussion).

\subsection{$J$-integral}

\begin{figure}
	\centering
	\subfloat[$q = 1.0 \times 10^3$]{
		\includegraphics[width=.5\textwidth]{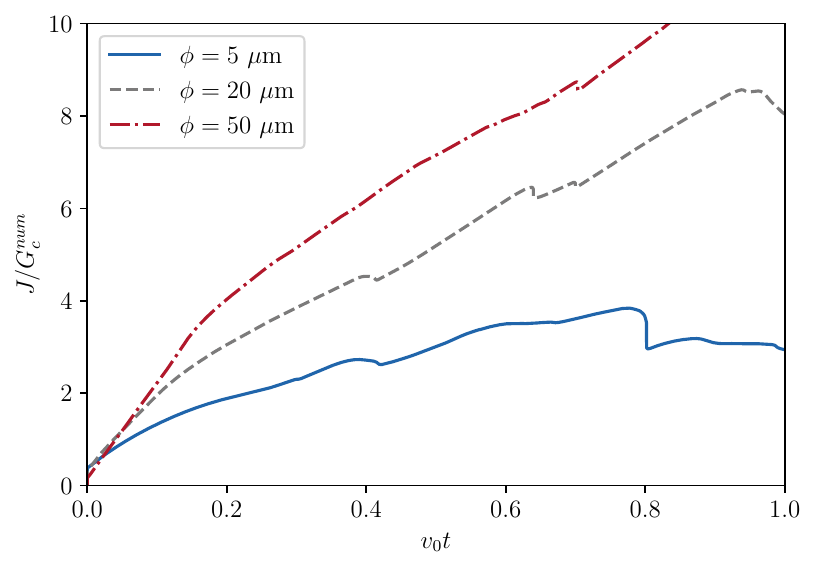}
		\label{subfig:surfing_grain_size_tau0_1e-3_jintegral}
	}
	\subfloat[$q = 5.0 \times 10^2$]{
		\includegraphics[width=.5\textwidth]{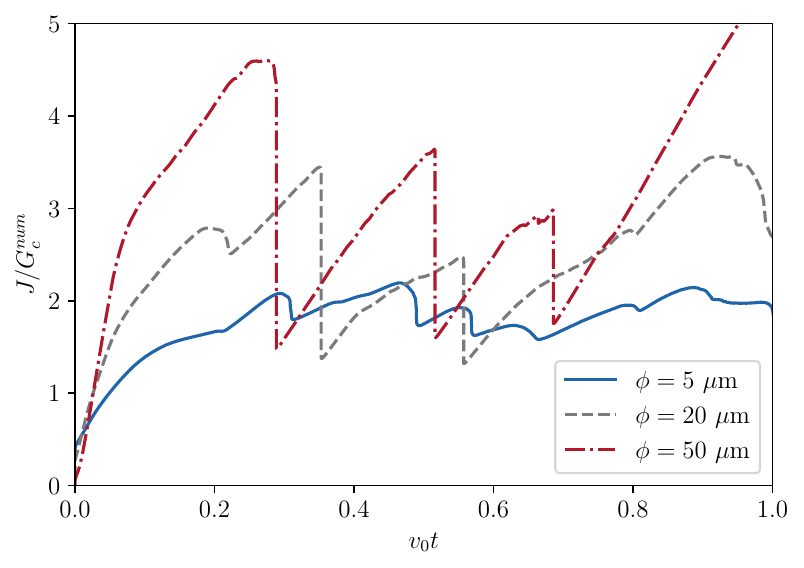}
		\label{subfig:surfing_grain_size_tau0_2e-3_jintegral}
	}\\
	\subfloat[$q = 2.5\times 10^2$]{
		\includegraphics[width=.5\textwidth]{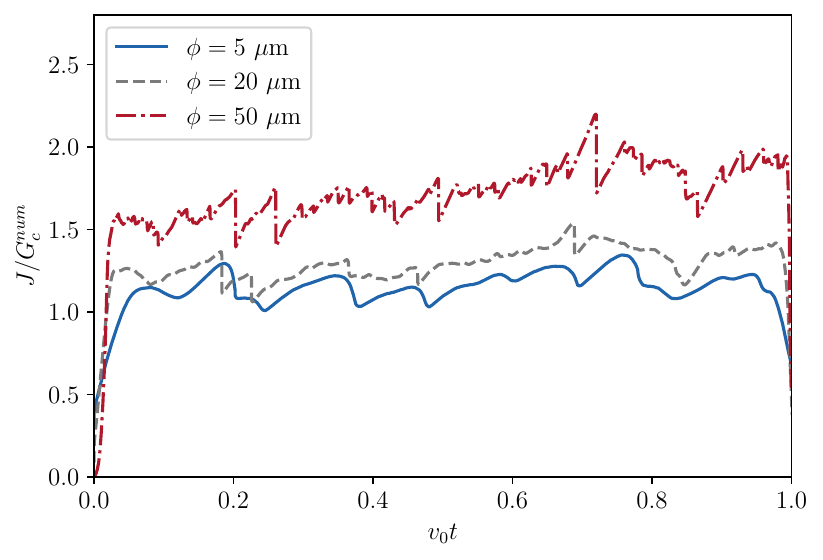}
		\label{subfig:surfing_grain_size_tau0_4e-3_jintegral}
	}
	\subfloat[$d = \SI{20}{\micro\meter}$]{
		\includegraphics[width=.5\textwidth]{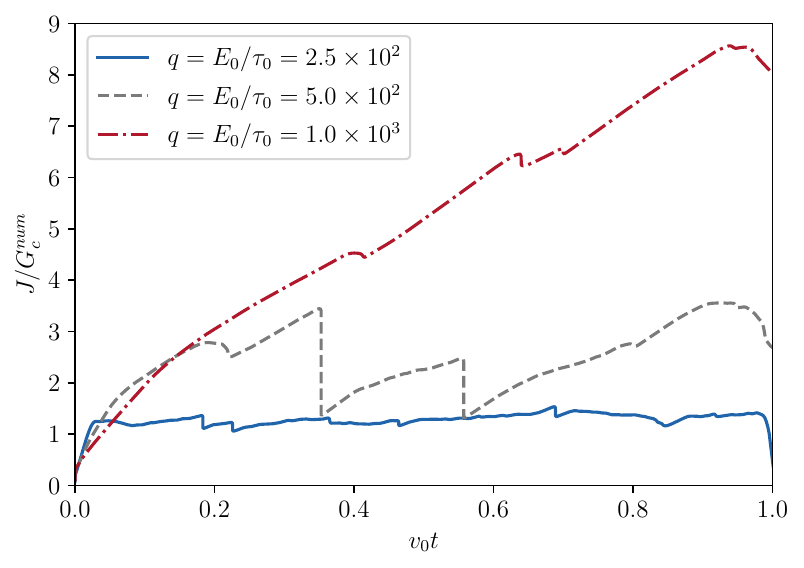}
		\label{subfig:surfing_ductility_jintegral}
	}
	\caption{Normalized $J$-integral as a function of the macroscopic crack length.
	In (a), (b) and (c) the ductility ratio $q$ is fixed at $1.0 \times 10^3$, $5.0 \times 10^2$ and $2.5 \times 10^2$ respectively and the grain size is varied. (c) The ductility ratio $q$ is varied and grain size fixed at $\SI{20}{\micro\meter}$.
	The $J$-integrals are normalized by the numerical fracture toughness $G_c^{num} = G_c(1+3h/8\ell)$.}
\end{figure}

First, we study the effect of the grain size $d$ and of the ductility ratio $q$ on the $J$-integral. The $J$-integral is plotted as a function of the macroscopic crack length for three grain sizes and the largest ductility ratio of $q = 1.0 \times 10^3$ in Figure~\ref{subfig:surfing_grain_size_tau0_1e-3_jintegral}. The $J$-integral is normalized by the numerical fracture toughness $G_c^{num} = G_c(1+3h/8\ell)$~\citep{bourdin2008variational}, where $h$ is the mesh size. $J/G_c^{num}$ is greater for the largest grain size and continues to increase until the end of the simulation. The driving force is not yet high enough to let the crack propagate in a steady manner over a long distance and therefore the $J$-integral continues to increase as the loading progresses. For smaller grain sizes, the $J$-integral tends to saturate, which indicates that the driving force reaches the necessary threshold to let the crack propagate continuously. For this relatively large ductility ratio, the crack propagation and corresponding $J$-integral are mainly governed by plastic dissipation, hence the regime can be called \textit{plasticity-driven}.

The same analysis is performed for the intermediate ductility ratio of $q = 5\times 10^2$ in Figure~\ref{subfig:surfing_grain_size_tau0_2e-3_jintegral}. The $J$-integral is plotted as a function of the macroscopic crack length for three grain sizes. The behaviour is very different from the previous case. The $J$-integrals are non-monotonic and display large serrations. These serrations are particularly significant for the largest grain size and milder for the lowest grain size. The sudden drops of the $J$-integral are due to the crack jumps within the microstructure. The phase that precedes each jump corresponds to a plasticity-driven phase during which the driving force and $J$-integral increase. On the other hand, as the crack jumps, limited plastic dissipation takes place. These jumps are hence mainly governed by the available elastic energy stored ahead of the crack tip. The crack jumps occurs up to the point where it becomes sub-critical again. This intermediate ductility can hence be characterized by an alternating sequence of \textit{plasticity-driven} and \textit{elasticity-driven} phases.

Figure~\ref{subfig:surfing_grain_size_tau0_4e-3_jintegral} shows the normalized $J$-integral as a function of the macroscopic crack length for the lowest ductility ratio of $q = 2.5\times 10^2$. The $J$-integral is plotted for three grain sizes. Similar to Figures~\ref{subfig:surfing_grain_size_tau0_1e-3_jintegral} and~\ref{subfig:surfing_grain_size_tau0_2e-3_jintegral}, the maximum peak values of the $J$-integral are reached for the largest grain size. However, for this low ductility case, the difference between the $J$-integral values for the different grain sizes is less pronounced. Except for the largest grain size, the normalized $J$-integral remains in the range of $[1.0,1.5]$. This indicates that the crack propagation is close to the theoretical limit of $G_c^{num}$ corresponding to purely brittle fracture. The crack propagation is mainly governed by the elastic energy release rate and this behaviour is mainly \textit{elasticity-driven}. The larger grain size displays rapid variations of the $J$-integral over distances shorter than the grain size. This is the indication of a sensitivity to the spatial discretization. For this particular grain size, the ratio of the phase field length scale and the mesh size $\ell / h$ is equal to 1. A refined mesh could be used to cancel the sensitivity to the mesh size, however this would not drastically change the crack path, nor the $J$-integral values.

The $J$-integral for an intermediate grain size of $\SI{20}{\micro\meter}$ is plotted as a function of the macroscopic crack length $v_0t$ in Figure~\ref{subfig:surfing_ductility_jintegral} for the three different ductility ratios $q$. For the lowest value of $q$, the $J$-integral is close to the theoretical value of $G_c^{num}$ corresponding to brittle fracture. Small drops in the $J$-integral correspond to short crack jumps at grain boundaries due to the elastic mismatch between neighbouring grains~\citep{ming1989crack}. The elastic anisotropy and the different crystal orientation in each grain induces a variation of the singularity exponent $\lambda$ in $\sigma \propto r^{-(1-\lambda)}$ at grain boundaries. When the crack reaches (from left to right) an interface where the apparent stiffness in the vertical direction is higher on the right of the interface than on the left, then the singularity exponent $\lambda$ increases, \textit{i.e.} $(1-\lambda)$ decreases~\citep{tanne2018crack,hsueh2018stress}. In this case, the singularity is sub-critical in the sense of Griffith, hence the crack is pinned at the boundary and needs to renucleate. On the contrary if the crack transitions from a stiff to a compliant grain, the crack will jump into the compliant grain as it approaches the interface. For the intermediate value of $q$, the $J$-integral reaches higher values. The plastic activity relaxes the stresses at the crack tip which are no longer singular~\citep{ritchie1973relationship}. A larger driving force is thus required to increase the stress sufficiently, over a large enough distance, to let the crack propagate. The $J$-integral shows large drops which are due to long crack jumps. As the crack jumps, little plastic dissipation occurs. The $J$-integral therefore decreases until it reaches the theoretical limit of $G_c^{num}$ corresponding to purely brittle fracture. For the highest value of $q$, the $J$-integral is the highest. The crack propagates continuously at a slow rate and hence the $J$-integral steadily increases. This steady increase is due to the significant amount of plastic dissipation throughout the crack propagation.

\subsection{Grain size effect}
\label{subsec:surfing_grain_size_effect}

\begin{figure}
	\centering
	\includegraphics[width=.9\textwidth]{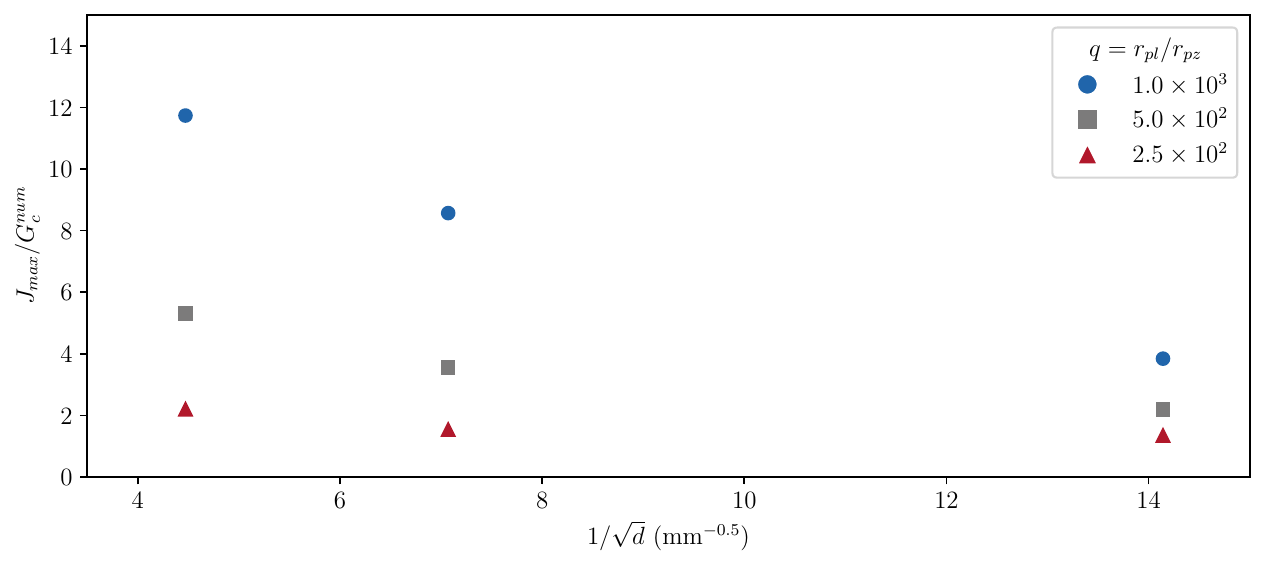}
	\caption{Normalized $J$-integral as a function of the inverse square root of the mean grain diameter.}
	\label{fig:surfing_grain_size_effect}
\end{figure}

Following~\cite{hossain2014effective,brach2019anisotropy,brodnik2021fracture}, the effective fracture toughness of a heterogeneous material can be characterized by the peak value of the $J$-integral over a sufficiently long crack propagation distance. Assuming that such a distance is reached in our simulations, we compute the peak values $J_{max}$ of the $J$-integral for the three grain sizes and three ductility ratios. Figure~\ref{fig:surfing_grain_size_effect} shows these values as a function of the inverse of the square root of the grain size. In contrast with the Hall-Petch size effect obtained for crack nucleation in plane strain tension in Section~\ref{subsec:tension_hall_petch}, the $J_{max}$ values display an inverse Hall-Petch size effect.

The grain size effect is particularly significant for the largest ductility ratio. The $J_{max}$ value is approximately 3.1 times larger for the largest grain size (\SI{50}{\micro\meter}) compared to the smallest grain size (\SI{5}{\micro\meter}). For the intermediate ductility ratio, the grain size effect is less pronounced, with a 2.4 ratio of the $J_{max}$ value between the largest and the lowest grain size. For the lowest ductility ratio, the grain size effect is even less pronounced, with the $J_{max}$ value being only 1.6 times larger for the largest grain size compared to the smallest grain size. This dependence of the grain size effect on the ductility ratio can be explained by the competition between plastic relaxation and crack propagation. As the ductility decreases, the plastic zone vanishes and therefore the grain size effect included in the plasticity model becomes less significant. In the pure brittle fracture limit, a grain size effect persists for grain sizes comparable to the phase field length scale~\citep{hsueh2018stress}. The effect of the ductility ratio decreases as the grain size decreases. This observation is consistent with the analysis proposed by~\cite{friedel1959propagation} who noted that the stress necessary to propagate a crack should be independent of work hardening for sufficiently fine grains (below $\sim$\SI{10}{\micro\meter}).

\begin{figure}
	\centering
	\subfloat[$d = \SI{5}{\micro\meter}$]{  
		\includegraphics[width=0.46\textwidth,trim=0cm 3cm 1.2cm 3cm,clip]{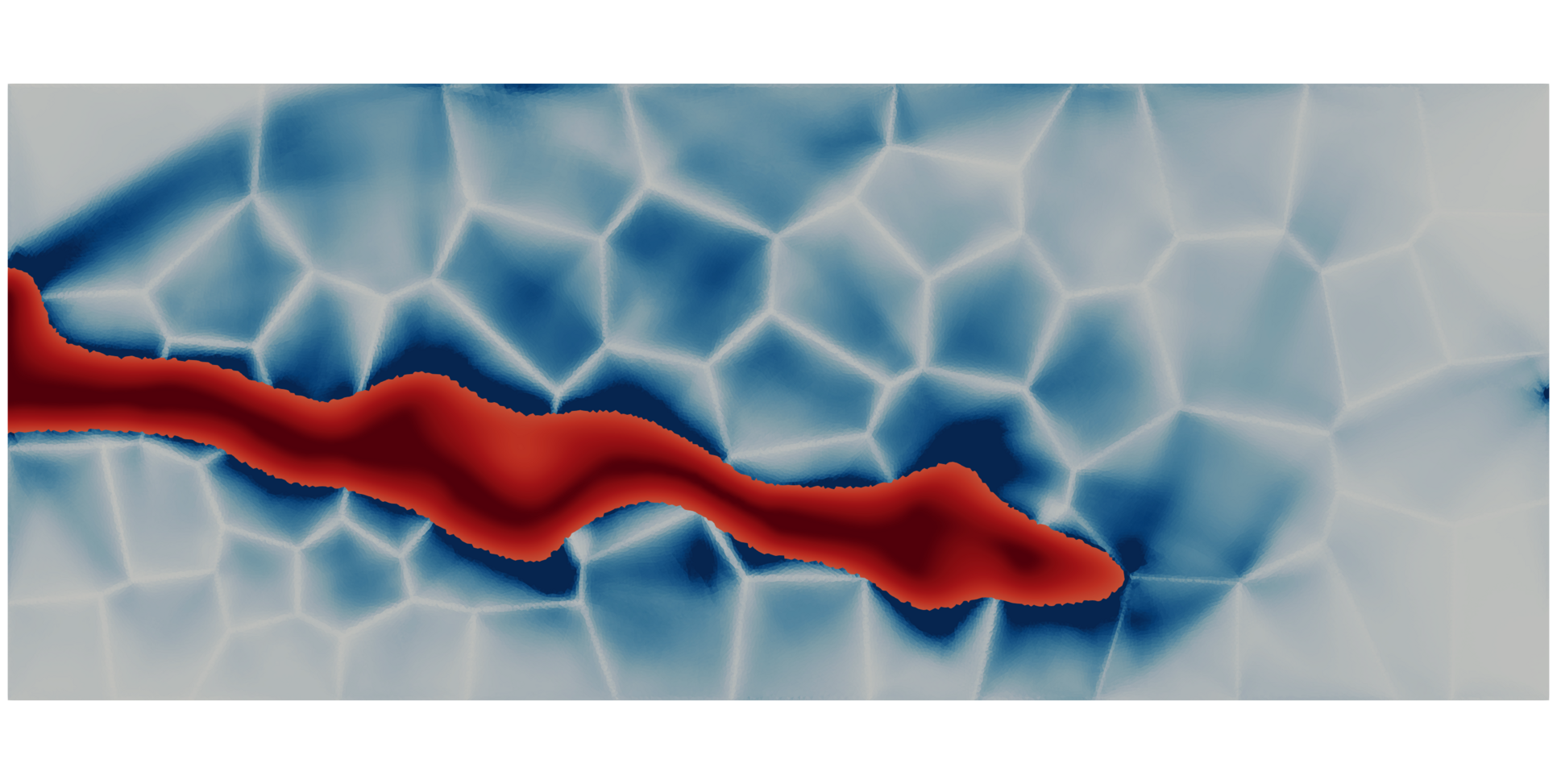}
		\label{subfig:surfing_tau0_1e-3_scale_0.5_damage}
	}
	\subfloat[$d = \SI{5}{\micro\meter}$]{  
		\includegraphics[width=0.46\textwidth,trim=0cm 3cm 1.2cm 3cm,clip]{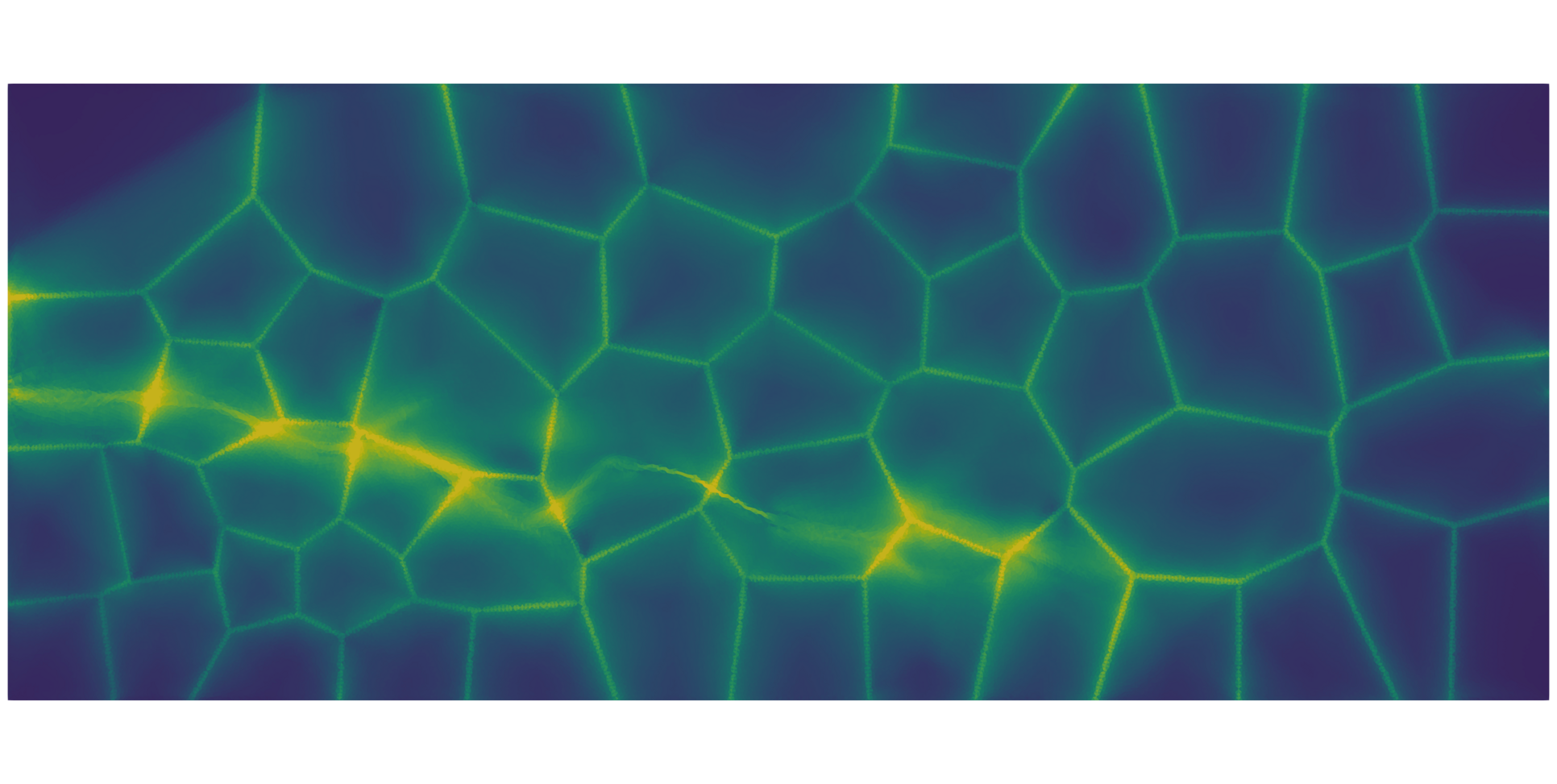}
		\label{subfig:surfing_tau0_1e-3_scale_0.5_dislocation_density}
	}
	\subfloat{  
	    \includegraphics[width=0.076\textwidth]{fig94.png}
	}\setcounter{subfigure}{2}
    \\
	\subfloat[$d = \SI{20}{\micro\meter}$]{  
		\includegraphics[width=0.46\textwidth,trim=0cm 3cm 1.2cm 3cm,clip]{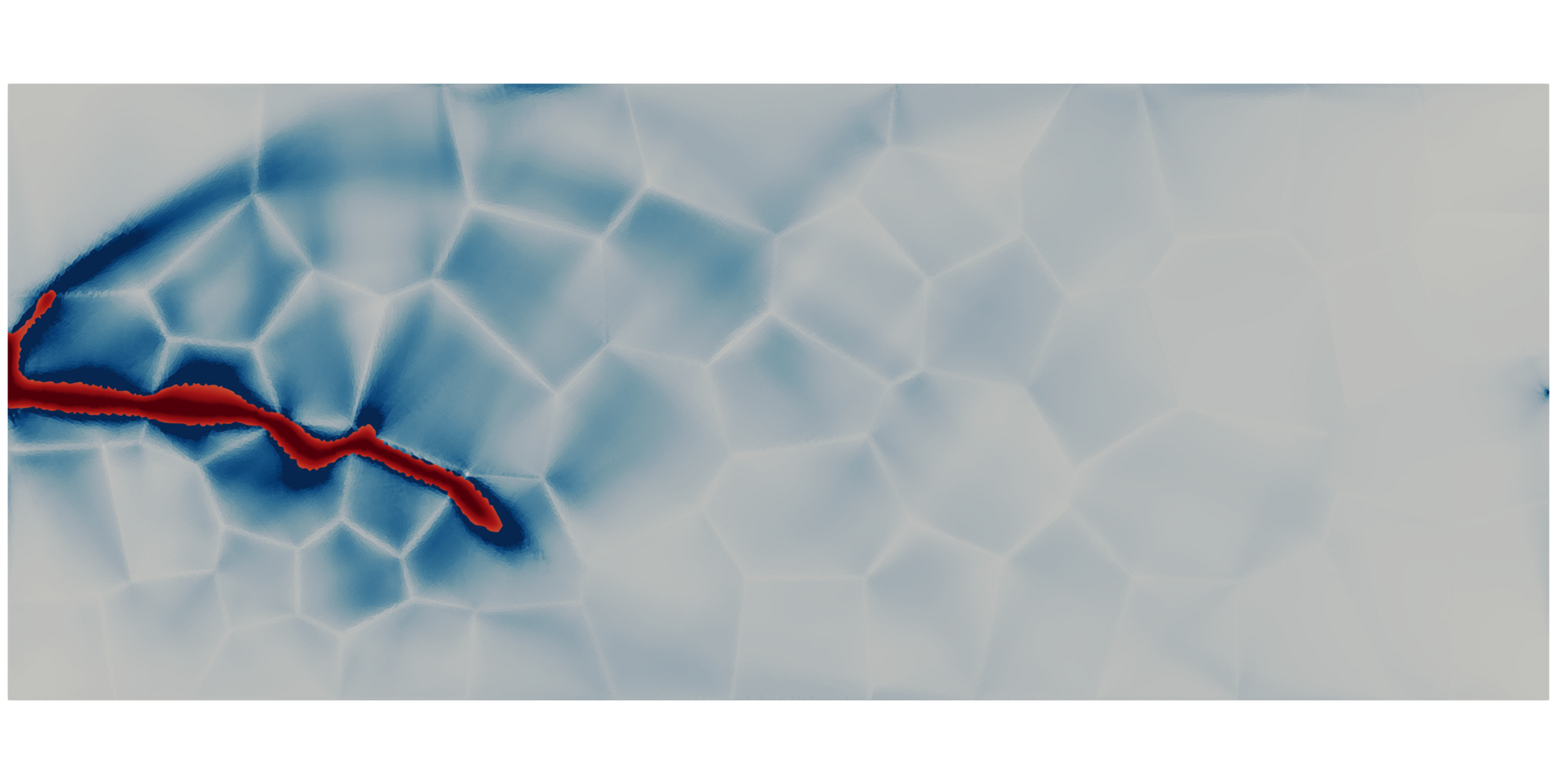}
		\label{subfig:surfing_tau0_1e-3_scale_2.0_damage}
	}
	\subfloat[$d = \SI{20}{\micro\meter}$]{  
		\includegraphics[width=0.46\textwidth,trim=0cm 3cm 1.2cm 3cm,clip]{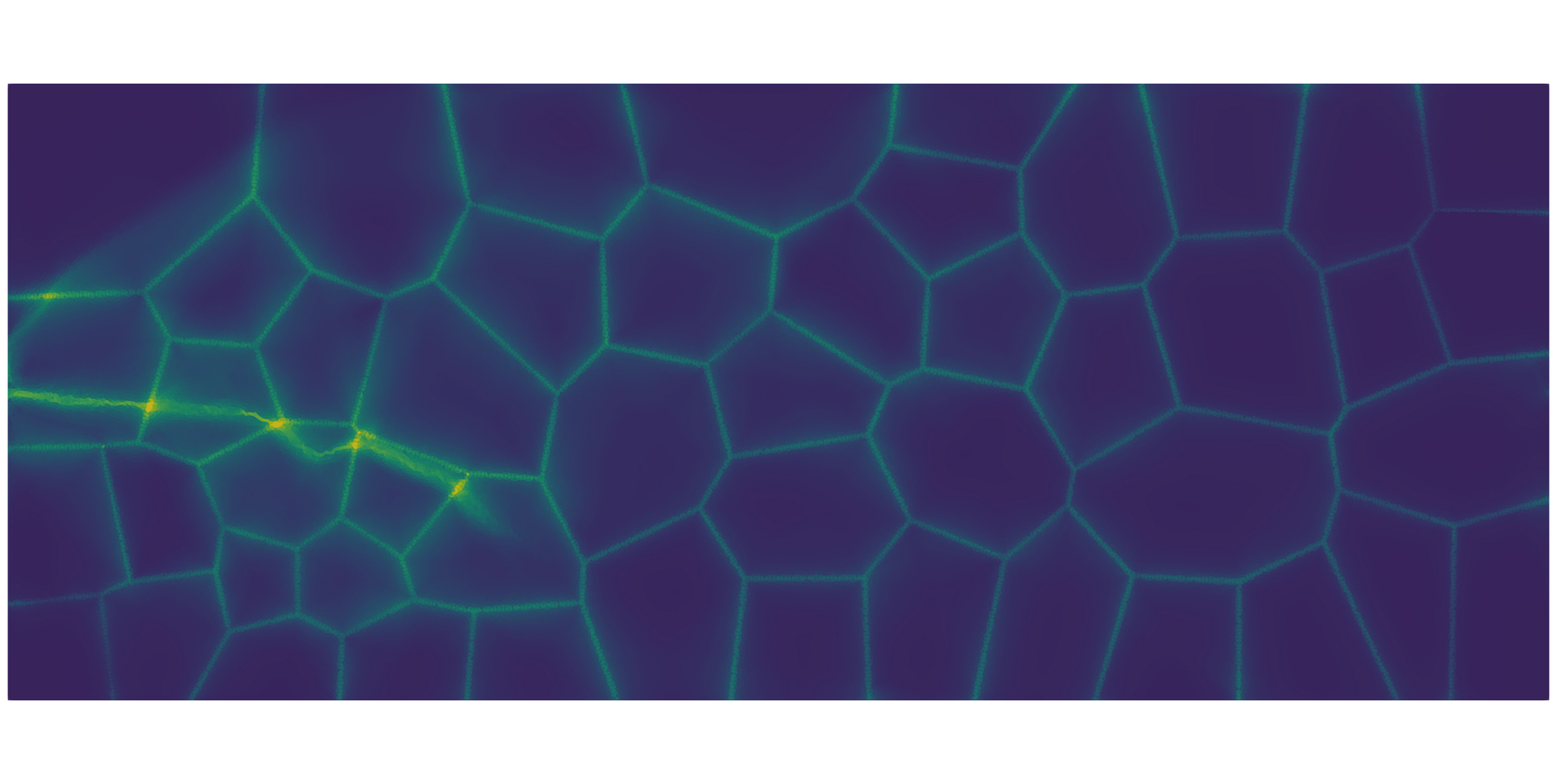}
		\label{subfig:surfing_tau0_1e-3_scale_2.0_dislocation_density}
	}
	\subfloat{  
	    \includegraphics[width=0.077\textwidth]{fig97.png}
	}\setcounter{subfigure}{4}
    \\
	\subfloat[$d = \SI{50}{\micro\meter}$]{  
		\includegraphics[width=0.46\textwidth,trim=0cm 3cm 1.2cm 3cm,clip]{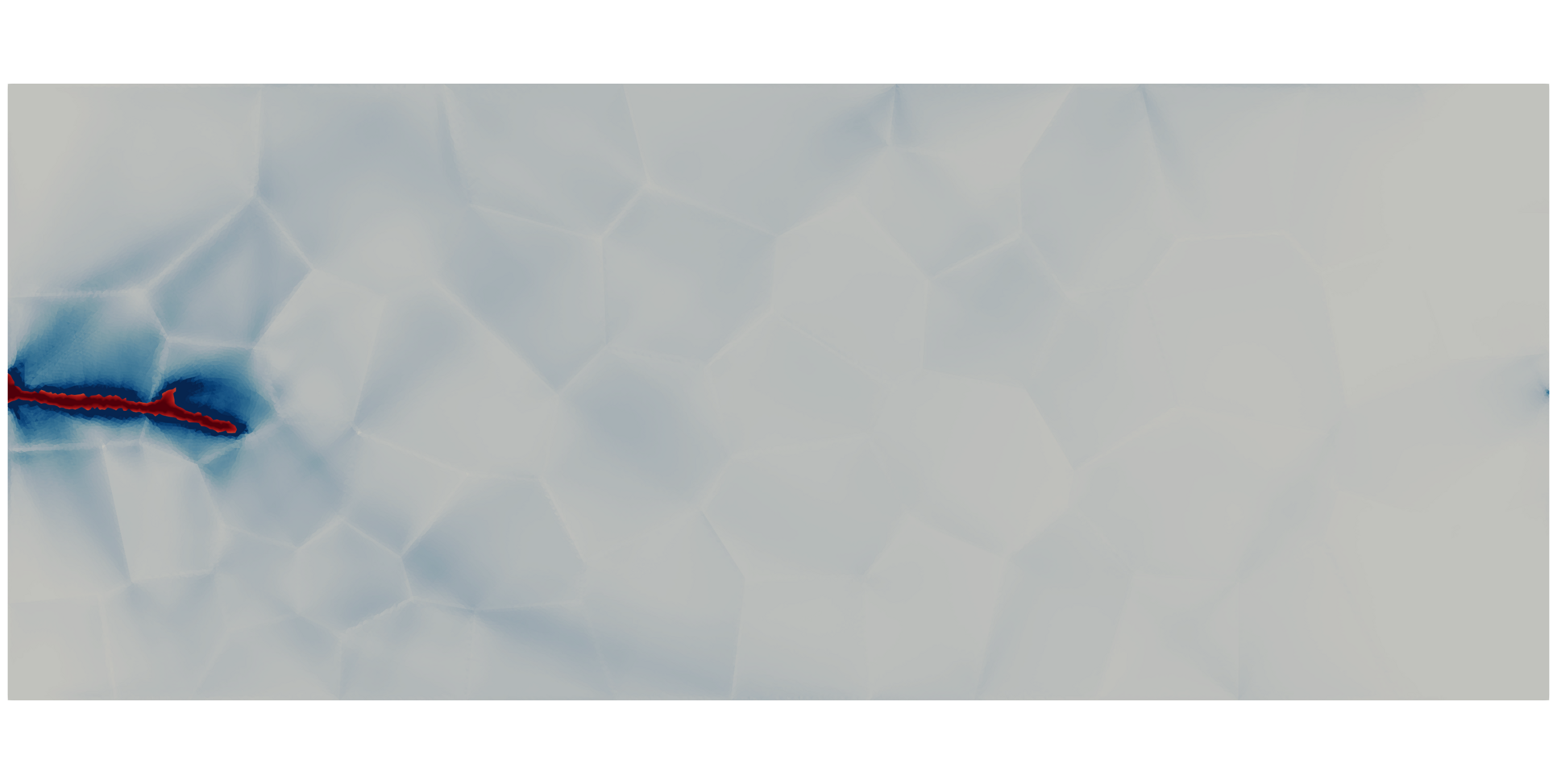}
		\label{subfig:surfing_tau0_1e-3_scale_5.0_damage}
	}
	\subfloat[$d = \SI{50}{\micro\meter}$]{  
		\includegraphics[width=0.46\textwidth,trim=0cm 3cm 1.2cm 3cm,clip]{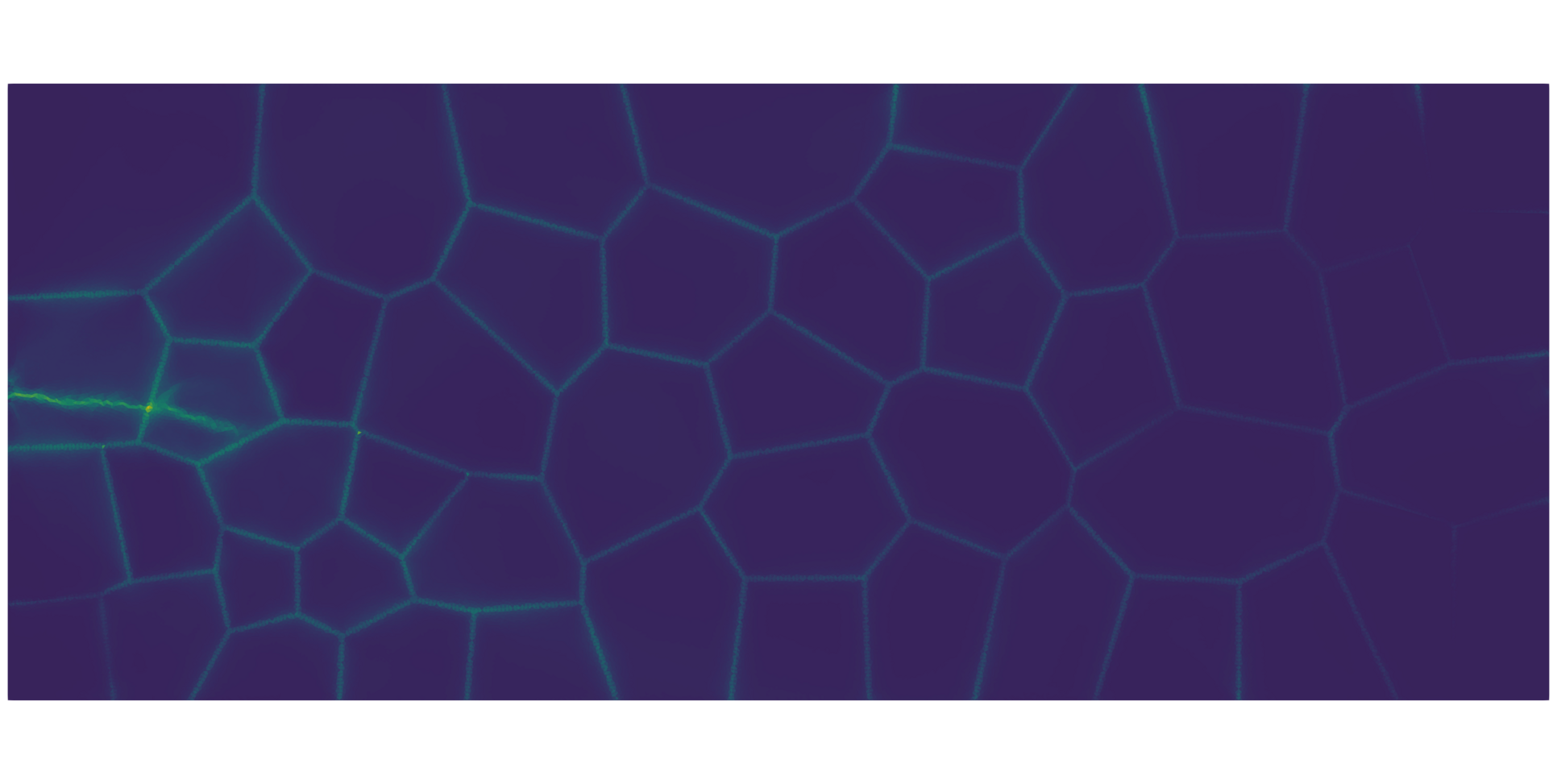}
		\label{subfig:surfing_tau0_1e-3_scale_5.0_dislocation_density}
	}
	\subfloat{  
	    \includegraphics[width=0.075\textwidth]{fig100.png}
	}
	\caption{Phase field, accumulated plastic slip field (a, c, e) and dislocation density field (b, d, f) in a polycrystal submitted to surfing boundary conditions. The ductility ratio $q = r_{pl}/r_{pz}$ is equal to $10^3$.}
	\label{fig:surfing_tau0_1e-3}
\end{figure}

Figures~\ref{subfig:surfing_tau0_1e-3_scale_0.5_damage},~\ref{subfig:surfing_tau0_1e-3_scale_2.0_damage} and~\ref{subfig:surfing_tau0_1e-3_scale_5.0_damage} show where the phase field $\alpha$ is greater than 0.5 (red colorscale), superimposed on the accumulated plastic slip field (blue colorscale) for three different grain sizes and a fixed value of the non-dimensional ductility ratio $q = 1.0 \times 10^{3}$. The Figures~\ref{subfig:surfing_tau0_1e-3_scale_0.5_dislocation_density},~\ref{subfig:surfing_tau0_1e-3_scale_2.0_dislocation_density} and~\ref{subfig:surfing_tau0_1e-3_scale_5.0_dislocation_density} show the corresponding total dislocation density. These plots are all shown at the same time step, \textit{i.e.} after the same loading history. We can observe that the crack propagates a longer distance in the microstructure with smaller grains and the length of propagation decreases with increasing grain size. As the grain size decreases, the Hall-Petch size effect increases and thus, the maximum stress and plastic strain amplitude reached ahead of the crack tip increase. Since the fracture toughness $G_c$ is kept independent of the grain size, the crack propagates more easily in the microstructure with smaller grains. Similarly to what was observed in Section~\ref{sec:nucleation}, the apparent width of the crack increases as the grain size decreases, because the non-dimensionalized phase field length $\ell_0$ is scaled with the factor $\eta L_0$. For the ductility ratio of $q = 1.0 \times 10^3$, the crack path is jagged because of the plastic activity ahead of the crack tip. Cracks bifurcate at grain boundaries and are attracted by triple junctions. These regions concentrate stresses, plastic strain gradients and dislocation density.

\subsection{Ductility effect}
\label{subsec:surfing_ductility_effect}

\begin{figure}
	\centering
	\subfloat[$d = \SI{5}{\micro\meter}$]{  
		\includegraphics[width=0.46\textwidth,trim=0cm 3cm 1.2cm 3cm,clip]{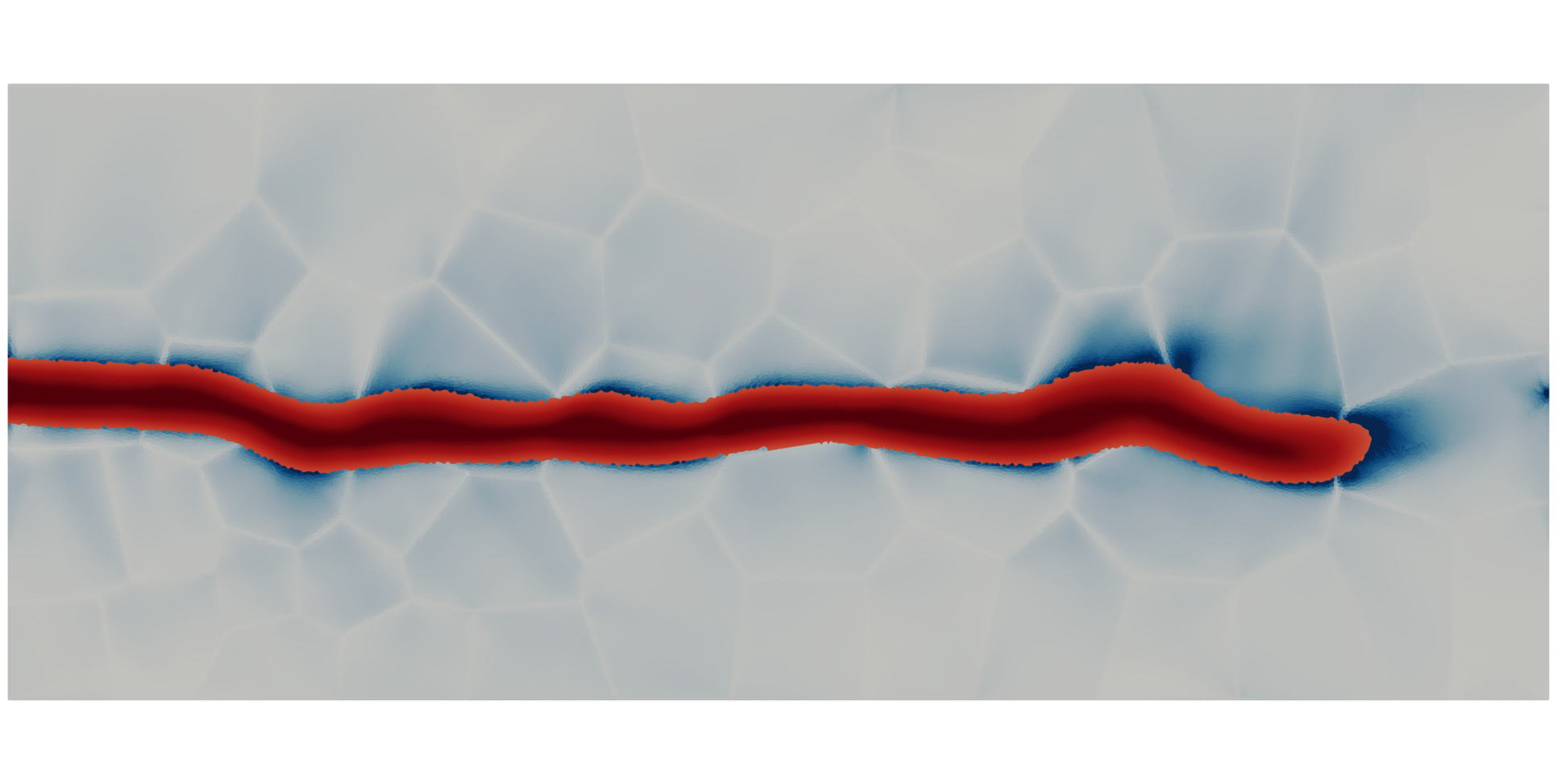}
		\label{subfig:surfing_tau0_2e-3_scale_0.5_damage}
	}
	\subfloat[$d = \SI{5}{\micro\meter}$]{  
		\includegraphics[width=0.46\textwidth,trim=0cm 3cm 1.2cm 3cm,clip]{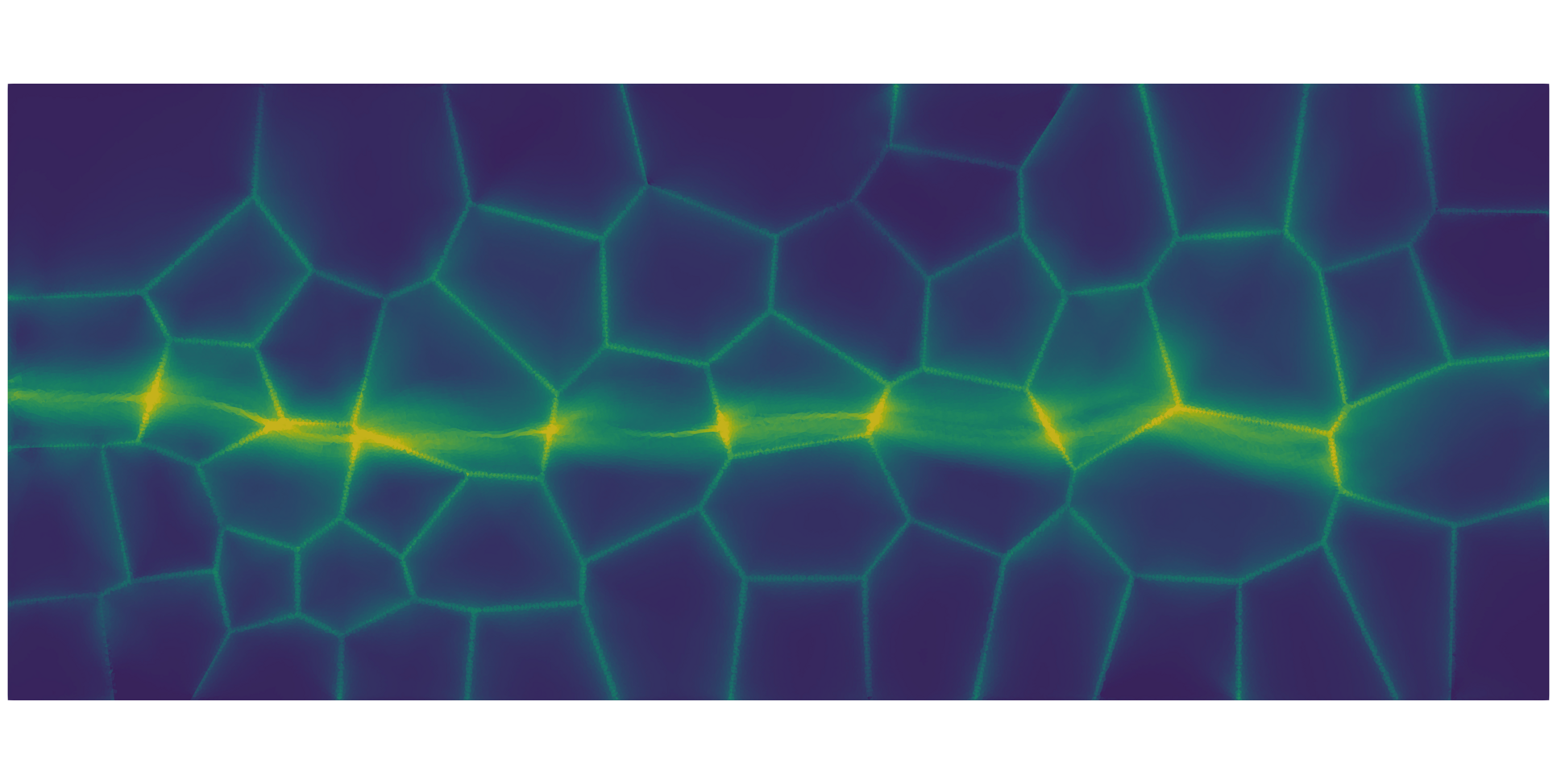}
		\label{subfig:surfing_tau0_2e-3_scale_0.5_dislocation_density}
	}
	\subfloat{  
	    \includegraphics[width=0.076\textwidth]{fig94.png}
	}\setcounter{subfigure}{2}
    \\
	\subfloat[$d = \SI{20}{\micro\meter}$]{  
		\includegraphics[width=0.46\textwidth,trim=0cm 3cm 1.2cm 3cm,clip]{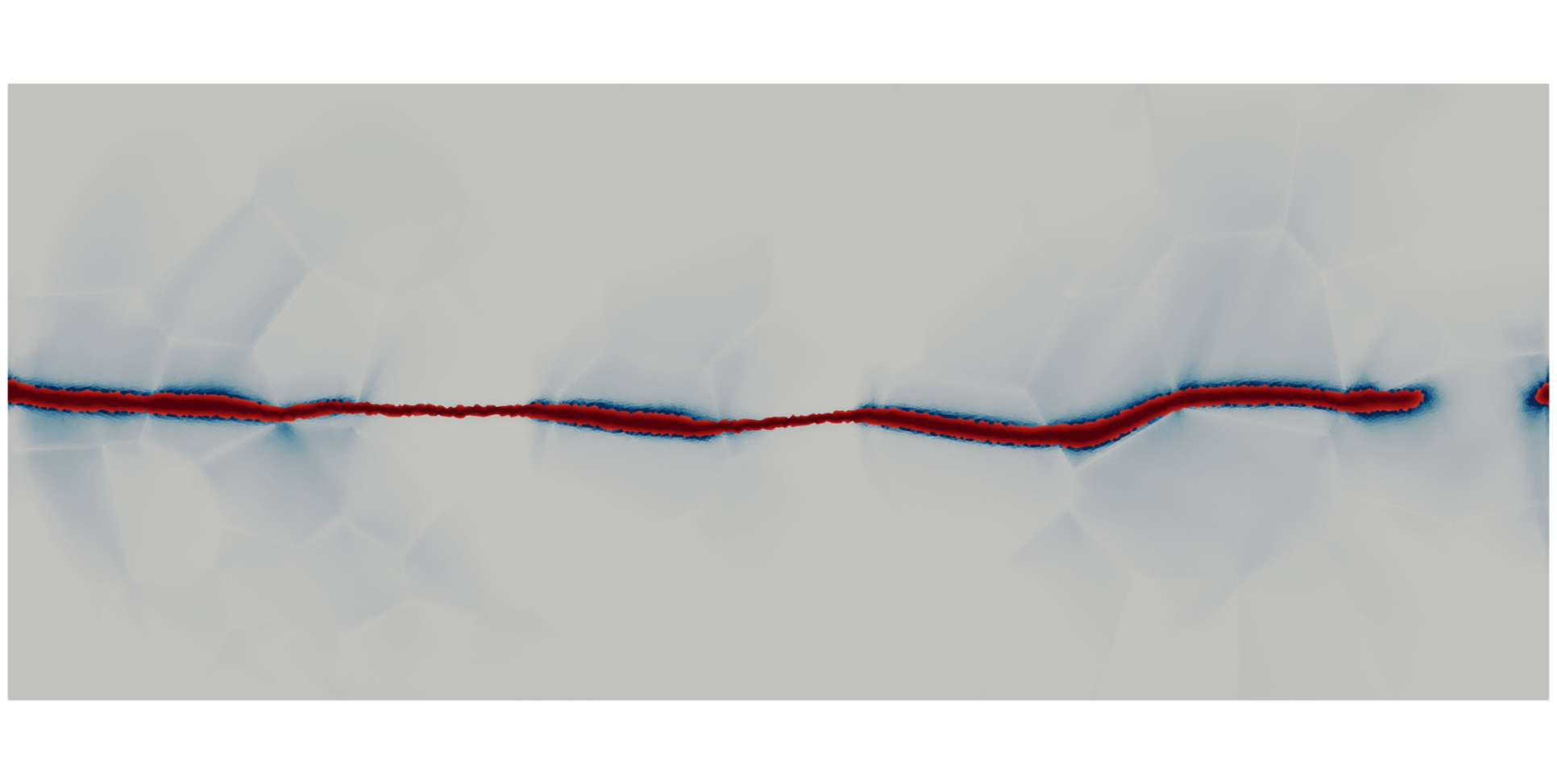}
		\label{subfig:surfing_tau0_2e-3_scale_2.0_damage}
	}
	\subfloat[$d = \SI{20}{\micro\meter}$]{  
		\includegraphics[width=0.46\textwidth,trim=0cm 3cm 1.2cm 3cm,clip]{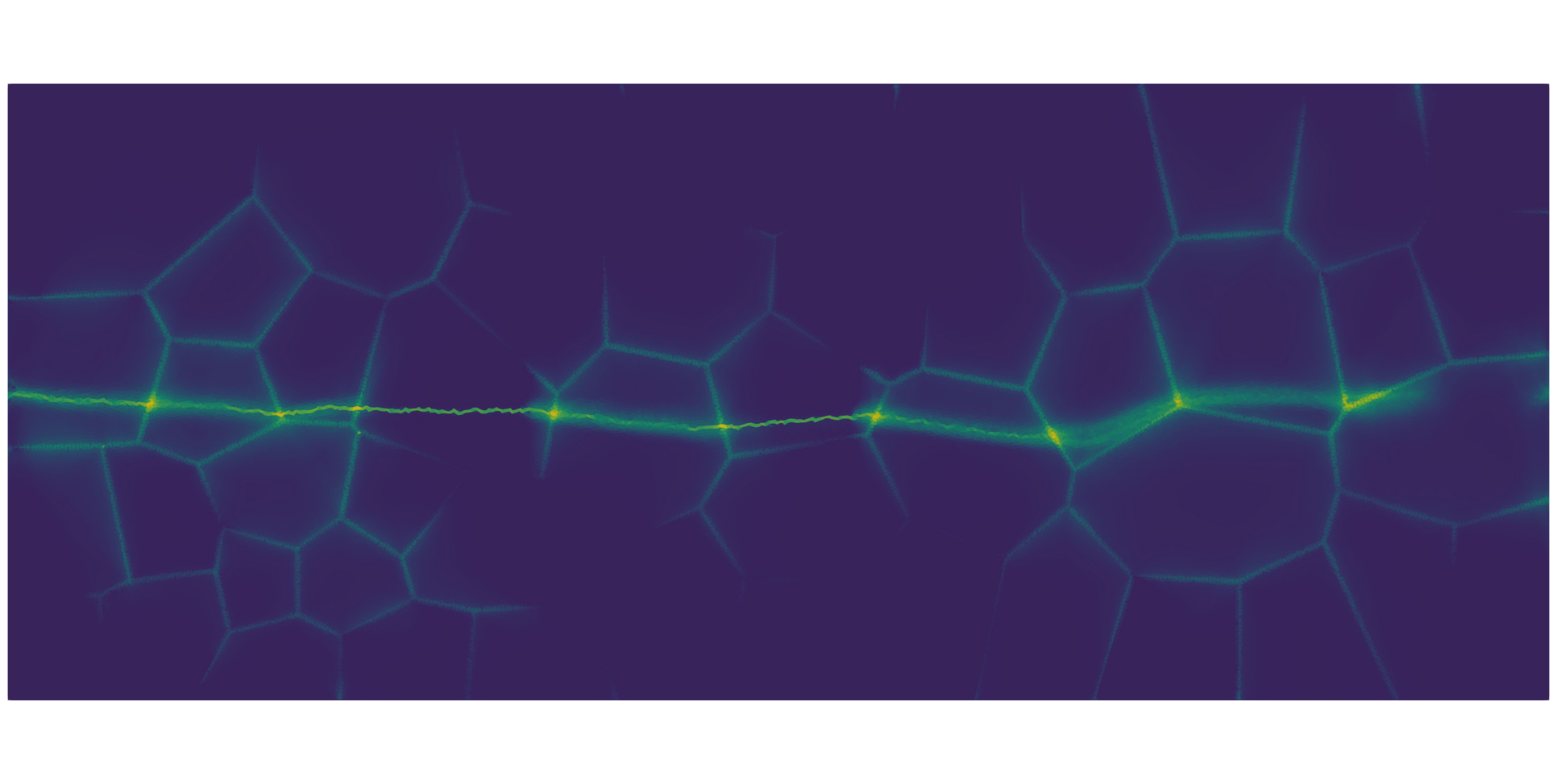}
		\label{subfig:surfing_tau0_2e-3_scale_2.0_dislocation_density}
	}
	\subfloat{  
	    \includegraphics[width=0.077\textwidth]{fig97.png}
	}\setcounter{subfigure}{4}
    \\
	\subfloat[$d = \SI{50}{\micro\meter}$]{  
		\includegraphics[width=0.46\textwidth,trim=0cm 3.cm 0cm 3.cm,clip]{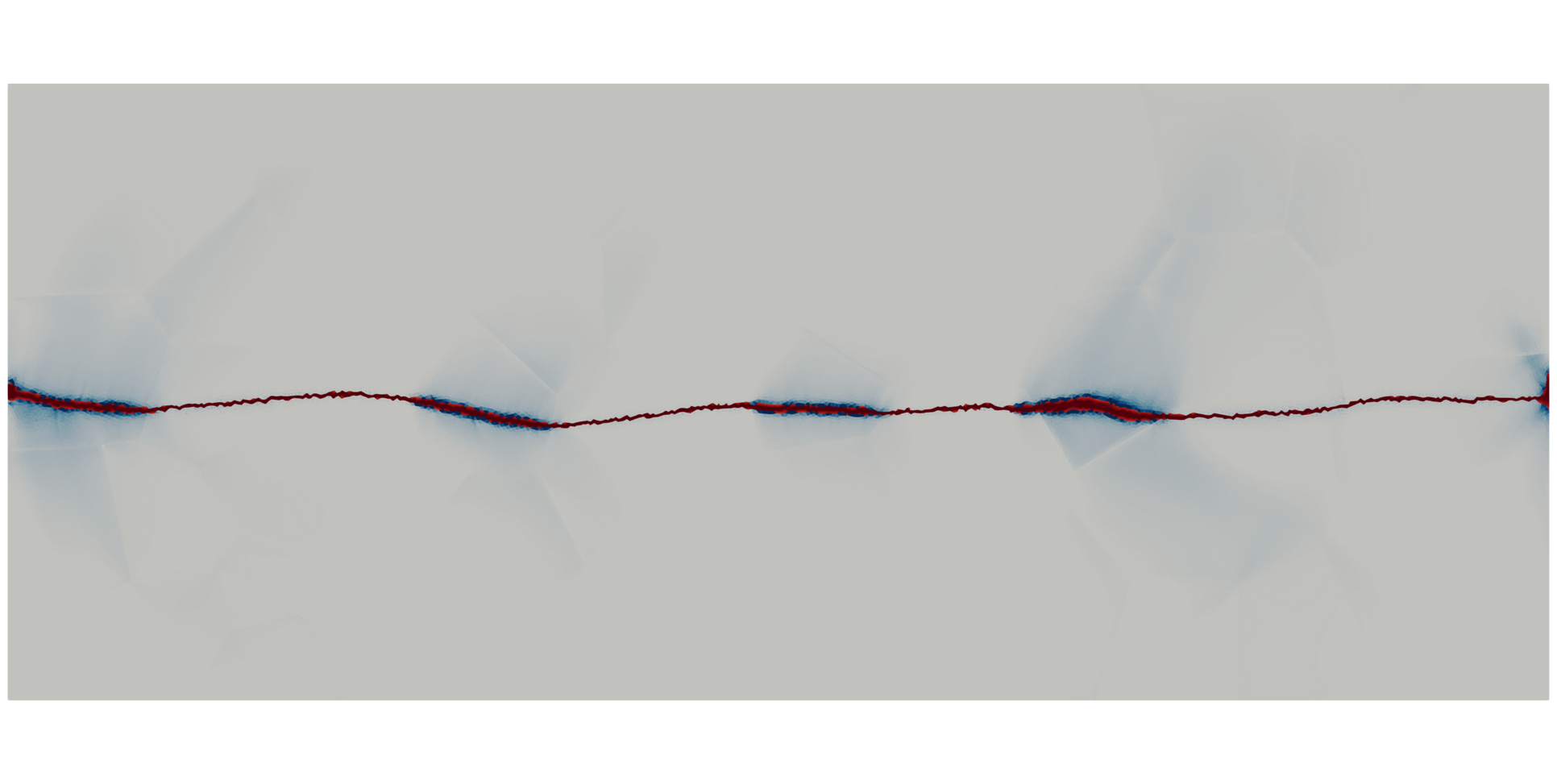}
		\label{subfig:surfing_tau0_2e-3_scale_5.0_damage}
	}
	\subfloat[$d = \SI{50}{\micro\meter}$]{  
		\includegraphics[width=0.46\textwidth,trim=0cm 3.cm 0cm 3.cm,clip]{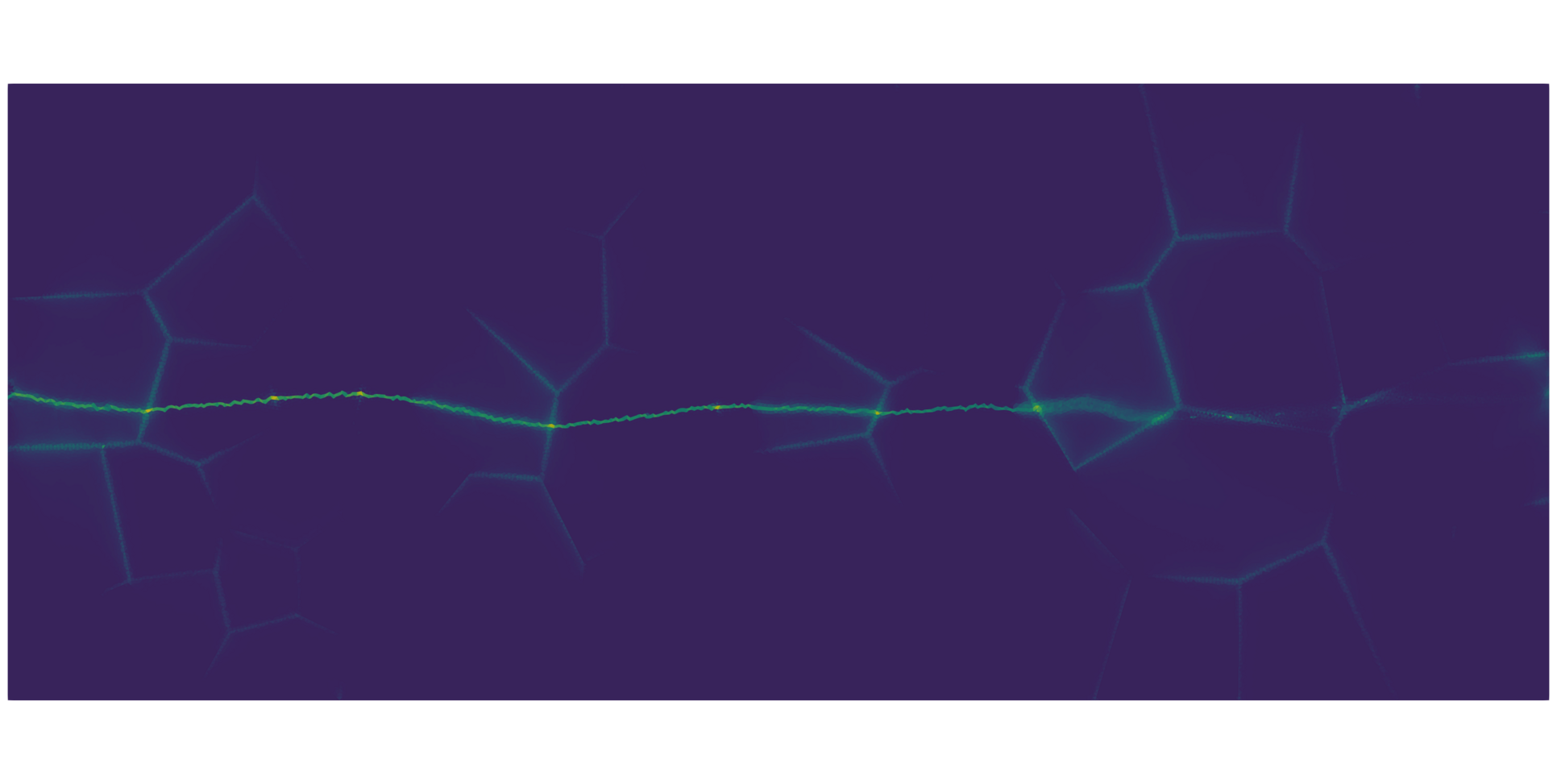}
		\label{subfig:surfing_tau0_2e-3_scale_5.0_dislocation_density}
	}
	\subfloat{  
	    \includegraphics[width=0.075\textwidth]{fig100.png}
	}
	\caption{Phase field, accumulated plastic slip field (a, c, e) and dislocation density field (b, d, f) in a polycrystal submitted to surfing boundary conditions. The ductility ratio $q = r_{pl}/r_{pz}$ is equal to $5\times 10^2$.}
	\label{fig:surfing_tau0_2e-3}
\end{figure}

\begin{figure}
	\centering
	\subfloat[$d = \SI{5}{\micro\meter}$]{  
		\includegraphics[width=0.46\textwidth,trim=0cm 3cm 1.2cm 3cm,clip]{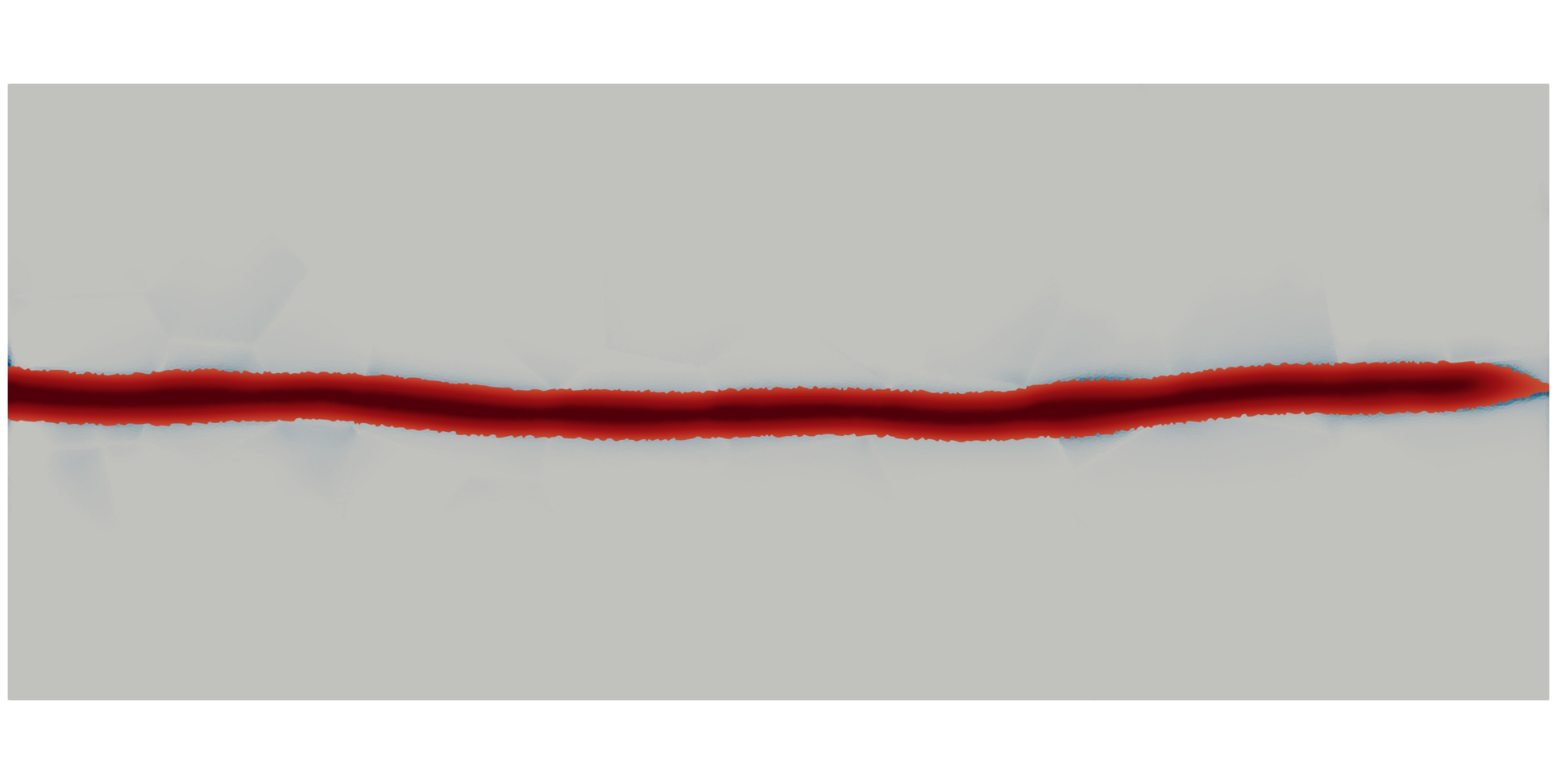}
		\label{subfig:surfing_tau0_4e-3_scale_0.5_damage}
	}
	\subfloat[$d = \SI{5}{\micro\meter}$]{  
		\includegraphics[width=0.46\textwidth,trim=0cm 3cm 1.2cm 3cm,clip]{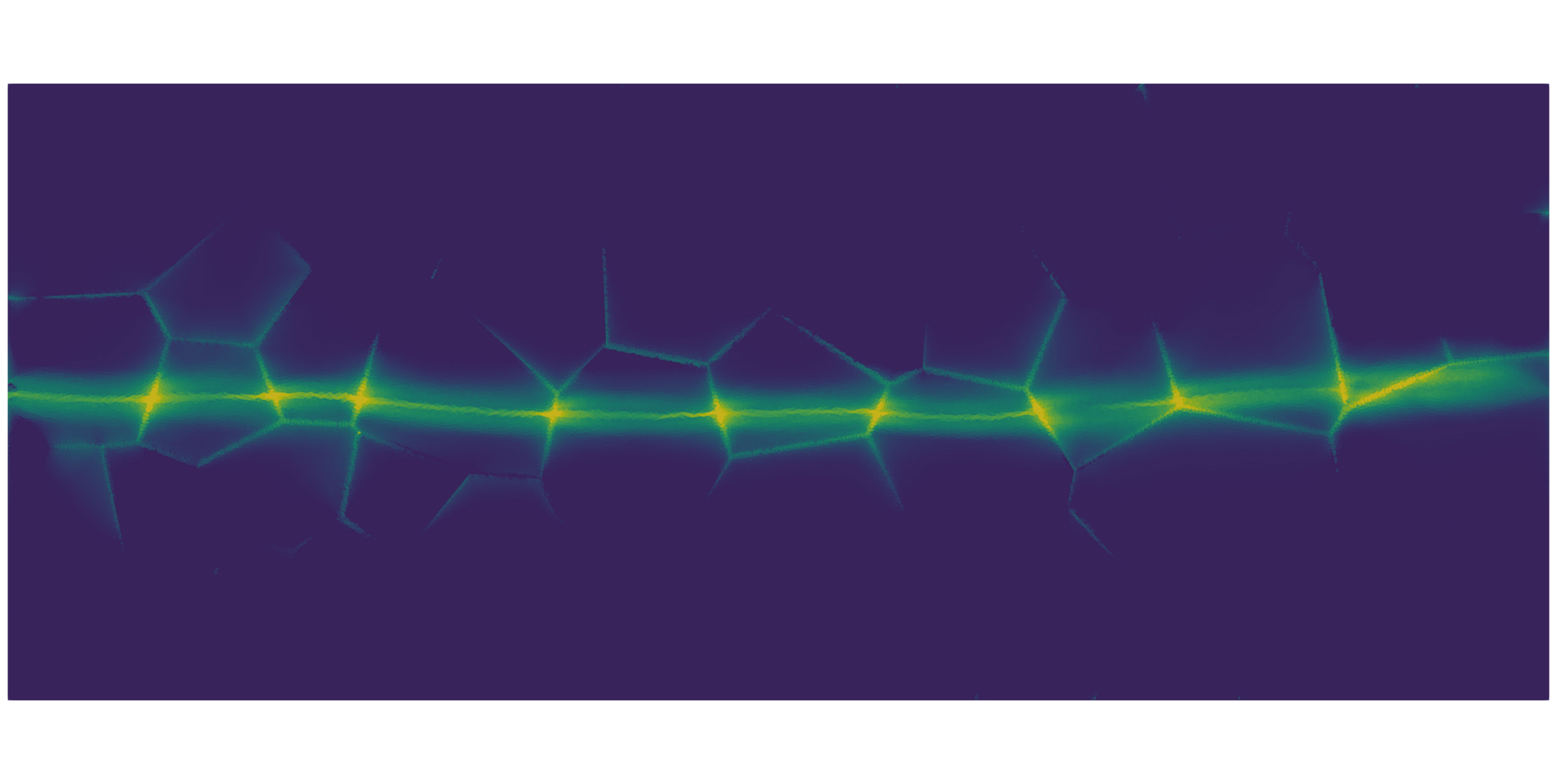}
		\label{subfig:surfing_tau0_4e-3_scale_0.5_dislocation_density}
	}
	\subfloat{  
	    \includegraphics[width=0.076\textwidth]{fig94.png}
	}\setcounter{subfigure}{2}
    \\
	\subfloat[$d = \SI{20}{\micro\meter}$]{  
		\includegraphics[width=0.46\textwidth,trim=0cm 3cm 1.2cm 3cm,clip]{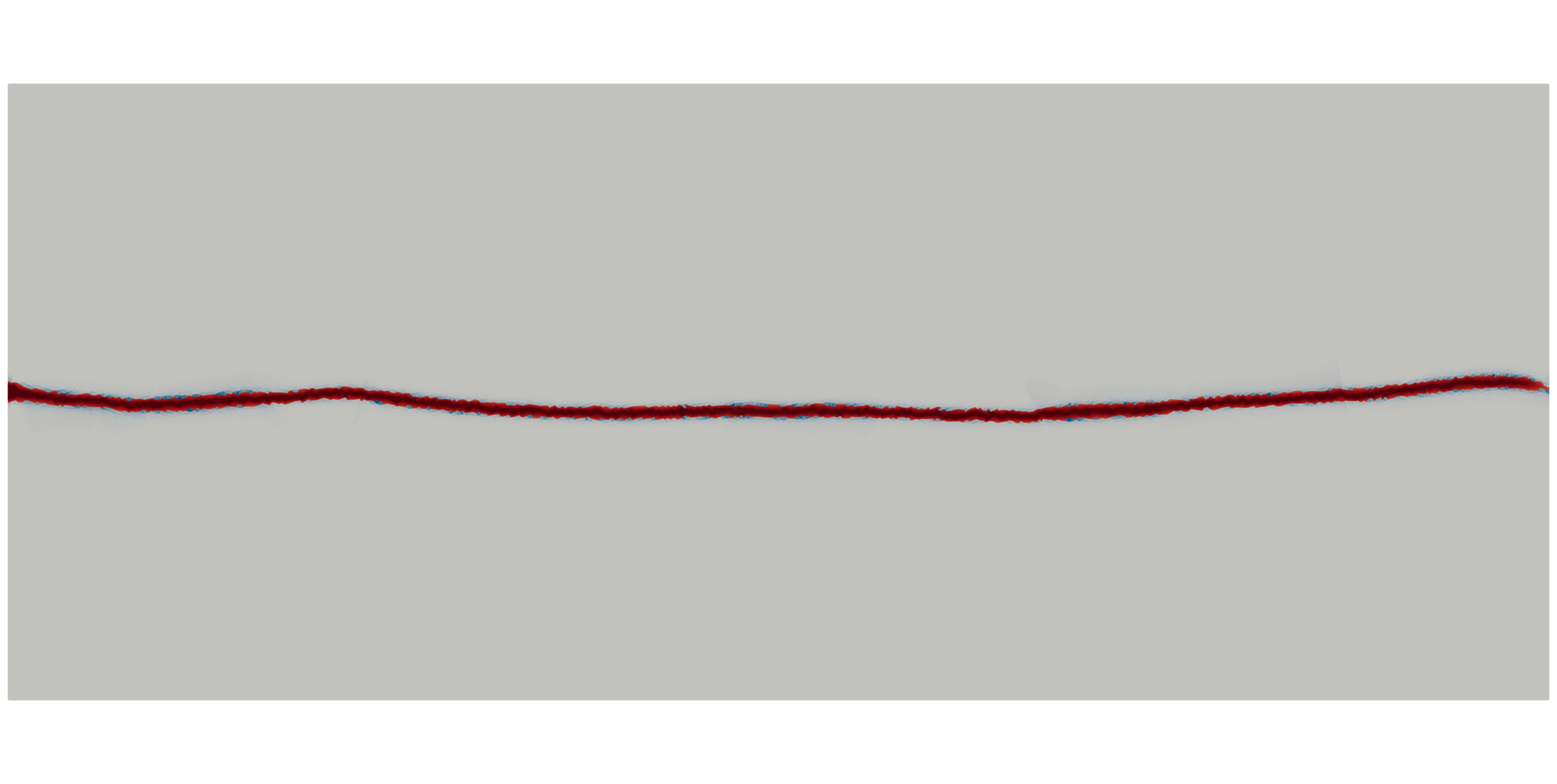}
		\label{subfig:surfing_tau0_4e-3_scale_2.0_damage}
	}
	\subfloat[$d = \SI{20}{\micro\meter}$]{  
		\includegraphics[width=0.46\textwidth,trim=0cm 3cm 1.2cm 3cm,clip]{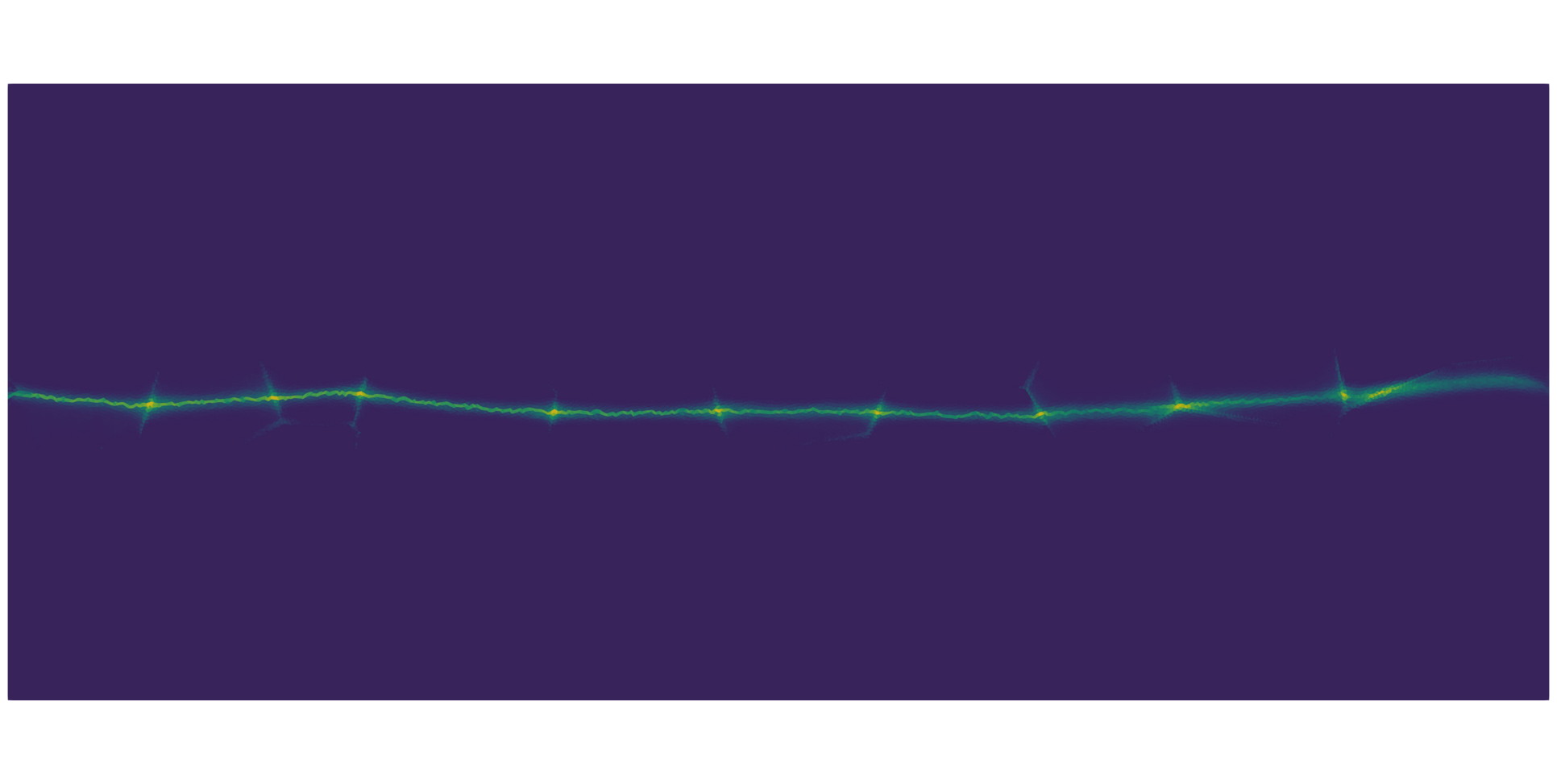}
		\label{subfig:surfing_tau0_4e-3_scale_2.0_dislocation_density}
	}
	\subfloat{  
	    \includegraphics[width=0.077\textwidth]{fig97.png}
	}\setcounter{subfigure}{4}
    \\
	\subfloat[$d = \SI{50}{\micro\meter}$]{  
		\includegraphics[width=0.46\textwidth,trim=0cm 3.cm 0cm 3.cm,clip]{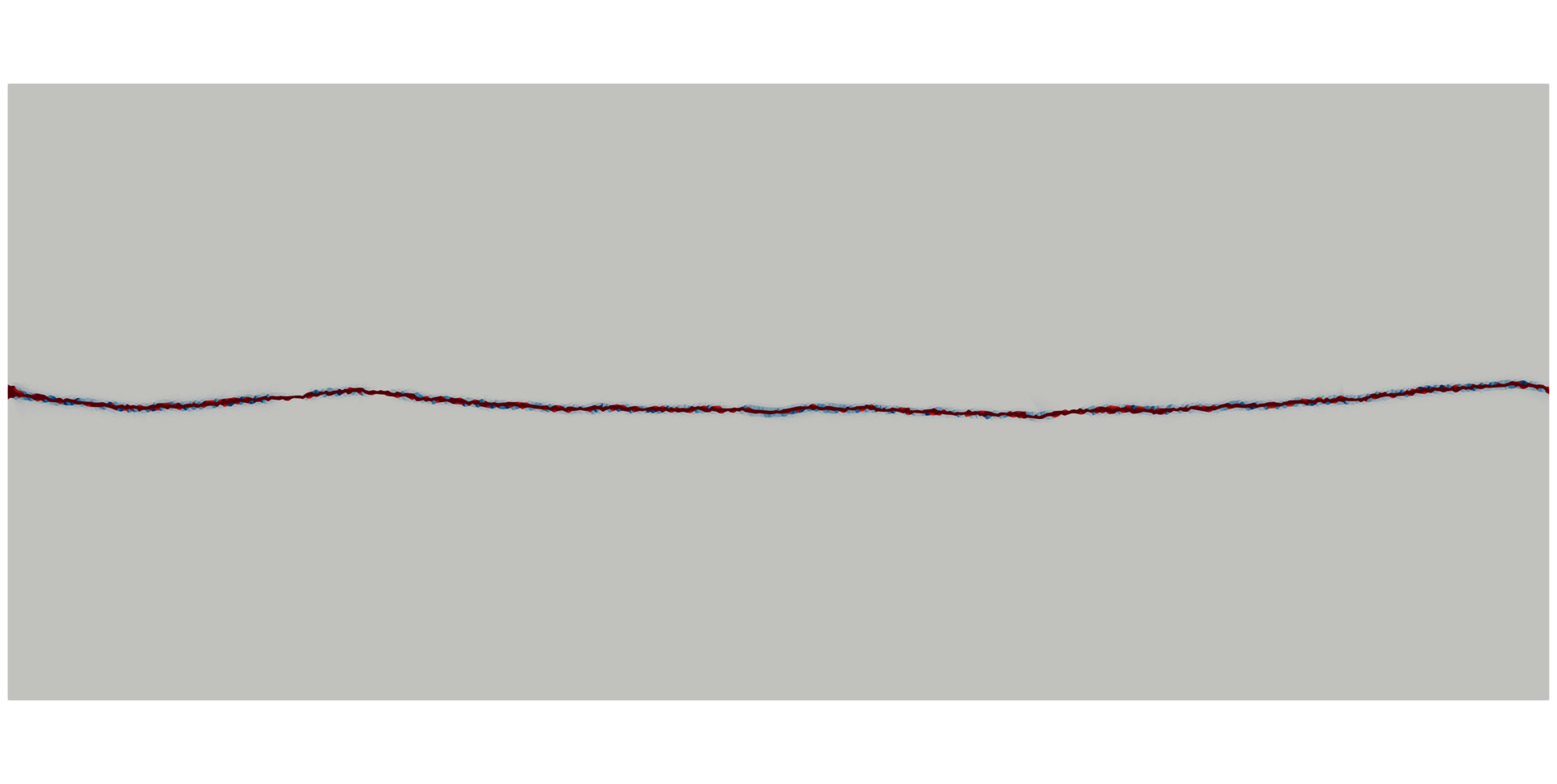}
		\label{subfig:surfing_tau0_4e-3_scale_5.0_damage}
	}
	\subfloat[$d = \SI{50}{\micro\meter}$]{  
		\includegraphics[width=0.46\textwidth,trim=0cm 3.cm 0cm 3.cm,clip]{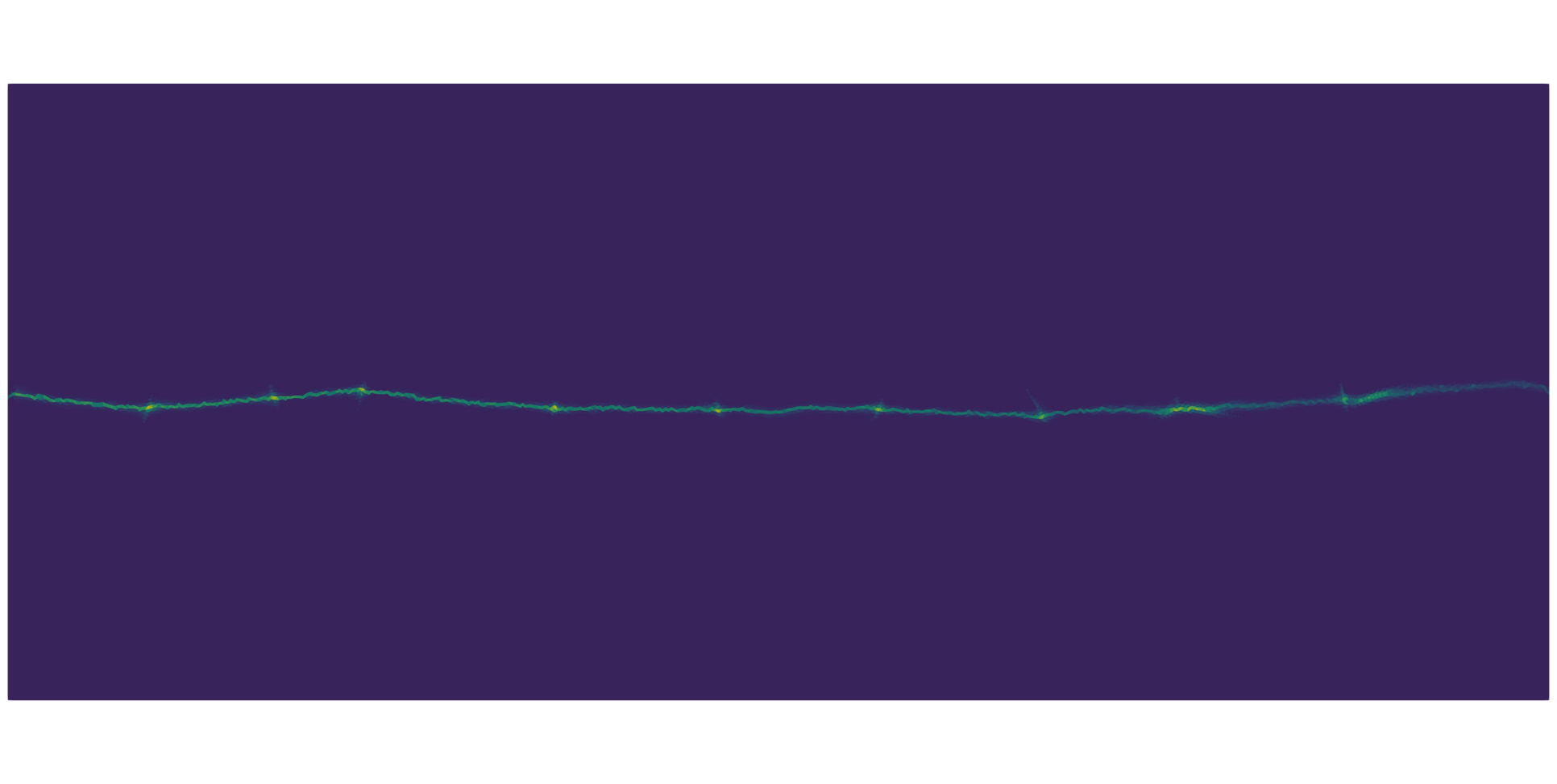}
		\label{subfig:surfing_tau0_4e-3_scale_5.0_dislocation_density}
	}
	\subfloat{  
	    \includegraphics[width=0.075\textwidth]{fig100.png}
	}
	\caption{Phase field, accumulated plastic slip field (a, c, e) and dislocation density field (b, d, f) in a polycrystal submitted to surfing boundary conditions. The ductility ratio $q = r_{pl}/r_{pz}$ is equal to $2.5\times 10^2$.}
	\label{fig:surfing_tau0_4e-3}
\end{figure}

Figure~\ref{subfig:surfing_ductility_jintegral} highlights the strong competition between plastic relaxation and crack propagation~\citep{rice1992dislocation}. The plastic activity, which corresponds in practice to dislocation emission, shields the crack tip and thus delays crack growth. The ductility ratio controls the balance between these two mechanisms. For the lowest value of $q$, the crack propagates in an almost brittle manner with little plastic dissipation. For the intermediate ductility, the crack propagates in an unsteady manner, with alternating phases of plastic dissipation and crack jumps~\citep{brach2019phase,brach2020effects}. For the highest ductility, the crack propagates continuously and the $J$-integral increases steadily as both mechanisms are active simultaneously. As the ductility increases, \textit{i.e.} when $\tau_0$ decreases and $q$ increases, with constant fracture properties, the crack propagation transitions from an \textit{elasticity-driven} to a \textit{plasticity-driven} behaviour and the crack path becomes more tortuous. Figure~\ref{fig:surfing_tau0_2e-3} and~\ref{fig:surfing_tau0_4e-3} show the phase field $\alpha$ and the dislocation density for two other values of the non-dimensional ductility ratio of $q = 5\times 10^{2}$ and $q = 2.5\times 10^{2}$ respectively. In both figures, the crack paths, plastic strains and total dislocation density are shown for three grain sizes.

At the intermediate ductility ($q = 5\times 10^2$), the crack propagation occurs in successive jumps separated by periods of slow crack growth accompanied with crack tip blunting. Blunting is particularly significant in the microstructures with larger grain sizes in Figures~\ref{subfig:surfing_tau0_2e-3_scale_2.0_damage},~\ref{subfig:surfing_tau0_2e-3_scale_2.0_dislocation_density},~\ref{subfig:surfing_tau0_2e-3_scale_5.0_damage} and~\ref{subfig:surfing_tau0_2e-3_scale_5.0_dislocation_density}. During the blunting phases, the crack growth rate is much slower than the imposed macroscopic crack growth velocity $v_0$ in Eq.~\eqref{eq:surfing_bc}. This is due to a significant amount of energy being dissipated by plastic slip ahead of the crack tip. As the crack lags behind, significant amount of elastic energy builds up ahead of the crack tip. This energy is then released in a sudden jump of the crack.~\cite{friedel2013dislocations} already noted that "\textit{a plastically relaxed crack can [...] move forward only by becoming elastic or nearly elastic again [and that] this is most easily done in fairly brittle materials [...]}".~\cite{tetelman1963direct} observed such alternating crack tip blunting and brittle crack propagation in hydrogenated 3\% silicon-iron single crystals. This effect was also observed by~\cite{brach2019phase,brach2020effects} in the phase field fracture modelling of isotropic plastic materials. In our polycrstal simulations, the length of these crack jumps  are comparable to the grain size. This highlights the importance of the grain boundaries for interrupting unstable crack growth. The crack propagation is more continuous in the microstructures with smaller grains (see Figure~\ref{subfig:surfing_tau0_2e-3_scale_0.5_damage} and~\ref{subfig:surfing_tau0_2e-3_scale_0.5_dislocation_density}). In this case, the crack path is smoother and the crack growth rate is more uniform.

For the lowest ductility ($q = 2.5\times 10^2$), the crack growth is continuous. Due to the larger yield strength, the plastic zone radius is smaller. The dislocation density plots in Figures~\ref{subfig:surfing_tau0_4e-3_scale_0.5_dislocation_density},~\ref{subfig:surfing_tau0_4e-3_scale_2.0_dislocation_density} and~\ref{subfig:surfing_tau0_4e-3_scale_5.0_dislocation_density} show that the plastic activity is contained in a single layer of grains above and below the crack path. The microstructures with larger grain sizes display an almost elastic-brittle behaviour. In this case, the small deflections of the crack at grain boundaries are due to the elastic anisotropy rather than plasticity.

\subsection{Bimodal grain size microstructures}
\label{subsec:surfing_bimodal}

\begin{figure}
	\centering
	\subfloat[$d_1/d_2 = 2$] { 
	    \includegraphics[width=0.7\textwidth,trim=0cm 0cm 0cm 1cm,clip]{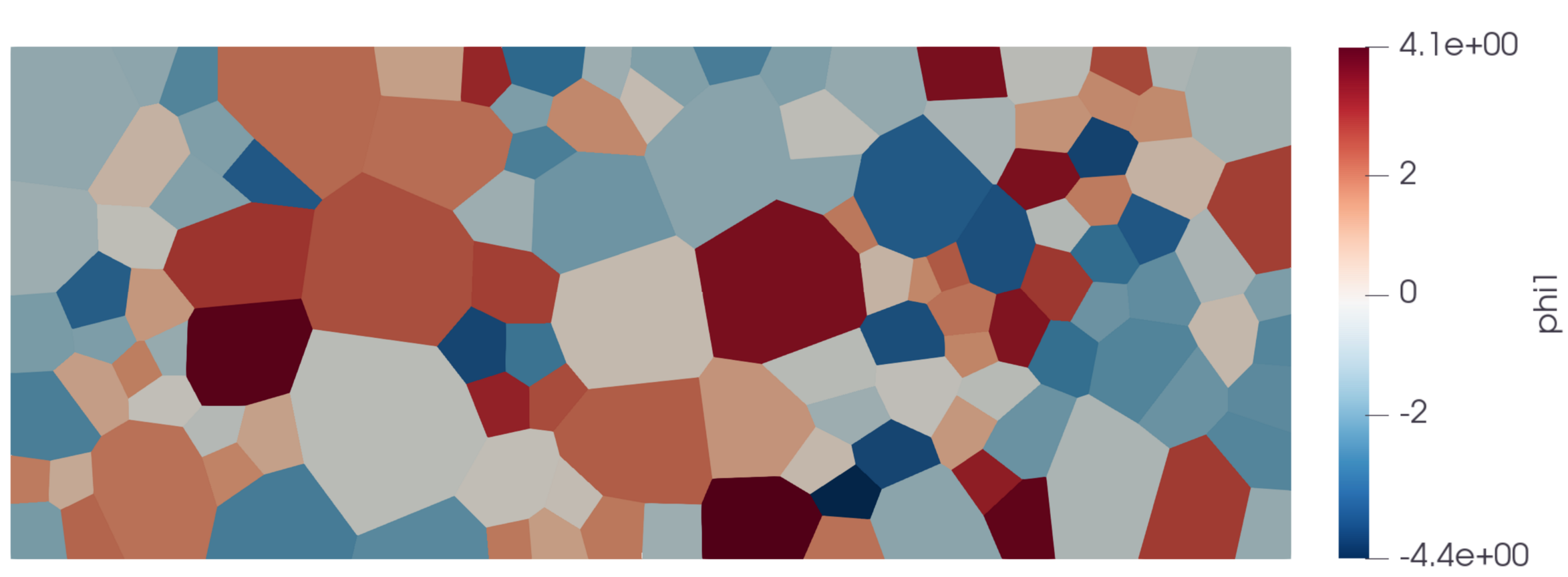}
	}\\
	\subfloat[$d_1/d_2 = 3$] {  
	    \includegraphics[width=0.7\textwidth,trim=0cm 0cm 0cm 1cm,clip]{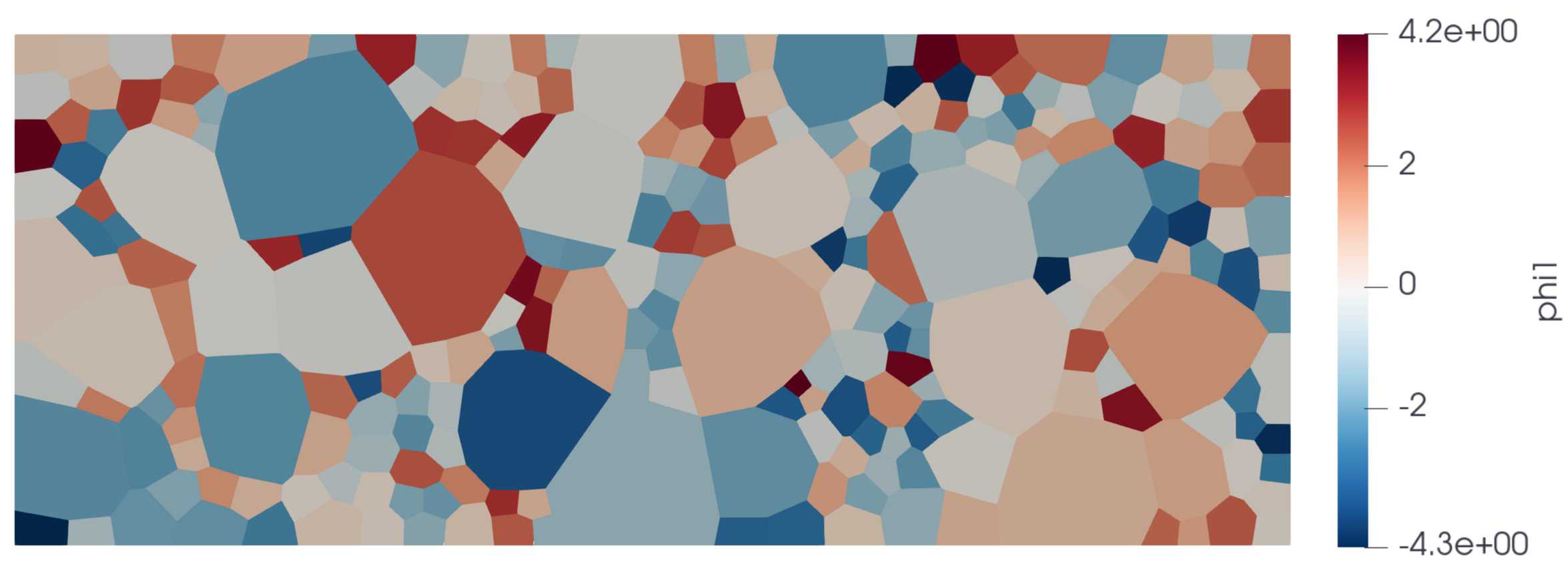}
	}
	\\
	\subfloat[$d_1/d_2 = 4$] { 
	    \includegraphics[width=0.7\textwidth,trim=0cm 0cm 0cm 1cm,clip]{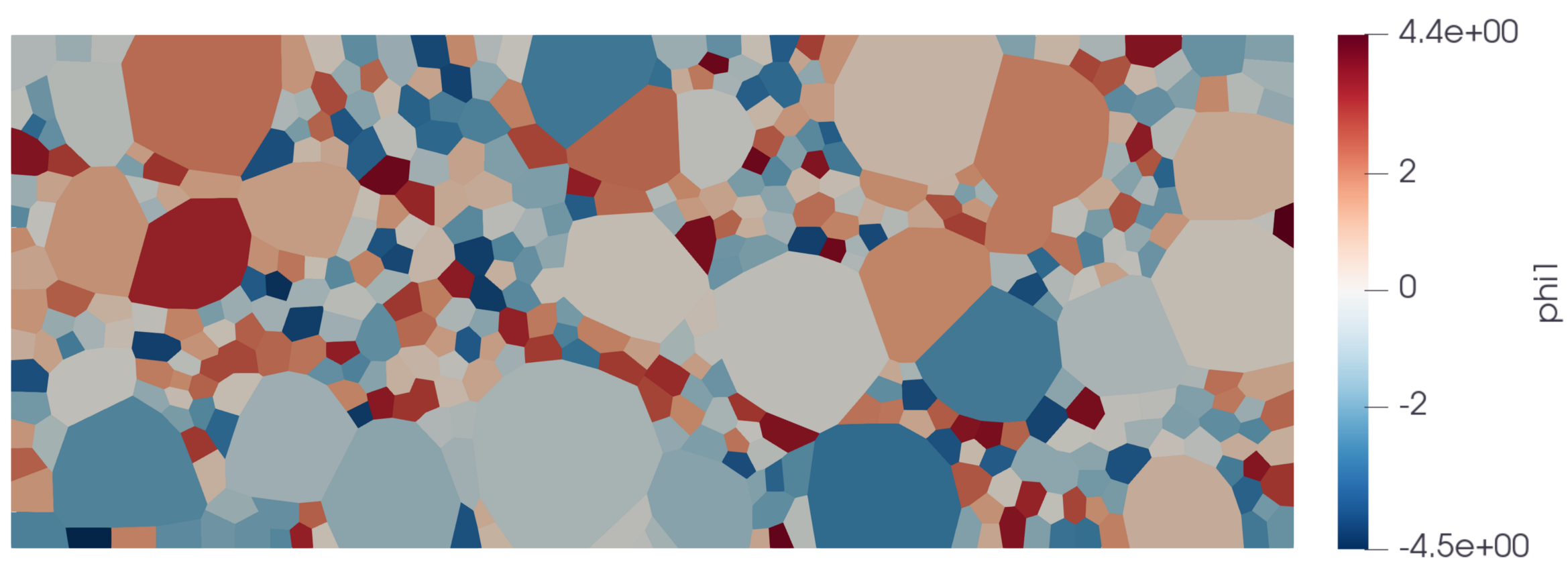}
	}
	\caption{Bimodal polycrystal microstructures, with two grain sizes $d_1$ and $d_2$, used to simulate crack propagation in plane strain conditions. The equivalent grain diameter of the first population is kept fixed at $d_1=\SI{20}{\micro\meter}$ while $d_2$ is varied. Each population of grains occupies half of the surface area. The colorscale represents the first Euler angle $\phi_1$.}
	\label{fig:surfing_microstructure_bimodal}
\end{figure}

We now investigate the propagation of cracks in polycrystals with a bimodal distribution of grain sizes. The microstructures used are shown in Figure~\ref{fig:surfing_microstructure_bimodal}. The average grain size of the two population of grains are denoted $d_1$ and $d_2$ respectively. Each population occupies half of the surface area of the whole domain. The ratio $d_1/d_2$ between the grain size of each population is varied from 1 to 4. The size of the first population of grains is kept constant at $d_1 = \SI{20}{\micro\meter}$ and the size of the second population of grains is varied accordingly. The value  $d_1/d_2 = 1$ corresponds to the equiaxed microstructure, with a single population of grains, shown in Figure~\ref{fig:surfing_microstructure}. The ductility ratio is set to $q = 5.0 \times 10^{2}$. 

\begin{figure}
	\centering
	\subfloat[$d_1 / d_2 = 2$]{  
		\includegraphics[width=0.46\textwidth,trim=0cm 3cm 1.2cm 3cm,clip]{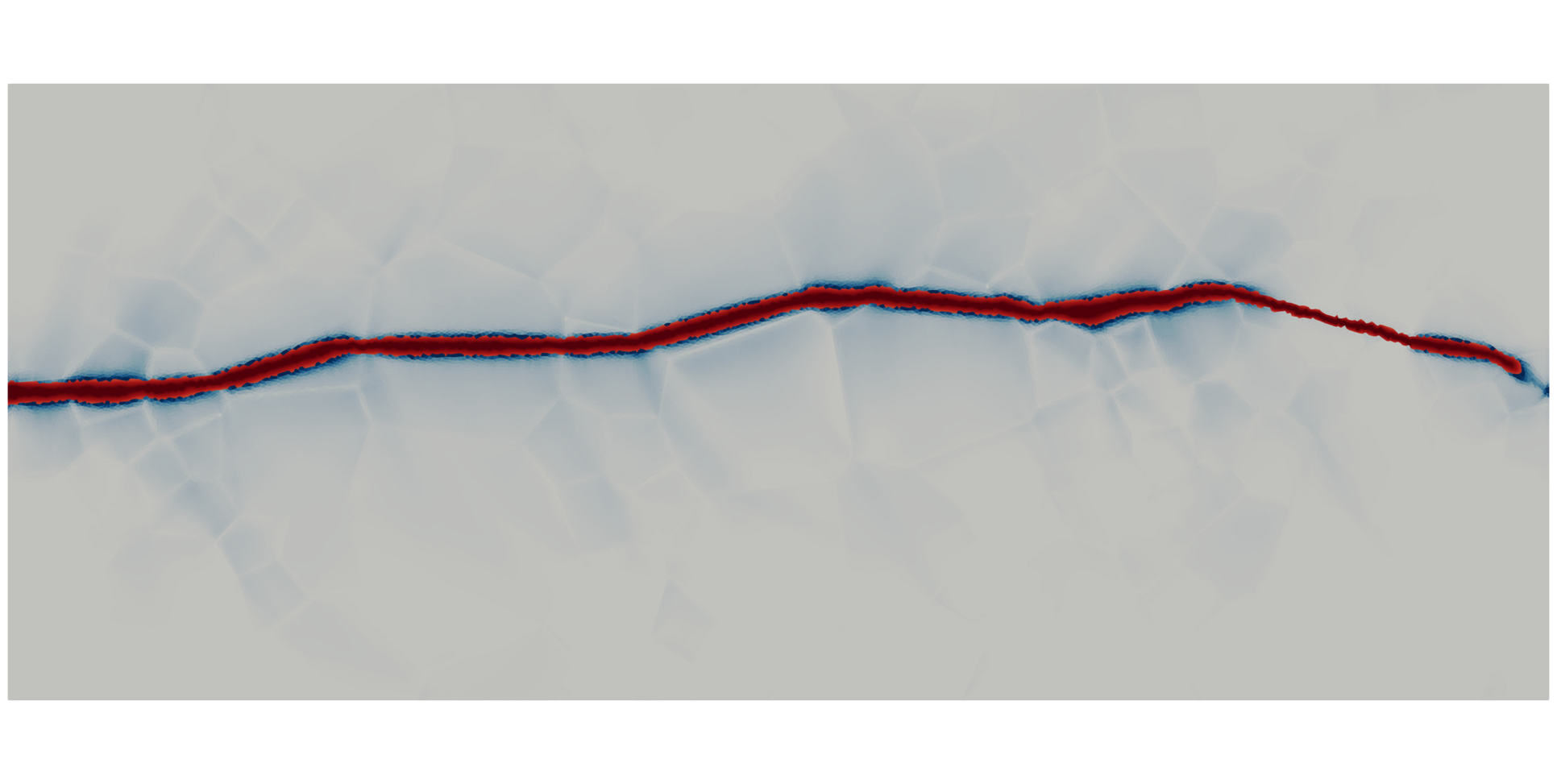}
		\label{subfig:surfing_tau0_2e-3_scale_2.0_mu1_0.1_mu2_0.05_damage}
	}
	\subfloat[$d_1 / d_2 = 2$]{  
		\includegraphics[width=0.46\textwidth,trim=0cm 3cm 1.2cm 3cm,clip]{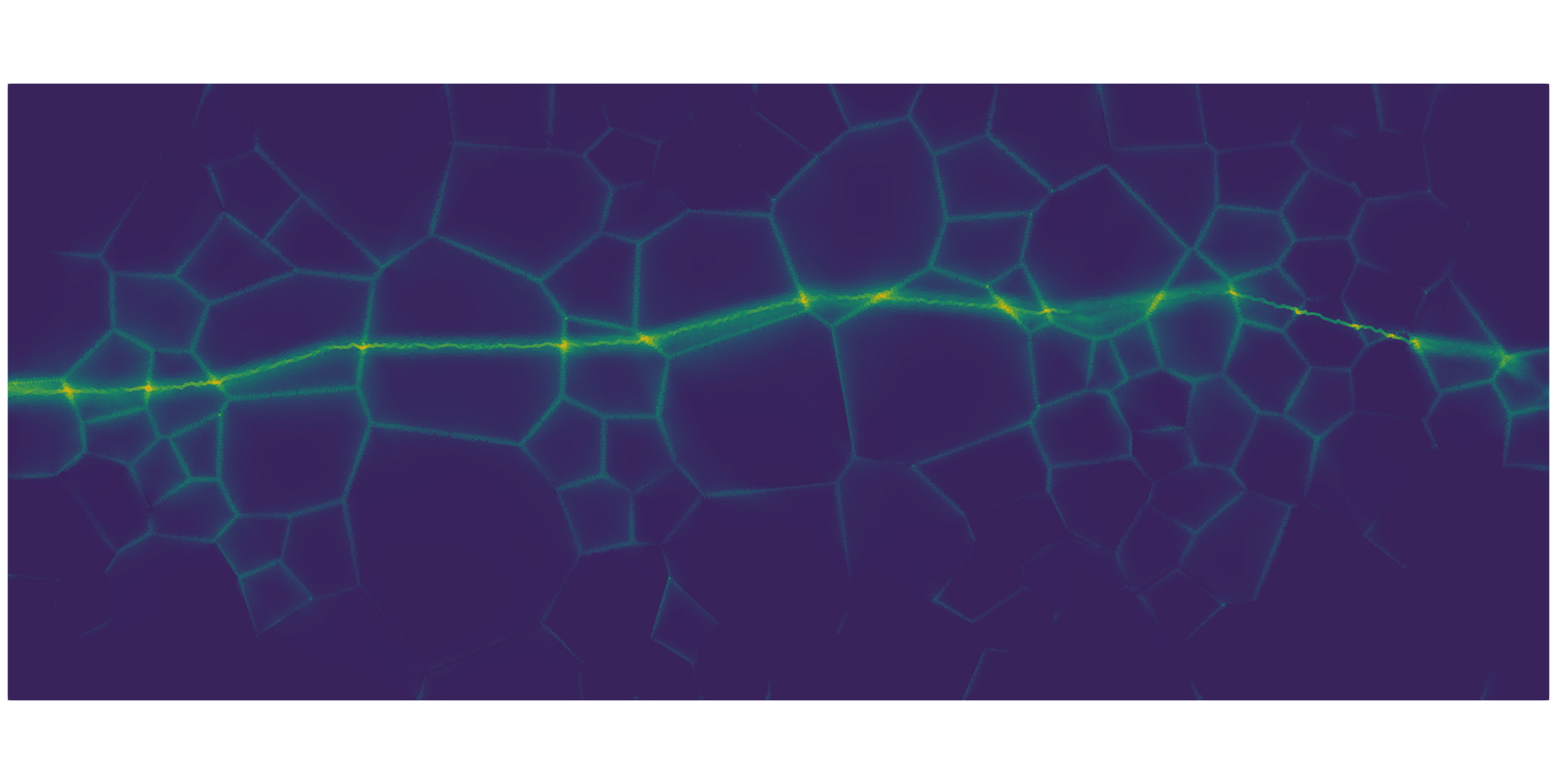}
		\label{subfig:surfing_tau0_2e-3_scale_2.0_mu1_0.1_mu2_0.05_dislocation_density}
	}
	\subfloat{  
	    \includegraphics[width=0.076\textwidth]{fig94.png}
	}\setcounter{subfigure}{2}
    \\
	\subfloat[$d_1 / d_2 = 3$]{  
		\includegraphics[width=0.46\textwidth,trim=0cm 3cm 1.2cm 3cm,clip]{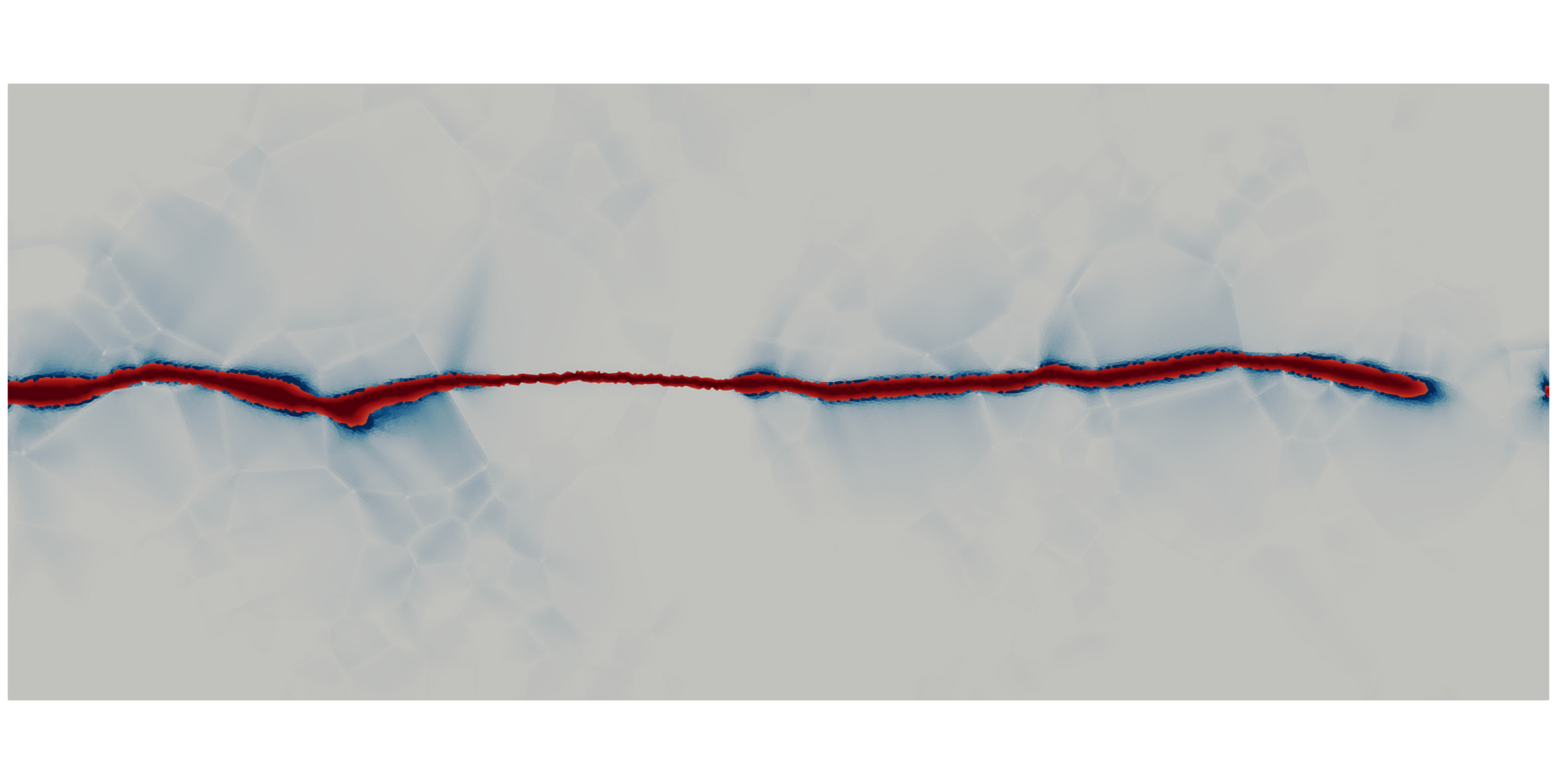}
		\label{subfig:surfing_tau0_2e-3_scale_2.0_mu1_0.1_mu2_0.033_damage}
	}
	\subfloat[$d_1 / d_2 = 3$]{  
		\includegraphics[width=0.46\textwidth,trim=0cm 3cm 1.2cm 3cm,clip]{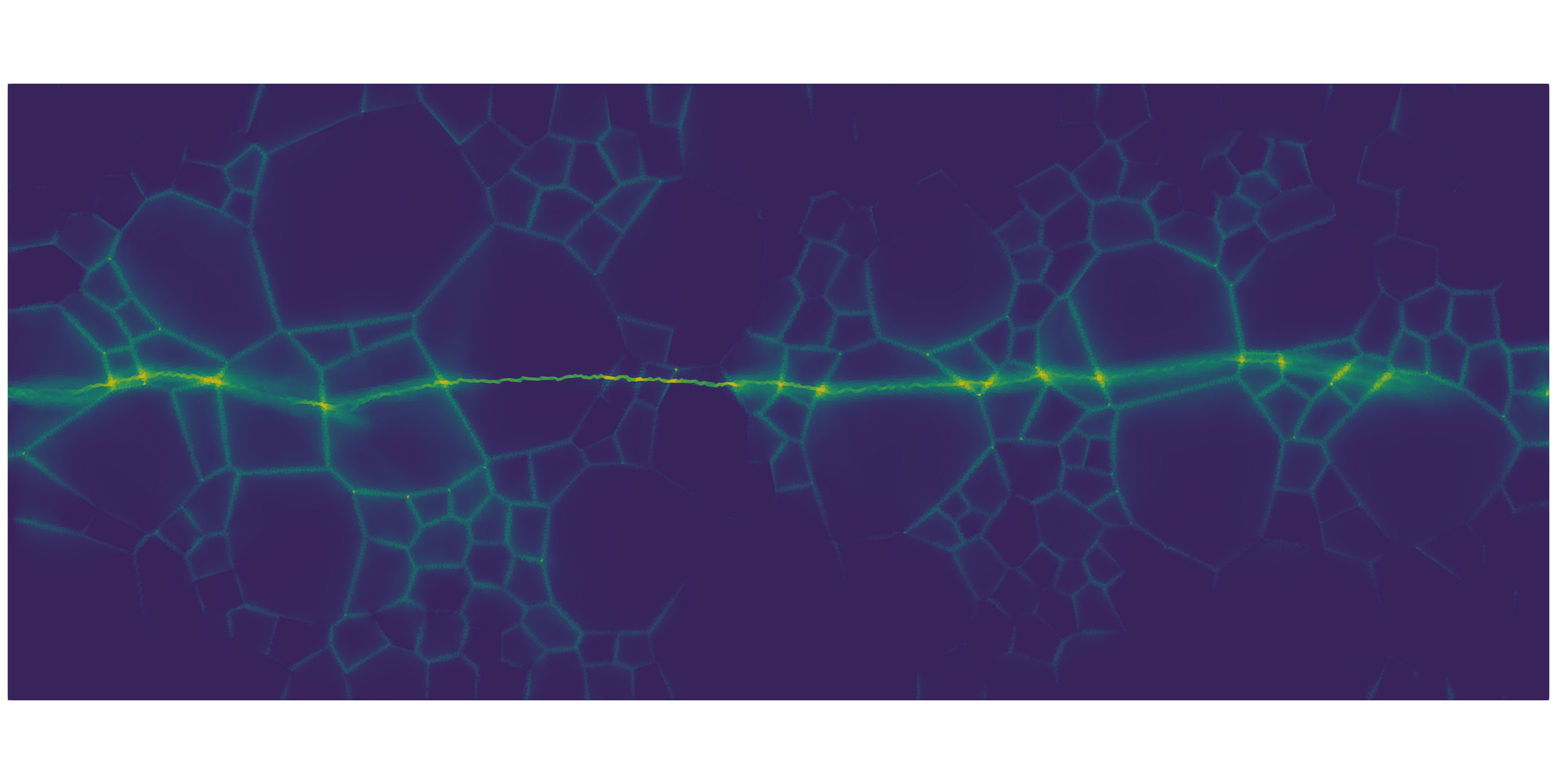}
		\label{subfig:surfing_tau0_2e-3_scale_2.0_mu1_0.1_mu2_0.033_dislocation_density}
	}
	\subfloat{  
	    \includegraphics[width=0.077\textwidth]{fig97.png}
	}\setcounter{subfigure}{4}
    \\
	\subfloat[$d_1 / d_2 = 4$]{  
		\includegraphics[width=0.46\textwidth,trim=0cm 3cm 1.2cm 3cm,clip]{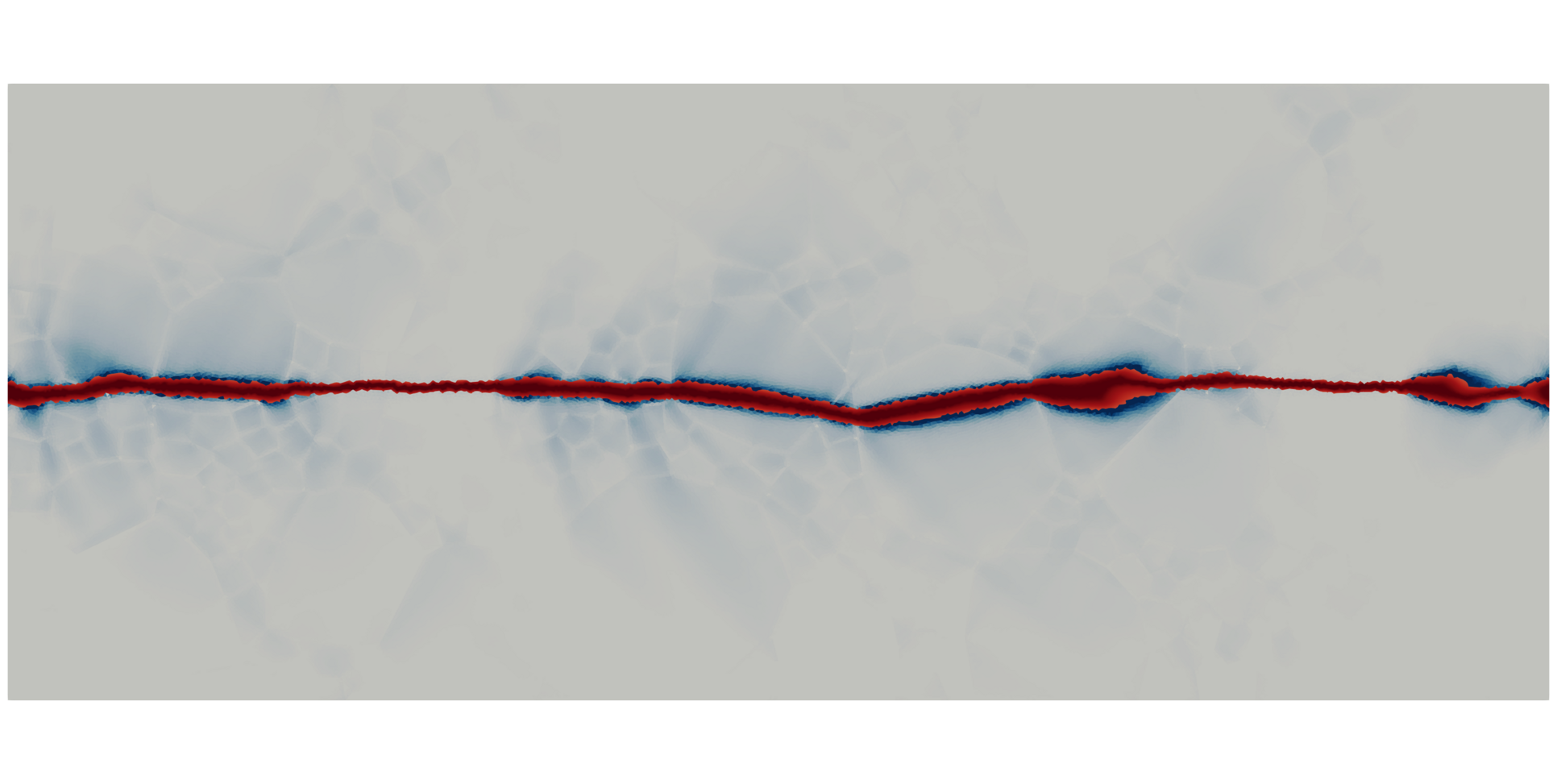}
		\label{subfig:surfing_tau0_2e-3_scale_2.0_mu1_0.1_mu2_0.025_damage}
	}
	\subfloat[$d_1 / d_2 = 4$]{  
		\includegraphics[width=0.46\textwidth,trim=0cm 3cm 1.2cm 3cm,clip]{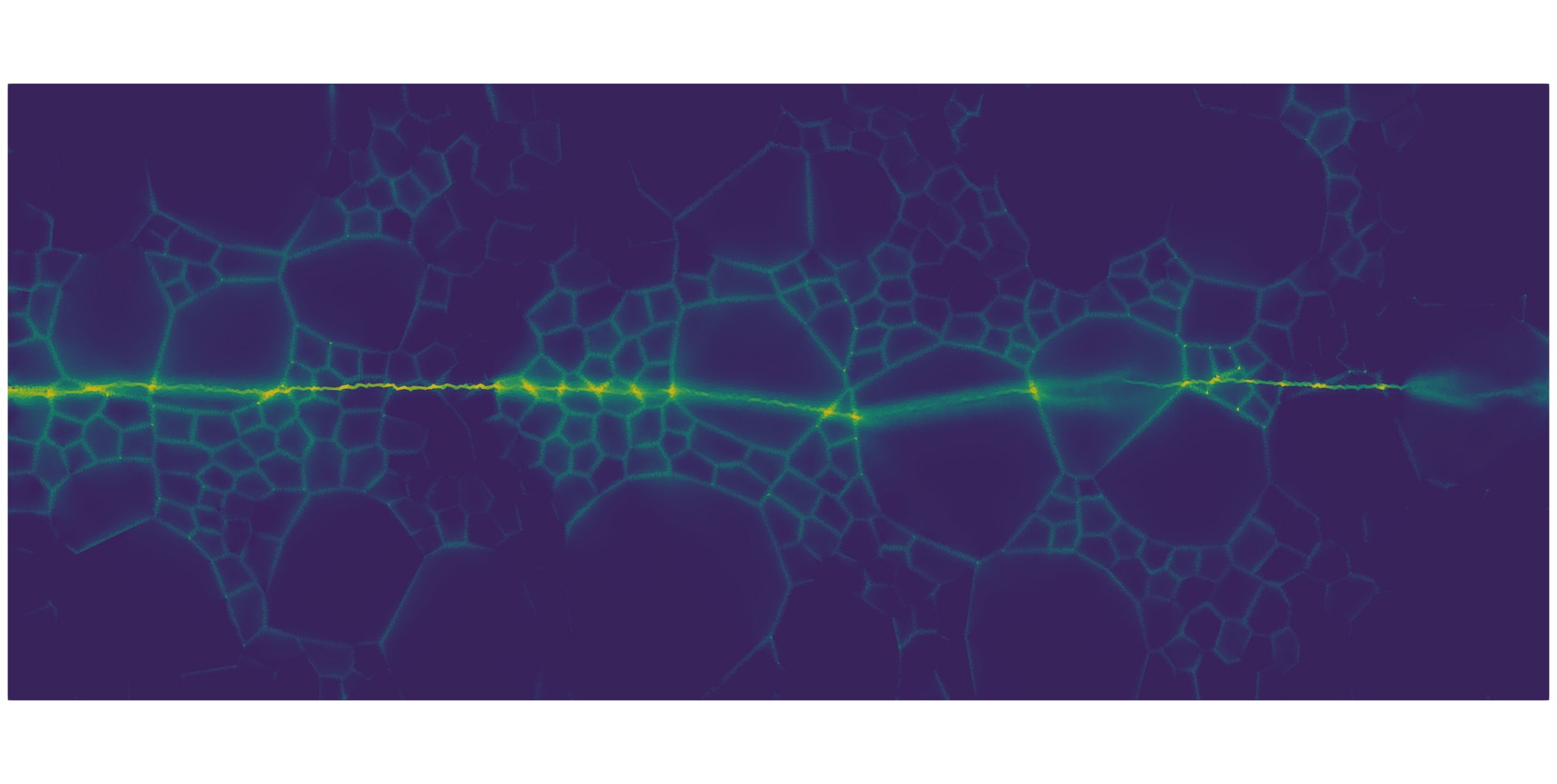}
		\label{subfig:surfing_tau0_2e-3_scale_2.0_mu1_0.1_mu2_0.025_dislocation_density}
	}
	\subfloat{  
	    \includegraphics[width=0.075\textwidth]{fig100.png}
	}
	\caption{Phase field, accumulated plastic slip field (a, c, e) and dislocation density field (b, d, f) in bimodal polycrystals submitted to surfing boundary conditions. The ductility ratio $q = r_{pl}/r_{pz}$ is equal to $5.0\times 10^2$ and the equivalent grain diameter of the first population is $d_1=\SI{20}{\micro\meter}$. Each population of grains occupies half of the surface area.}
	\label{fig:surfing_tau0_2e-3_scale_2.0_bimodal}
\end{figure}

The crack path, accumulated plastic slip and dislocation density fields are shown in Figure~\ref{fig:surfing_tau0_2e-3_scale_2.0_bimodal} for the three bimodal microstructures with $d_1/d_2 > 1$. The three microstructures display crack propagation with at least one crack jump. These unstable crack growth events systematically occur at grain boundaries, as the crack exits a grain belonging to the population with the larger grain size. Except for the microstructure with $d_1/d_2 = 3$ (Figure~\ref{subfig:surfing_tau0_2e-3_scale_2.0_mu1_0.1_mu2_0.033_damage}), the crack jumps at the interface between a large and a small grain. The crack then propagates through several grains of the smaller population. It is then pinned at a grain boundary either between two small grains (first jump in Figure~\ref{subfig:surfing_tau0_2e-3_scale_2.0_mu1_0.1_mu2_0.05_damage} and~\ref{subfig:surfing_tau0_2e-3_scale_2.0_mu1_0.1_mu2_0.025_damage}), two large grains (second jump in Figure~\ref{subfig:surfing_tau0_2e-3_scale_2.0_mu1_0.1_mu2_0.025_damage}) or a large and a small grain (Figure~\ref{subfig:surfing_tau0_2e-3_scale_2.0_mu1_0.1_mu2_0.033_damage}). 

All the crack jump events are preceded by phases of intense plastic dissipation at the crack tip. In Figure~\ref{subfig:surfing_tau0_2e-3_scale_2.0_mu1_0.1_mu2_0.033_damage} and~\ref{subfig:surfing_tau0_2e-3_scale_2.0_mu1_0.1_mu2_0.025_damage}, two of these crack jumps are preceded by the separation of the crack into two branches nuclei. After inititation of these branches, a single branch continues to propagate while the other stops. The resulting plastic activity relaxes the stresses at the crack tip and thus slows down the crack propagation. As a constant macroscopic crack tip velocity is imposed through the surfing boundary conditions (Eq.~\eqref{eq:surfing_bc}), the phase-field crack lags behind the macroscopic crack tip. The elastic energy built up ahead of the crack tip is eventually released in a sudden crack jump, thereby reducing the distance between the phase-field crack tip and the macroscopic crack tip positions.

In the microstructure with $d_1/d_2 = 2$ (Figure~\ref{subfig:surfing_tau0_2e-3_scale_2.0_mu1_0.1_mu2_0.05_damage}), the crack path exits the horizontal middle plane which is the macroscopically imposed propagation direction. This is because, locally, the crack path is slanted in each grain and bifurcates only at grain boundaries. As the grains are larger in the $d_1/d_2 = 2$ microstructure, the crack propagation direction is more prone to have a significant vertical component than for microstructures with smaller grains. 

The dislocation density fields show the difference of behaviour between the two populations of grains in each microstructure. The term $K_s / \delta$ in the dislocation densities evolution equations~\eqref{eq:dislocation_density}, leads to a larger dislocation density in the smaller grains. This effect is particularly visible in the microstructure with the smaller second population of grains, \textit{i.e.} $d_1/d_2 = 4$ (Figure~\ref{subfig:surfing_tau0_2e-3_scale_2.0_mu1_0.1_mu2_0.025_dislocation_density}). This larger dislocation density in smaller grains induce higher local stresses. This might be the reason why the earliest crack jump is observed in the microstructure with $d_1/d_2 = 4$ (Figure~\ref{subfig:surfing_tau0_2e-3_scale_2.0_mu1_0.1_mu2_0.025_damage}). 

In all three microstructures, the accumulated plastic slip is largest in the larger grains. On the contrary, the dislocation density is largest in the smaller grains. This suggests that the presence of a secondary population of larger grains in a fine grained microstructure might have a toughening effect on the material behaviour as the resulting plastic activity has a shielding effect on the crack tip.

\begin{figure}
	\centering
	\includegraphics[width=\textwidth]{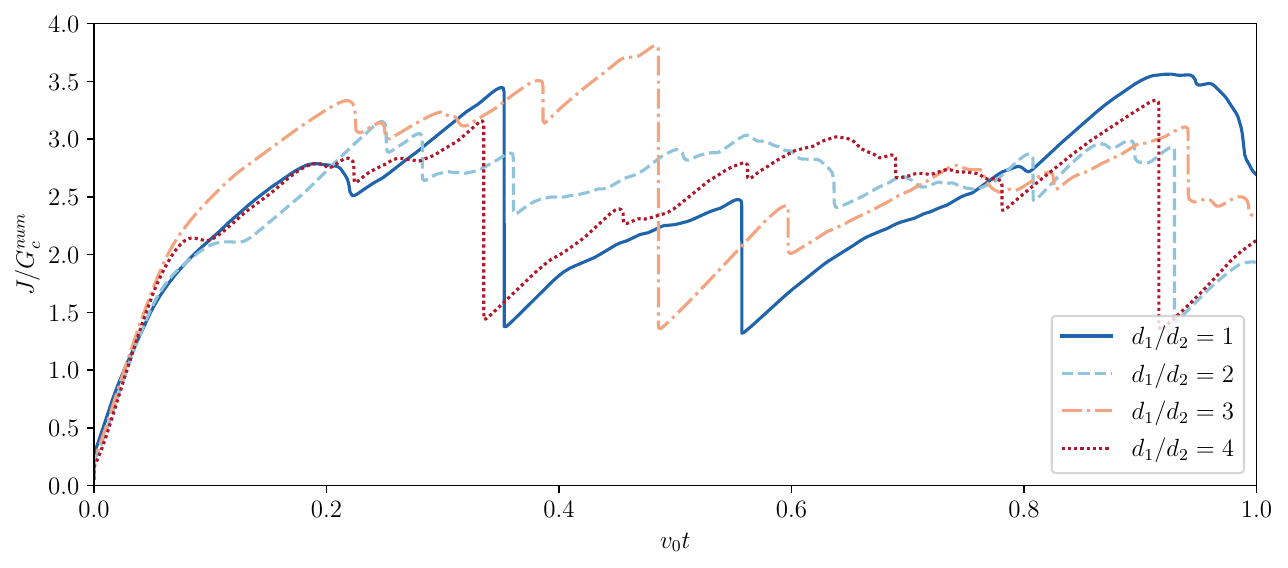}
	\caption{Normalized $J$-integral as a function of the macroscopic crack length for four different bimodal microstructures $d_1/d_2$, with $d_1=\SI{20}{\micro\meter}$ and a ductility ratio of $5.0 \times 10^2$. The $J$-integral is normalized by the numerical fracture toughness $G_c^{num} = G_c(1+3h/8\ell)$.}
	\label{fig:surfing_tau0_2e-3_scale_2.0_bimodal_jintegral}
\end{figure}

The $J$-integrals corresponding to the bimodal microstructures are shown in Figure~\ref{fig:surfing_tau0_2e-3_scale_2.0_bimodal_jintegral}. As discussed previously, sudden drops of the $J$-integral are associated to crack jumps. The peak values reached in the different microstructures are in the range $[3.2; 3.8]G_c^{num}$. The maximum peak value is reached in the microstructure with $d_1/d_2 = 3$ and the minimum peak value for $d_1/d_2 = 2$. In Figure~\ref{fig:surfing_grain_size_effect} we had noted that, in a microstructure with a single population of grains, decreasing the grain size leads to a decrease of the peak $J$-integral value. For an equiaxed microstructure, with a grain size $d = \SI{5}{\micro\meter}$, and $q = 5.0 \times 10^{2}$, the peak $J$-integral value is approximately $2.2G_c^{num}$. In a bimodal microstructure with half of the surface area occupied by grains of size $d_1 = \SI{20}{\micro\meter}$ and the other half by grains of size $d_2 = \SI{5}{\micro\meter}$ (\textit{i.e.} $d_1/d_2 = 4$), the peak $J$-integral value is approximately $3.3G_c^{num}$. This value is comparable to the peak $J$-integral value obtained for the unimodal microstructure with $d = \SI{20}{\micro\meter}$ in which $J_{max} \approx 3.6G_c^{num}$. The bimodal microstructure with $d_1=\SI{20}{\micro\meter}$ and $d_2=\SI{5}{\micro\meter}$ thus shows a 67\% increase of the peak $J$-integral with respect to the unimodal microstructure with a single population of grains with $d=\SI{5}{\micro\meter}$. This confirms the significant toughening effect of larger grains embeded in a fine grained microstructure.

In Appendix~\ref{sec:surfing_texture} we investigate the effect of the texture on the crack propagation. The effect of elongated grains in the longitudinal and transverse directions is compared to the equiaxed microstructure. The main findings are that the texture has a limited impact on the peak $J$-integral value. However, the crack propagation is smoother in the longitudinal textured microstructure compared to the equiaxed and transverse textured microstructures, where crack jumps are observed.

\section{Conclusion}
\label{sec:conclusion}

In this work, we address the non-monotonic behavior of the fracture resistance observed in plastic-brittle non-face-centered-cubic metals and alloys (see Figure~\ref{subfig:reiser_hartmaier} adapted from~\citep{reiser2020elucidating}).  We develop a computational model including both crystal plasticity and fracture, and focus on body centered cubic materials by combining crystal visco-plasticity through dislocation density evolution equations~\cite{hoc2001polycrystal} with  variational phase field approach ~\citep{francfort1998revisiting,bourdin2000numerical,bourdin2008variational} in finite deformation.  Our main results are summarized in Figures~\ref{subfig:hall_petch} and~\ref{subfig:inverse_hall_petch}, and these are consistent with the observed non-monotonic behavior.

We subject a representative volume consisting of a number of grains to uniaxial tension  in plane strain, and observe observe that both the initial yield and the peak stress display the classical Hall-Petch relation where the strength decreases with the (square-root of the) grain size -- see Figure~\ref{subfig:hall_petch}.  Importantly, we find that the peak stress is associated with the nucleation of cracks.  Thus, the crack nucleation threshold follows the Hall-Petch relationship.  Once a crack has nucleated it has to propagate through multiple grains, and we study this using surfing boundary conditions~\citep{hossain2014effective}.  We observe that the fracture toughness characterized by the critical energy release rate for propagation follows the inverse Hall-Petch relation where the toughness increases with the (square-root of the) grain size.  Thus, crack propagation follows the inverse Hall-Petch relationship.  Together, we conclude that it is difficult to nucleate cracks when the grain size is small, and to propagate cracks when the grain size is large, with easy nucleation and propagation at intermediate grain-size.  This explains the experimentally observed non-monotonic behavior in fracture.

Our computations also provide insights into the origin of this behavior.  As a polycrystal deforms, the anisotropy of the grains leads to inhomogeneous stress and plastic slip activity.  This inhomogeneous plastic slip gives rise to stress concentrations in regions where slip bands kink and this leads to crack nucleation.  The modified dislocation evolution equations and cross-hardening induce increased strain hardening in the vicinity of grain boundaries.  This leads to  a Hall-Petch grain size effect in both yield and failure stresses since inhomogeneous slip increases significantly as the grain size increases.  
This relationship has been widely  verified experimentally both for yield and failure strength~\citep{sasaki1975grain,schulson1983brittle}. We also observe an increase in ductility with decreasing the grain size, and this is also consistent with experimental evidence~\citep{hull1961effect,worthington1966slip,schulson1983brittle}.

In the case of propagation, the grain boundaries act as interfaces across which elastic and plastic properties change and can thus trap the cracks.  Therefore, crack propagation is characterized by alternating sequence of slow crack growth episodes interrupted by sudden crack jumps.   This is most pronounced when the grain size is large, and this leads to the inverse Hall-Petch relation ship.  Further, this effect is enhanced when the ductility ratio of the material is high and diminished when the ductility ratio is small.  This is again consistent with experimental observations~\citep{reiser2020elucidating}.

Finally, the paper also explores the role of bimodal grain size distribution and texture on crack propagation. Our results show that, transverse and longitudinal texture do not have a significant effect on the fracture toughness when the ratio between the longest and shortest size of the grain does not exceed 4. The crack propagation is more intermittent and tortuous for the transverse texture, while it is more continuous for the longitudinal texture. Significant toughening is achieved when a secondary population of coarse grains is embeded in a fine grain microstructure. These larger grains efficiently relax the stresses at the crack tip and hence increase the crack propagation resistance. As a result, materials with bimodal microstructures can exhibit high strength (Hall-Petch effect) while maintaining good fracture toughness.

\section*{Acknowledgement}
We gratefully acknowledge useful discussions with Sara Gorske, Peter Voorhees and Kathy Faber.  We gratefully acknowledge the financial support of the US Office of Naval Research (grant number: N00014-21-1-2784).  The work of BB is also supported by the Natural Sciences and Engineering Research Council of Canada (NSERC) grant RGPIN-2022-04536 and the Canada Research Chairs program.  The simulations reported here were conducted on the Resnick High Performance Computing Cluster at the California Institute of Technology.

\appendix

\section{Effect of texture}
\renewcommand{\thefigure}{A\arabic{figure}}
\setcounter{figure}{0}
\label{sec:surfing_texture}

\begin{figure}
	\centering
	\subfloat[$d_x/d_y = 0.25$ (transverse)] {
	    \includegraphics[width=0.7\textwidth,trim=0cm 0cm 0cm 1cm,clip]{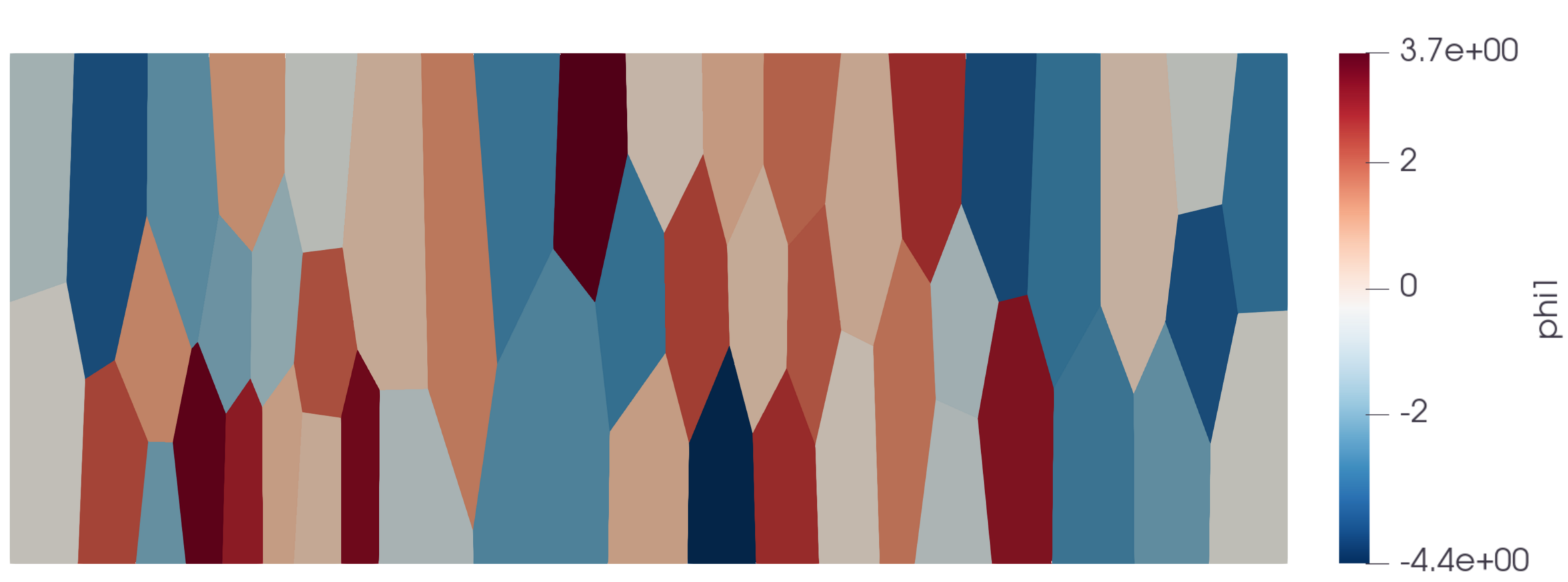}
		\label{subfig:surfing_microstructure_texture_0.25}
	}\\
	\subfloat[$d_x/d_y = 1$ (equiaxed)] {
	    \includegraphics[width=0.7\textwidth,trim=0cm 0cm 0cm 1cm,clip]{fig90.png}
		\label{subfig:surfing_microstructure_texture_1.0}
	}
	\\
	\subfloat[$d_x/d_y = 4$ (longitudinal)] {
	    \includegraphics[width=0.7\textwidth,trim=0cm 0cm 0cm 1cm,clip]{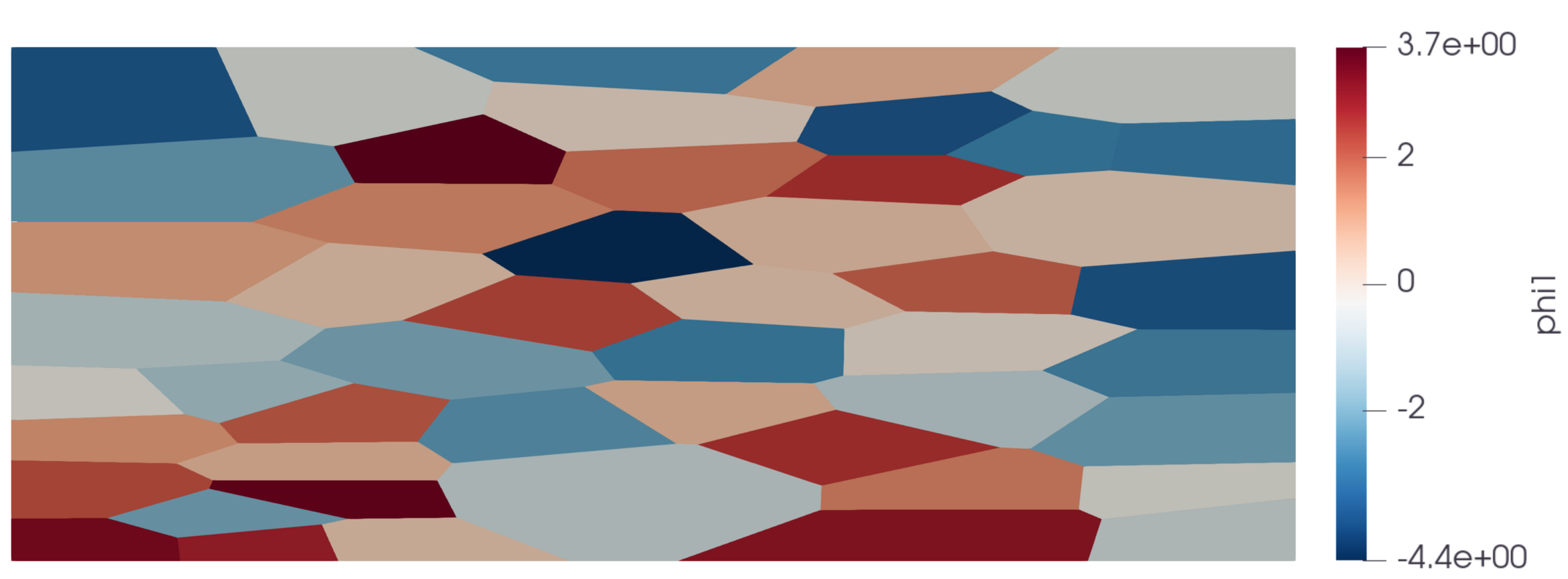}
		\label{subfig:surfing_microstructure_texture_4.0}
	}
	\caption{Textured polycrystal microstructures, with grain sizes $d_x$ in horizantal direction and $d_y$ in vertical direction, used to simulate crack propagation in plane strain conditions. The colorscale represents the first Euler angle $\phi_1$.}
	\label{fig:surfing_microstructure_texture}
\end{figure}

In this section, we investigate the effect of the texture of the polycrystal on crack propagation. In addition to the equiaxed microstructure shown in Figure~\ref{fig:surfing_microstructure}, two additional textured microstructures are considered with grains elongated in the horizontal and in the vertical directions respectively (see Figures~\ref{subfig:surfing_microstructure_texture_0.25} and~\ref{subfig:surfing_microstructure_texture_4.0}). The texture can be characterized by the ratio of the grain sizes in the horizontal and vertical directions as $d_x/d_y$. This ratio is varied from 0.25 to 4.0. The equiaxed microstructure, \textit{i.e.} $d_x/d_y=1$, used in previous sections was shown in Figure~\ref{fig:surfing_microstructure} and is replotted in Figure~\ref{subfig:surfing_microstructure_texture_1.0} to aid the comparison with textured microstructures. The equivalent grain diameter $d$ corresponding to a circular grain with identical surface area, is set to \SI{20}{\micro\meter}. Crack propagation simulations were carried out for these microstructures with a ductility ratio $q = 5.0 \times 10^{2}$.

\begin{figure}
	\centering
	\subfloat[$d_x / d_y = 0.25$ (transverse)]{  
		\includegraphics[width=0.46\textwidth,trim=0cm 3cm 1.2cm 3cm,clip]{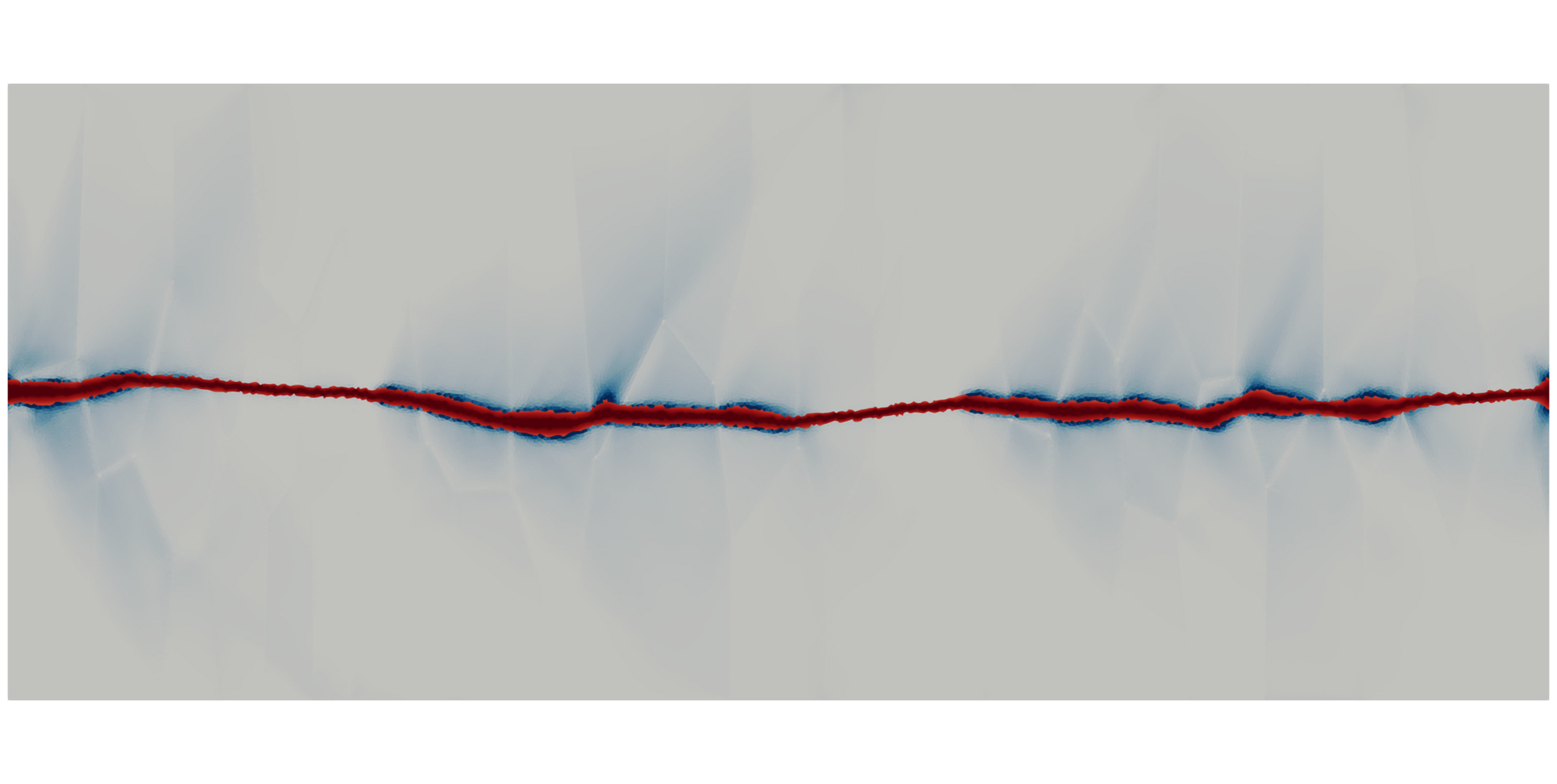}
		\label{subfig:surfing_r0.25_tau0_2e-3_scale_2.0_damage}
	}
	\subfloat[$d_x / d_y = 0.25$ (transverse)]{  
		\includegraphics[width=0.46\textwidth,trim=0cm 3cm 1.2cm 3cm,clip]{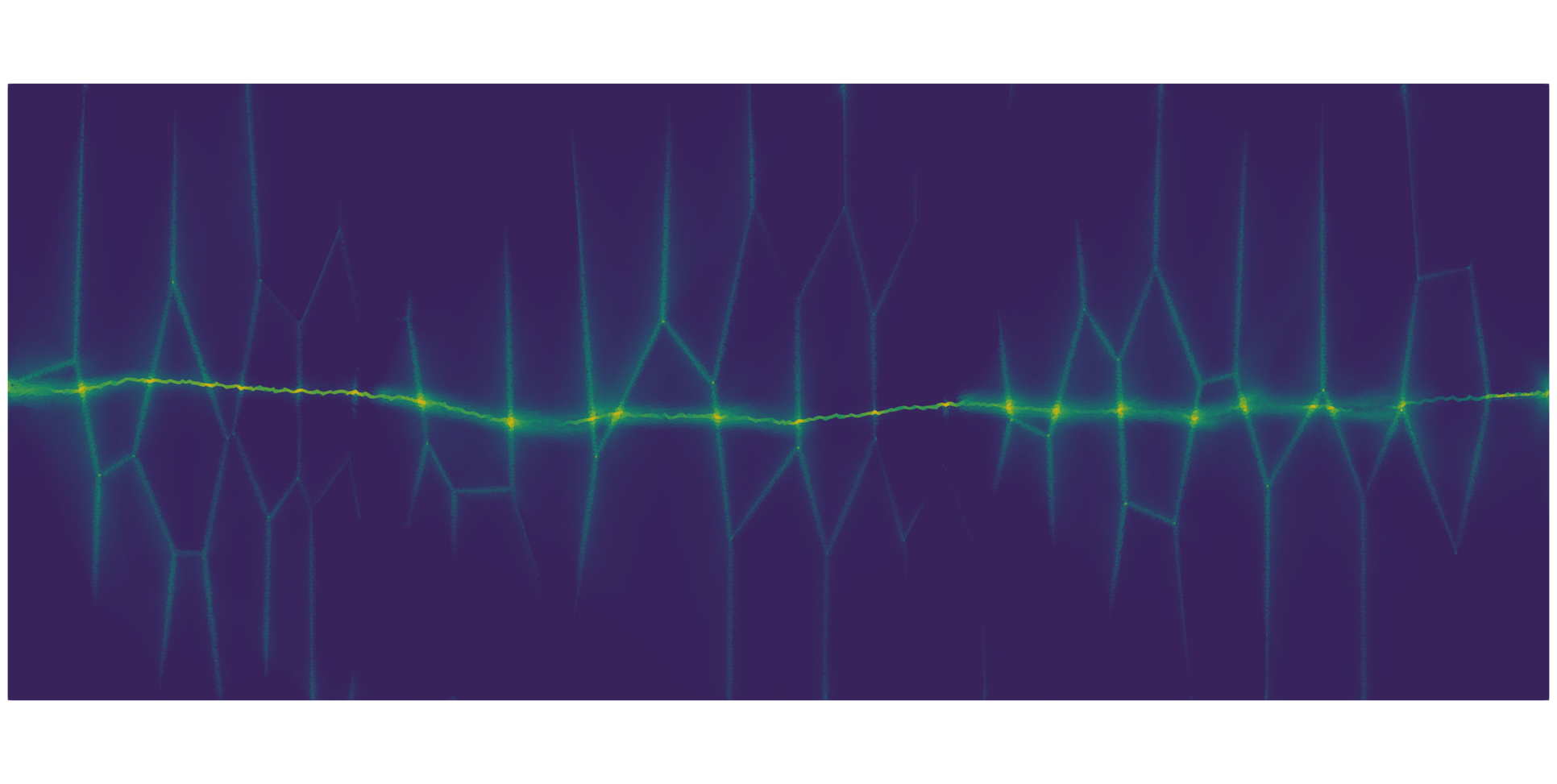}
		\label{subfig:surfing_r0.25_tau0_2e-3_scale_2.0_dislocation_density}
	}
	\subfloat{  
	    \includegraphics[width=0.076\textwidth]{fig94.png}
	}\setcounter{subfigure}{2}
    \\
	\subfloat[$d_x / d_y = 1$ (equiaxed)]{  
		\includegraphics[width=0.46\textwidth,trim=0cm 3cm 1.2cm 3cm,clip]{fig95.png}
		\label{subfig:surfing_r1.0_tau0_2e-3_scale_2.0_damage}
	}
	\subfloat[$d_x / d_y = 1$ (equiaxed)]{  
		\includegraphics[width=0.46\textwidth,trim=0cm 3cm 1.2cm 3cm,clip]{fig96.png}
		\label{subfig:surfing_r1.0_tau0_2e-3_scale_2.0_dislocation_density}
	}
	\subfloat{  
	    \includegraphics[width=0.077\textwidth]{fig97.png}
	}\setcounter{subfigure}{4}
    \\
	\subfloat[$d_x / d_y = 4$ (longitudinal)]{  
		\includegraphics[width=0.46\textwidth,trim=0cm 3cm 1.2cm 3cm,clip]{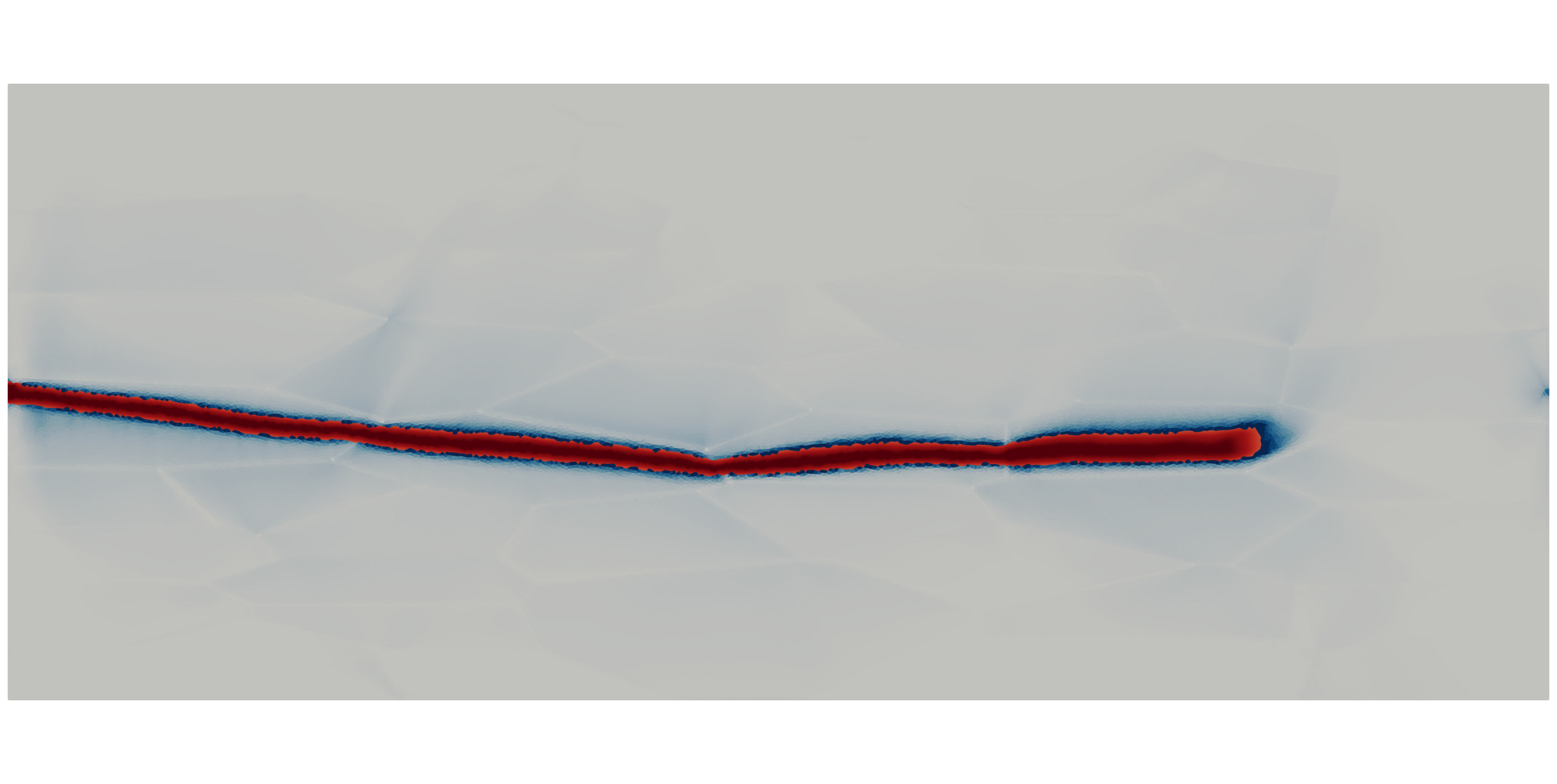}
		\label{subfig:surfing_r4.0_tau0_2e-3_scale_2.0_damage}
	}
	\subfloat[$d_x / d_y = 4$ (longitudinal)]{  
		\includegraphics[width=0.46\textwidth,trim=0cm 3cm 1.2cm 3cm,clip]{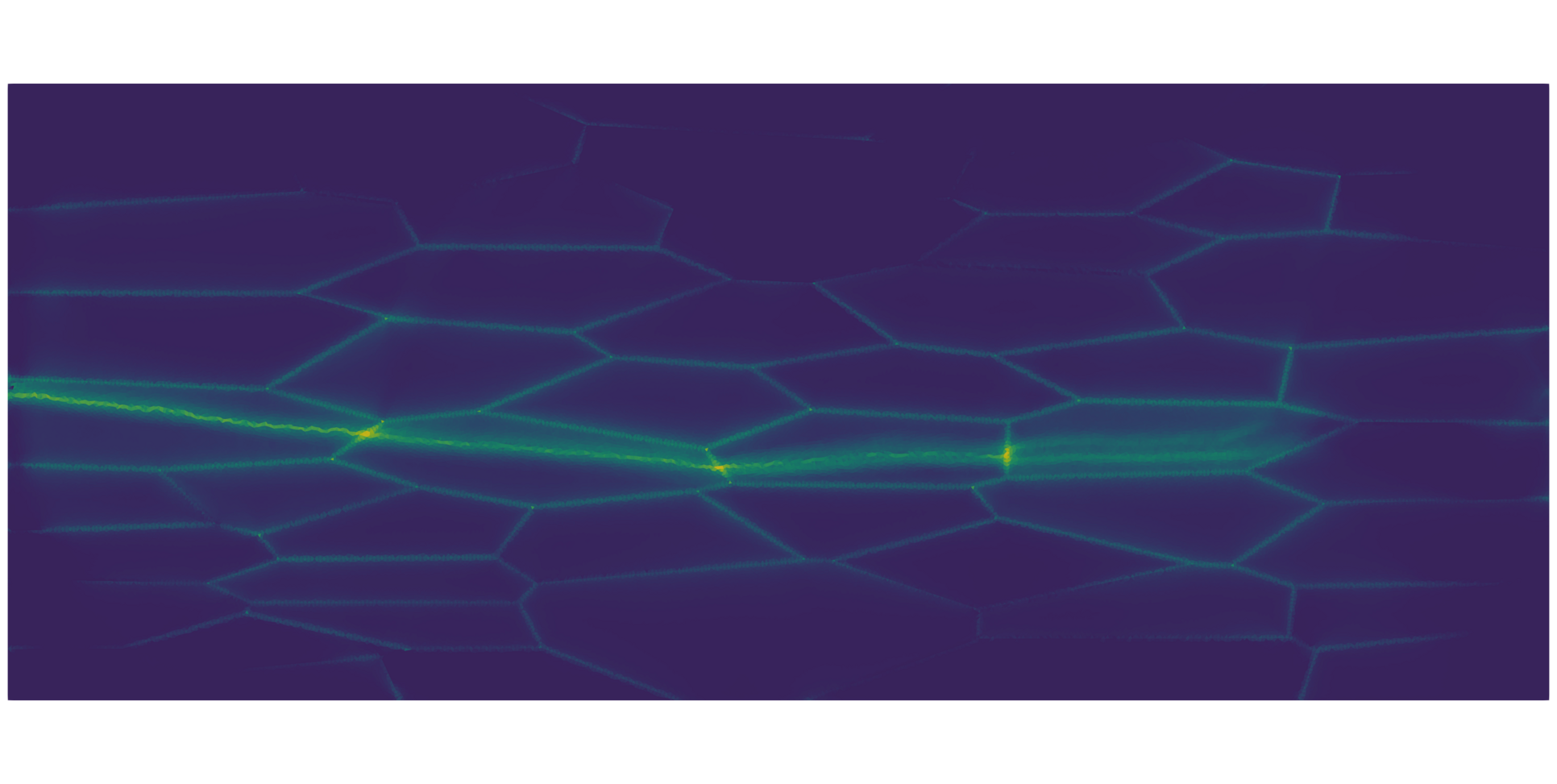}
		\label{subfig:surfing_r4.0_tau0_2e-3_scale_2.0_dislocation_density}
	}
	\subfloat{  
	    \includegraphics[width=0.075\textwidth]{fig100.png}
	}
	\caption{Phase field, accumulated plastic slip field (a, c, e) and dislocation density field (b, d, f) in textured polycrystals submitted to surfing boundary conditions. The ductility ratio $q = r_{pl}/r_{pz}$ is equal to $5.0\times 10^2$ and the equivalent grain diameter $d=\SI{20}{\micro\meter}$.}
	\label{fig:surfing_tau0_2e-3_scale_2.0_texture}
\end{figure}

The crack path, accumulated plastic slip and disloctation density fields are shown in Figure~\ref{fig:surfing_microstructure_texture} for the three microstructures. For the microstructure with grains elongated in the vertical direction, \textit{i.e.} $d_x/d_y = 0.25$ (transverse), the crack propagation is composed of periods of slow crack growth interrupted by sudden crack jumps. A similar crack growth mode was already observed for the equiaxed microstructure with $d_x/d_y = 1$ in Figure~\ref{subfig:surfing_tau0_2e-3_scale_2.0_damage} and is replotted in Figure~\ref{subfig:surfing_r1.0_tau0_2e-3_scale_2.0_damage} to aid the comparison with textured microstructures. The dislocation density plots in Figure~\ref{subfig:surfing_r0.25_tau0_2e-3_scale_2.0_dislocation_density} and~\ref{subfig:surfing_r1.0_tau0_2e-3_scale_2.0_dislocation_density} reveal a difference between the transverse texture and the equiaxed microstructure. For the latter, the crack jumps from a grain boundary to the next, crossing only a single grain. For the former, the crack jumps across two or three grains without getting arrested by the first grain boundary on its path. This suggest that grain boundaries might not always be strong enough barriers to arrest unstable crack growth. In addition, in the transverse textured microstructure, the crack path is more tortuous than in the equiaxed microstructure. This is because the grain size is smaller in the crack propagation direction and thus the crack could bifurcate each time it intersects with a new grain boundary. For the microstructure with grains elongated in the horizontal direction, \textit{i.e.} $d_x/d_y = 4.0$ (longitudinal), the crack propagation is more continuous. In contrast with the other two microstructures, unstable crack growth events are not observed. Despite the elongated shape of the grains, it is interesting to note that the crack remains predominantly transgranular. Since the grains are elongated in the horizontal direction, the crack crosses only 4 grain boundaries, while in the equiaxed microstructure it crossed 10 of them. The crack might have not yet encountered a grain that is not well aligned for plastic slip on its path. This could explain the continuous crack propagation and the absence of crack jumps.

\begin{figure}
	\centering
	\includegraphics[width=\textwidth]{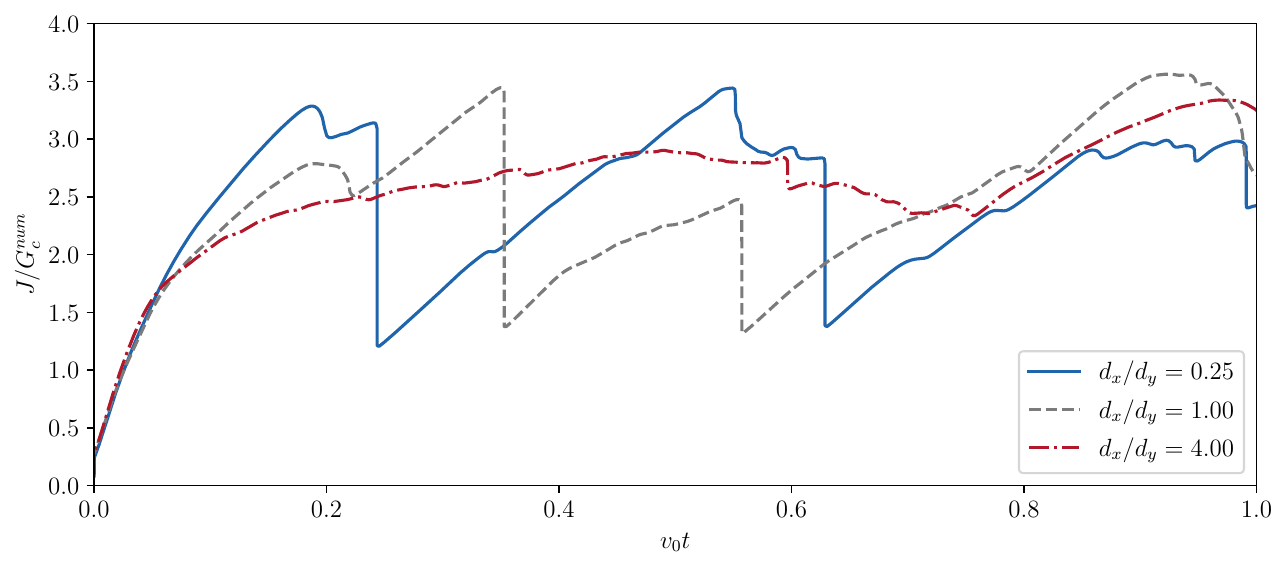}
	\caption{Normalized $J$-integral as a function of the macroscopic crack length for three different textures $d_x / d_y$, with and equivalent grain diameter $d=\SI{20}{\micro\meter}$ and a ductility ratio of $5.0 \times 10^2$. The $J$-integral is normalized by the numerical fracture toughness $G_c^{num} = G_c(1+3h/8\ell)$.}
	\label{fig:surfing_tau0_2e-3_scale_2.0_texture_jintegral}
\end{figure}

The normalized $J$-integral for these three microstructures are plotted in Figure~\ref{fig:surfing_tau0_2e-3_scale_2.0_texture_jintegral}. The texture does not seem to have a strong impact on the maximum value of the $J$-integral. However, the equiaxed and transverse textured microstructures display abrupt drops of the $J$-integral which are associated with crack jumps and renucleation. Conversly, the longitudinal textured microstructure shows a steadier evolution of the $J$-integral, because the crack propagates continuously without being pinned at grain boundaries. 

\bibliographystyle{abbrvnat}
\bibliography{bibliography.bib}

\end{document}